\documentclass [12pt,twoside] {book}

\usepackage{epsfig}

\usepackage{amssymb}
\usepackage{isolatin1}
\usepackage{geometry}
\usepackage{fancyhdr}
\usepackage{amsmath}
\usepackage{amsbsy}
\usepackage{amstext}
\usepackage{amsfonts}
\usepackage{tabularx}
\usepackage{hhline}
\usepackage{fancybox}
\usepackage{bbm}

\usepackage{diplom} 
\newcommand{\contract}{\negthickspace\negthickspace\,\slash\:}
\newcommand{\tr}{\text{\, tr}}
\newcommand{\projectorl}{\text{P}_\text{L}}
\newcommand{\projectorr}{\text{P}_\text{R}}
\newcommand{\Dslash}{D\negthickspace\negthickspace\slash\:}

\begin{document}

\thispagestyle{empty}

\begin{center}

$\left. \right.$

\Huge

Spontaneous CP Symmetry Breaking\\
in\\
Top-condensation Models\\
\vspace{4.0cm}
\large
Von der Fakult\"{a}t f\"{u}r Mathematik, Informatik und Naturwissenschaften der Rheinisch-Westf\"{a}lischen Technischen Hochschule Aachen zur Erlangung des akademischen Grades eines Doktors der Naturwissenschaften genehmigte Dissertation \\
\vspace{1cm}
vorgelegt von\\
\Large
Cristi\'{a}n Valenzuela Roubillard\\
\vspace{1cm}
\large
Mag\'{\i}ster en Ciencias\\
aus Santiago de Chile\\
\vspace{2cm}
\normalsize
\begin{tabular}{cl}
Berichter: & Universit\"{a}tsprofessor Dr. W.~Bernreuther  \\
 & Professor Dr. L.~M.~Sehgal \\
\end{tabular} \\
\medskip
Tag der m\"{u}ndlichen Pr\"ufung: 13.10.2004\\
\medskip
Diese Dissertation ist auf den Internetseiten der Hochschulbibliothek online verf\"{u}gbar\\
\end{center}

\cleardoublepage
\thispagestyle{empty}

\setcounter{page}{1}

\small
\tableofcontents
\listoffigures
\normalsize

\chapter{Introduction}

\setcounter{page}{1}

One of the major problems of present-day particle physics is to understand the mechanism which is responsible for electroweak gauge symmetry breaking (EWSB).
The standard model of particle physics (SM) assumes that this phenomenon is caused by the condensation of one type of Higgs bosons.
In view of the phenomenological success of the SM its prediction that there exists a neutral Higgs boson must be taken very seriously.
Nevertheless it is a fact that within the SM electroweak symmetry breaking is only parametrized but not explained.
Therefore theorists have been investigating extensions of the SM which offer a more satisfactory treatment of this phenomenon.
These include supersymmetric models and models of dynamical gauge symmetry breaking
-- see Section \ref{section:alternatives}.
It is believed that the experiments which will be made at future accelerators,
especially at the {\it Large Hadron Collider} and at the future planned electron-positron collider,
will permit a qualitative improvement in the understanding of this mechanism in the near future.

In this thesis we investigate a class of models known as top-condensation
models \cite{NAMBU89,Miransky:1988xi,Miransky:1989ds,Bardeen:1989ds,Cvetic:1997eb}.
Top-condensation models are special cases of models that aim at providing a dynamical reason for electroweak symmetry breaking \cite{Hill:2002ap}.
The aim of this work is to formulate a phenomenologically acceptable model within this class, in which the dynamical breaking of the electroweak symmetry causes also the breaking of the $CP$ symmetry.

The content of the thesis is as follows.
We start in this Chapter introducing the standard model, discuss its limitations, and thereby motivate its extentions.
In Chapter \ref{chapter:spontaneous_cpv} we introduce the concepts of spontaneous and dynamical $CP$ breaking.
Then in Chapters \ref{chapter:min}, \ref{chapter:eff_potential}, and \ref{chapter:1fam} minimal top-condensation models are studied within the Nambu-Jona-Lasinio approach.
Chapter \ref{chapter:3fam} contains the central part of this thesis.
Here, a three-generation model of dynamical gauge and $CP$ symmetry breaking is presented.
It is shown that this class of models leads,
for a range  of four-quark couplings,
to the correct ground state.
Thus, dynamical generation of gauge boson and quark masses,
quark flavor mixing angles,
and in particular,
dynamical generation of the $CP$-violating phase of the Cabibbo-Kobayashi-Maskawa matrix is realized.
In Chapter \ref{chapter:composite_condition} we make use of a complementary method based on the renormalization group in order to compute for a class of top-condensation models the values of the top-quark and spin-zero bound state (composite Higgs bosons) masses at low energies.
Finally in Chapter \ref{chapter:conclusion} we give a summary and present our conclusions.

\section{The Standard Model}

In this Section we review the standard model  of electroweak (EW) interactions.
In the first Subsection we discuss the electroweak gauge bosons and fermions.
Then in the second Subsection we review the mechanism of electroweak symmetry breaking.
For this purpose the SM postulates a scalar particle, the Higgs boson, which so far has not been observed in laboratory experiments.
The SM model does not predict the value of the Higgs boson mass.
It predicts, however, the couplings of this particle to gauge bosons, fermions, and its self-couplings in terms of the measured masses.
A major experimental effort at future colliders is therefore to search for the Higgs particle and, if it is found, to measure its couplings.
After that in Section \ref{section:alternatives} we briefly discuss some unsatisfactory aspects of the SM which motivate extensions of it, in particular, 
alternative models of electroweak symmetry breaking.

\subsection{Gauge Bosons and Fermions}\label{subsection:sm1}

The EW interactions are described with an impressive precision by a weakly-coupled gauge theory with gauge group $SU(2)_L\times U(1)_Y$.
(The indices $L$ and $Y$ refer to left-handed and the weak hypercharge, respectively.)
The four gauge bosons are the photon $\gamma$, the neutral $Z^0$ boson, and the charged $W^\pm$ bosons.
These particles are described by the four vector fields $A_\mu (x)$, $Z^0_\mu (x)$, and 
$W^\pm_\mu (x)$, respectively.
The experimental values of the gauge boson masses are\footnote{All experimental values are taken from \cite{PDG} unless stated otherwise.}

\begin{equation}\label{sm1} 
\begin{split}
m_\gamma &= 0\qquad (<6\times 10^{-17}\; eV),\\
m_Z   &= 91.1876\pm 0.0021 \; GeV,\\
m_W   &= 80.425\;\pm 0.038\; \; GeV.
\end{split}
\end{equation}

The fact that the $W^\pm$ and $Z^0$ bosons are massive is at first sight not compatible with gauge invariance because
mass terms for gauge bosons in the Lagrangian of the theory are not invariant under local gauge transformations.
In order to see this in more detail we present below the infinitesimal $SU(2)_L\times U(1)_Y$ gauge transformations of gauge fields.
We shall see in the next Subsection how the concept of a spontaneously broken symmetry solves this apparent incompatibility.

Vector fields appear naturally in a theory by demanding it to be invariant under local transformations.
These fields are used to construct gauge-invariant kinetic terms for fermions and bosons.
In the SM we associate to the gauge groups  $SU(2)$ and  $U(1)$ the
vector fields $A^a_\mu(x)$, $a=1,2,3$, and $B_\mu(x)$, respectively.
As we see below, linear combinations of these fields correspond to the mass eigenstate fields $A_\mu (x)$, $Z^0_\mu (x)$, and $W^\pm_\mu (x)$.
Under infinitesimal $SU(2)\times U(1)$ gauge transformations the gauge fields transform as

\begin{equation}\label{sm2} 
\begin{split}
B_\mu(x) &\longrightarrow B_\mu(x) +\frac{1}{g_1} \partial_\mu \beta(x), \\
A^a_\mu(x) &\longrightarrow A^a_\mu(x) +\frac{1}{g_2} \partial_\mu \alpha^a(x)
			+\epsilon^{abc} A^b_\mu \alpha^c(x),
\end{split}
\end{equation} 
where $g_1$ and $\beta(x)$ are the gauge coupling constant and the gauge transformation parameter related to the $U(1)$ group.
For the non-abelian $SU(2)$ group the corresponding expressions are given by
$g_2$ and $\alpha^a(x)$, respectively.
In this case the gauge group possesses three generators and hence three gauge fields and three transformation parameters numerated by $a=1,2,3$.
Besides, the Lie algebra of the $SU(2)$ group is given by 

\begin{equation}\label{sm3} 
[T^a,T^b]=i\,\epsilon^{abc}\, T^c,
\end{equation} 
where $T^a$, with $a=1,2,3$, denote the generators of the group and
the structure constants $\epsilon^{abc}$
are chosen completely antisymmetric with the normalization $\epsilon^{123}=+1$.
Now we can see from Eqs. (\ref{sm2}) that gauge boson mass terms, for instance $\frac{1}{2}\, m_B^2 B^\mu B_\mu$, are not gauge invariant.

Gauge fields allow to construct covariant derivatives and with them gauge-invariant kinetic terms for fermions and bosons. 
The $SU(2)\times U(1)$ gauge transformation of a fermionic or bosonic field $\Psi$ in the fundamental representation  of the $SU(2)$ and with a $U(1)$ charge $Y$ is given by

\begin{equation}\label{sm4a} 
\Psi(x) \longrightarrow 
e^{i \alpha^a \sigma^a/2}\; e^{i Y\beta(x)} \; \Psi(x),
\end{equation}
which for infinitesimal transformation parameters can be approximated to

\begin{equation}\label{sm4} 
\Psi(x) \longrightarrow 
\left( 1+i \alpha^a \sigma^a/2+i Y\beta(x) \right) \; \Psi(x),
\end{equation} 
where $\sigma^a$, with  $a=1,2,3$, are the Pauli matrices 
($\sigma^a/2$ fulfill Eq. (\ref{sm3})).
The covariant derivatives acting on these fields are defined as:

\begin{equation}\label{sm5} 
D_\mu=\partial_\mu -ig_2 A^a_\mu(x) \sigma^a /2 - ig_1 Y B_\mu(x).
\end{equation} 

Covariant derivatives of the field $\Psi$ are objects which contain a derivative of the field $\Psi$ and transform under gauge transformations like the field itself:

\begin{equation}\label{sm6} 
D_\mu\Psi(x) \longrightarrow 
\left( 1+i \alpha^a \sigma^a/2+i Y\beta(x) \right) \; D_\mu\Psi(x).
\end{equation} 
Using these objects it is possible to write gauge-invariant kinetic terms 

\begin{equation}\label{sm7} 
\bar{\Psi}\,i\Dslash\Psi(x),
\qquad
(D_\mu\Psi)^\dagger (D^\mu\Psi),
\end{equation} 
for fermions and scalars, respectively.

In the SM there are three generations of fermions. 
Each generation contains leptons, which do not feel the strong interactions directly, and quarks that do.
In each lepton generation there is a left-handed $SU(2)$ doublet and a right-handed $SU(2)$ singlet 
(we ignore here the recently established fact that at least two of the three neutrinos are massive).
Similarly, each quark generation contains one left-handed $SU(2)$ doublet and two right-handed $SU(2)$ singlets.
In addition quarks transform under the $SU(3)_\text{color}$ gauge group in the fundamental representation and couple to gluon fields as described by Quantum Chromodynamics (QCD).
The weak $U(1)$ hypercharges $Y$ of the SM fermions are given in Table \ref{table:hypercharges}.
Apparently no logical structure governs them. 
However they appear in nature in such a way that the theory has no anomalies related to the gauge symmetry currents.
A weak hypercharge assignment like this can naturally be implemented in grand unified theories (GUTs).

\begin{table}[tbp]
\setlength{\extrarowheight}{4pt}
\begin{equation*}
\begin{array}{|!{\;}c!{\;}|!{\;}c!{\;}|!{\;}c!{\;}|}
\hline 
 & Y & Q=T^3-Y    \\
\hline\hline
\psi_{iL} & -1/6 &   \begin{matrix}
            \;\;\: 2/3     \\ 
              -1/3
            \end{matrix}       \\   
\hline
u_{iR} &  -2/3 &   \;\;\: 2/3   \\
\hline
d_{iR} &  \;\;\: 1/3 &   -1/3   \\
\hline
L_{iL} &  1/2 &   \begin{matrix}
             \;\;\: 0      \\ 
               -1
             \end{matrix}          \\
\hline
e_{iR} &  1 &   -1   \\
\hline
\end{array} 
\end{equation*}
\caption{Table showing the weak hypercharges $Y$ and the corresponding electric charges $Q$ of the left- and right-handed fermions of the SM.}
\label{table:hypercharges}
\end{table}

We also quote the masses of the SM fermions.
The experimental values of the lepton masses are (without error bars) 

\begin{equation}\label{sm8}
\begin{matrix}
m_e    &=& \qquad\quad 0.511\;  MeV, \\
m_\mu  &=& \;\: 106 \;  MeV,  \\
m_\tau &=& 1777 \;   MeV,
\end{matrix}
\end{equation}
and those of the quark masses are\footnote{The value of the top-quark mass was taken from a recent reanalysis of direct observation of top events \cite{mtop}.}$^{,}$\footnote{The values of the $u$, $d$, and $s$ quark masses are estimates of so-called ``current-quark masses", in a mass-independent subtraction scheme such as $\overline{\sf{MS}}$ at a scale $\mu\approx 2\; GeV$. The $c$ and $b$ quark masses  are the ``running" masses at $\mu=m_c,\; m_b$ in the $\overline{\sf{MS}}$ scheme.}

\newcommand{\ttaux}{\;\text{to}\;}

\begin{equation}\label{sm9}
\begin{matrix}
m_u &=& 1.5 \ttaux 4\; MeV,  & m_c &=& 1.15 \ttaux 1.35\;  GeV, &  
m_t &=& 178.0 \pm 4.3\; GeV, \\
m_d &=& 4 \ttaux 8  \; MeV,  & m_s &=& 80 \ttaux 130\;MeV, &  
m_b &=& 4.1 \ttaux 4.4 \;  GeV.
\end{matrix}
\end{equation}
Note that the heaviest fermion is about 5 orders of magnitude heavier than the lightest one 
(without considering neutrino masses which would increase this ratio).

Like the gauge boson mass terms, SM fermion mass terms are not invariant under $SU(2)_L\times U(1)_Y$ gauge transformations.
Fermion mass terms are invariant only under transformation which treat the right- and left-handed fermion fields in the same way.
The four-dimensional $SU(2)_L\times U(1)_Y$ gauge group has a one-dimensional subgroup
of this type, namely the $U(1)$ electromagnetic one.
This is related to the long-range electromagnetic force,
mediated by the massless photon and is an exact symmetry of nature.

Using the gauge-invariant fermionic kinetic term given in Eq. (\ref{sm7}),
with the weak hypercharges taken from Table \ref{table:hypercharges} we obtain the SM fermion kinetic terms (without considering the strong interactions):

\begin{equation}\label{sm9a}
\mathcal{L} \supset 
\bar{L}_{jL}\,i\Dslash L_{jL} + \bar{e}_{jR}\,i\Dslash e_{jR} +
\bar{\psi}_{jL}\,i\Dslash\psi_{jL} + \bar{u}_{jR}\,i\Dslash u_{jR}
						+ \bar{d}_{jR}\,i\Dslash d_{jR},
\end{equation} 
where $L_{jL}=(\nu_{jL},\, e_{jL})^T$ and $\psi_{jL}=(u_{jL},\, d_{jL})^T$,
with $j=1,2,3$, are the left-handed leptons and quarks, respectively.
The right-handed leptons are denoted by $e_{jR}$, while $u_{jR}$ and $d_{jR}$ correspond to the right-handed quarks. 
The previous quark fields are in the weak basis.
As we see later, 
they are related to the mass eigenstate quark fields by a chiral transformation.

Before turning to the spontaneous symmetry breaking mechanism of the SM we consider the kinetic terms for the gauge fields $A^a_\mu(x)$ and $B_\mu(x)$.
The abelian $U(1)_Y$ and non-abelian $SU(2)_L$ groups possess the following Yang-Mills terms:

\begin{equation}\label{sm10}
\mathcal{L}_{\text{gauge}} = -\frac{1}{4} H^{\mu\nu}H_{\mu\nu}
						-\frac{1}{4} F^{a\mu\nu}F^a_{\mu\nu},
\end{equation}    
with

\begin{equation}\label{sm11}
\begin{split}
H_{\mu\nu} &= \partial_\mu B_\nu-\partial_\nu B_\mu  ,    \\
F^a_{\mu\nu} &= \partial_\mu A^a_\nu-\partial_\nu A^a_\mu 
				+g_2\:\epsilon^{abc} A^b_\mu A^c_\nu.
\end{split}
\end{equation} 
The non-abelian part of Eq. (\ref{sm10}) contains triple gauge couplings (proportional to $g_2$) and quartic gauge couplings (proportional to $g_2^2$).
There is a strong evidence, 
remarkably from electron-positron scattering experiments at LEP \cite{lep-gc}, 
that the gauge bosons with masses given in Eq. (\ref{sm1}) self-interact as dictated by the Lagrangian given in Eq. (\ref{sm10}).
The relations between the mass eigenstates fields
$A_\mu (x)$, $Z^0_\mu (x)$, and $W^\pm_\mu (x)$
and the fields $A^a_\mu(x)$ and $B_\mu(x)$ are given by

\begin{equation}\label{sm12}
\begin{split}
A_\mu     &=\frac{1}{\sqrt{g_1^2+g_2^2}}\: (g_1 A_u^3- g_2 B_\mu), \\
Z_\mu^0   &=\frac{1}{\sqrt{g_1^2+g_2^2}}\: (g_2 A_u^3+ g_1 B_\mu),     \\
W_\mu^\pm &=\frac{1}{\sqrt{2}}\: (A_u^1\mp i A_u^2).
\end{split}
\end{equation} 
The Yang-Mills Lagrangian (Eq. (\ref{sm10})) predicts the structure and strength
(in terms\footnote{Approximate values of the gauge coupling constants at scales $\mu\approx 0$ are given by $g_1^2/4\pi = 0.01$ and $g_2^2/4\pi = 0.03$.}
of $g_1$ and $g_2$, or equivalently, in terms of $e$ and $\sin\theta_w$)
of the interactions of the electroweak gauge bosons among each other.
The following vertices exist at tree level:

\begin{equation}\label{sm13}
\begin{array}{c}
W^+ W^- \gamma ,\; W^+ W^- Z^0 ,\; W^+ W^- \gamma \gamma, \\
W^+ W^- Z^0 Z^0  ,\; W^+ W^- \gamma Z^0 ,\; W^+ W^-  W^+ W^-.
\end{array} 
\end{equation} 
The $W^+ W^- \gamma$ and $W^+ W^- Z^0$ interactions and thus the non-abelian nature of the $W^\pm$, $Z^0$ boson interactions were confirmed by experiments at LEP
which investigated $W$-boson pair production: $e^+\,e^- \rightarrow W^+ W^-$
\cite{lep-gc}.

\subsection{Standard Model Higgs Mechanism}\label{subsection:sm2}

We reviewed in the last Subsection that
the $SU(2)_L\times U(1)_Y$ gauge symmetry,
which dictates the dynamics of the EW sector,
is spontaneously broken to the electromagnetic $U(1)$ group.
This can be seen from the fact that the weak gauge bosons and the quark and leptons are not massless.
Classically speaking, a spontaneously broken symmetry is a symmetry of the equations of motion but not a symmetry of the solution of these equations.
A well known example of this phenomenon
from condensed matter is:
Although the Lagrangian of a ferromagnetic material is invariant under spatial rotations,
at temperatures below a critical one
the spins of the molecules tend to point in the same direction, breaking spontaneously the rotational symmetry of the Lagrangian.
The symmetry breaking takes place because the non-symmetric configuration is energetically favorable.
A similar situation occurs in the SM where the vacuum (or ground state) of the theory does not respect three of the four $SU(2)_L\times U(1)_Y$ gauge symmetries of the Lagrangian.

The SM model is a renormalizable weakly-coupled theory where
the gauge boson and fermion mass terms appear in the gauge-invariant Lagrangian as a consequence of the condensation of an elementary scalar field, the Higgs field $H$.
This field $H$ has a weak hypercharge $Y_H=1/2$, is a doublet under the $SU(2)_L$ group, and a singlet under the $SU(3)_\text{color}$ group transformations.
A vacuum expectation value (VEV) of this field different from zero,
i.e. $<\Omega|\:  H\: |\Omega>\neq 0$, causes exactly the symmetry breaking pattern described in the last Subsection.

All possible gauge-invariant hermitian operators of dimension four or less which involve the Higgs field form part of the Higgs Lagrangian:

\begin{equation}\label{sm14}
\mathcal{L}_{\text{Higgs}}= (D_\mu H)^\dagger (D^\mu H)
	   +\mathcal{L}_{\text{Yukawa}} - V.
\end{equation} 
The first is the Higgs kinetic term, the second contains the Higgs-fermion interaction, and $V$ is the Higgs potential, a function only of the Higgs field:

\begin{equation}\label{sm15}
V(H)=-\mu^2\, H^\dagger H+\frac{\lambda}{2}\, (H^\dagger H)^2.
\end{equation} 
The potential possesses two parameters, $\mu^2$ and $\lambda$.
In order for $V(H)$ to be bounded from below $\lambda$ must be bigger than zero.
If $\mu^2<0$ the minimum of the potential is located at $H=0$, which is a field configuration
invariant under $SU(2)_L\times U(1)_Y$ gauge transformations.
In this case the EW symmetry remains unbroken.
On the other hand for $\mu^2>0$ the minimum is located at 
$H^\dagger H=\mu^2/\lambda$.
A VEV of $H$ different from zero is not invariant under general EW gauge transformations but only under a $U(1)$ subgroup of them.
Three of the four symmetry generators of the group are broken in this way.
This corresponds to the situation observed in nature.

We consider in the following the theory in the unitarity gauge where three of the four real components of the field $H$ are gauged away:

\begin{equation}\label{sm16}
H(x)=\frac{1}{\sqrt{2}}
\begin{pmatrix}
v+h(x) \\ 
0
\end{pmatrix} ,
\end{equation}
where the factor $1/\sqrt{2}$ provides the real field $h(x)$ with the conventional normalization.
The constant $v=\sqrt{2\mu^2/\lambda}$ corresponds to the VEV of the Higgs field and the field $h(x)$,
with VEV $<\Omega|\, h(x)\, |\Omega>= 0$,
is the only remaining scalar degree of freedom in this gauge.
Inserting the Higgs VEV into the kinetic term given in Eq. (\ref{sm14}) we obtain the gauge boson mass terms:

\begin{equation}\label{sm17}
\mathcal{L}_{\text{gauge-mass}} = \frac12\: m_Z^2\; Z^0_\mu  Z^{0\mu}
									+ m_W^2\; W^+_\mu  W^{-\mu},
\end{equation}
where

\begin{equation}\label{sm18}
\begin{split}
m_Z &= \sqrt{g_1^2+g_2^2}\;\frac{v}{2}, \\
m_W &= g_2\;\frac{v}{2}.
\end{split}
\end{equation} 
In order to get the observed gauge boson masses the Higgs VEV must be 
$v=246 \; GeV$.
This can be seen as the only dimensionful parameter of the SM and is responsible for setting the EW scale.
The whole Higgs kinetic term is given by

\begin{equation}\label{sm17a}
(D_\mu H)^\dagger (D^\mu H)=\frac12\, \partial_\mu h\: \partial^\mu h+
\left[\frac12\: m_Z^2\: Z^0_\mu  Z^{0\mu}+ m_W^2\: W^+_\mu  W^{-\mu}\right] 
\left( 1+\frac{h}{v}\right)^2.
\end{equation} 
Besides the gauge boson mass terms this includes an ordinary kinetic term for the real field $h(x)$ and gauge boson-Higgs interactions.

We turn to the Higgs potential in the unitarity gauge.
Inserting Eq. (\ref{sm16}) into Eq. (\ref{sm15}) we get the Higgs potential as a function of the field $h(x)$ which describes displacements around the potential's minimum

\begin{equation}\label{sm15a}
V(h)=\,\frac12\, m^2_h\, h^2 + \frac{\lambda}{2} v h^3 +\frac{\lambda}{8} h^4,
\end{equation}
where we neglected a constant term.
In this tree level treatment the mass of the Higgs particle is given by

\begin{equation}\label{sm15b}
m_h=v \sqrt{\lambda}.
\end{equation} 
The other two terms of the potential are the triple and quartic Higgs self-couplings.
The mass of the Higgs particle, together with $v$, completely fixes the Higgs potential. 
Once one knows the mass of the SM Higgs boson, the triple and quartic Higgs self-couplings are predictions of the theory.

What is known about the SM Higgs particle from experiment?
The direct search for the production of the Higgs particle at LEP led to the lower bound \cite{PDG}:
$m_h>114.4 \; GeV$.
On the other hand, calculating electroweak precision observables including quantum corrections within the SM and using these formulae in global fits to the corresponding experimental data yields an allowed range for the SM Higgs boson mass.
With the recently updated value of the mass of the top quark such a fit yields the upper bound \cite{mtop}:
$m_h<251 \; GeV$ at the 95\% confidence level.

Finally, the Yukawa term of Eq. (\ref{sm14}) is given by

\begin{equation}\label{sm20}
\mathcal{L}_{\text{Yukawa}} = -
(\;g_{kl}\:\bar{\psi}_{kL}u_{lR}\:H +
 h_{kl}\:\epsilon^{ab}\:\bar{\psi}_{kL}^a d_{lR}\:H^{b*} +
 l_{kl}\:\epsilon^{ab}\:   \bar{L}_{kL}^a e_{lR}\:H^{b*}
\;+\; h.c.\;),
\end{equation}
where the Yukawa couplings $g_{kl}$, $h_{kl}$, and $l_{kl}$ are 
three completely arbitrary $3\times 3$ matrices.
Inserting Eq. (\ref{sm16}) into $\mathcal{L}_{\text{Yukawa}}$ we get mass terms for fermions.
Because the Yukawa matrices are arbitrary, the fermion mass terms are in general non-diagonal in flavor space.
In order to diagonalize the fermion mass matrix 
it is necessary to perform unitary rotations of the chiral fields.
For the quarks these rotations are given by

\begin{equation}\label{sm21}
\begin{matrix}
u_{iR} = W_{ij}^u \;u_{jR}^{'}, & & u_{iL} = U_{ij}^u \;u_{jL}^{'},\\
&&\\
d_{iR} = W_{ij}^d \;d_{jR}^{'}, & & d_{iL} = U_{ij}^d \;d_{jL}^{'},
\end{matrix}
\end{equation}
where $U^u$, $U^d$, $W^u$, $W^d$ are basis transformation matrices, and
the primed fields denote the quark fields in the mass basis.
In the leptonic sector we diagonalize  $l_{kl}$ by means of similar transformations.
After have diagonalized the Yukawa matrices 
Eq. (\ref{sm20}) is given by

\begin{equation}\label{sm20a}
\mathcal{L}_{\text{Yukawa}} = 
-\sum_f\; m_f\, \bar{f}^{'}f^{'} \left(1+\frac{h}{v}\right),
\end{equation}
where $-m_f\, \bar{f}^{'}f^{'}$ are mass terms for the SM fermions.
The values of the masses are given by $m_f=\lambda_f\, v/\sqrt{2}$,
where $\lambda_f$ denote the eigenvalues of the three Yukawa matrices.
Besides, in Eq. (\ref{sm20a}) we see fermion-Higgs interaction terms with coupling constants proportional to the fermion masses.\\

The Lagrangian of the SM is not invariant under the chiral rotation we defined in Eq.
(\ref{sm21}).
In addition to the  quark mass terms, there are two other terms in the
Lagrangian which are not invariant under this transformation,
the QCD $\theta$-term
and the term containing the charged current.

The QCD $\theta$-term is a dimension 4 operator which is non-invariant under a discrete $P$ (parity) transformation and under $CP$ (charge conjugation together with parity) transformations:

\begin{equation}\label{sm22}
\mathcal{L}_{\theta-\text{term}}=\theta\;\frac{\alpha_\text{strong}}{8\pi}\:
G_{a\mu\nu}\tilde{G}^{\mu\nu}_a,
\end{equation} 
where the parameter $\theta$ sets the strength of the interaction, $\alpha_\text{strong}$ is the strong coupling constant, 
$G^{\mu\nu}_a$ is the gluon field strength tensor, and $\tilde{G}^{\mu\nu}_a$ its dual tensor.
This operator reflects the non-trivial structure of the of the QCD vacuum.

As a consequence of the chiral anomaly, the chiral transformation of the quark fields given in Eq. (\ref{sm21}) modifies the parameter $\theta$:

\begin{equation}\label{sm23}
\theta\longrightarrow \bar{\theta}=\theta +\arg\det \mathcal{M},
\end{equation} 
where $\mathcal{M}$ is the quark mass matrix in the weak basis.
The QCD $\theta$-term induces $CP$-violating effects in flavor diagonal quark amplitudes, in particular an electric dipole moment of the neutron. 
For this observable a very stringent upper bound has been obtained experimentally.
This gives a very stringent upper bound for the parameter $\bar{\theta}$,
namely $|\bar{\theta}|\leq 10^{-9}$.

Why the value of $\bar{\theta}$ is so small or maybe zero, including an interplay between $\theta$ and $\mathcal{M}$ in Eq. (\ref{sm23}), is still an open question.
This is called the {\it strong $CP$ problem} \cite{STRONGCP}.\\

The other part of the Lagrangian which is not invariant under chiral transformations of the quark fields is the charged current.
From Eq. (\ref{sm9a}) we get the interaction term between the charged $W^\pm$ bosons and the left-handed quarks.
The gauge boson $W^+$ couples to the current

\begin{equation}\label{sm24a}
J^{\mu}_W = \frac{1}{\sqrt{2}}\;\bar{u}_{iL}\gamma^\mu d_{iL}.
\end{equation}
Using Eq. (\ref{sm21}) one obtains the charged current in the quark mass basis:

\begin{equation}\label{sm24b}
J^{\mu}_W = 
       \frac{1}{\sqrt{2}}\;\bar{u}_{iL}^{'}\gamma^\mu V_{ij} d_{iL}^{'},
\end{equation}
where the unitary matrix $V$ is given by

\begin{equation}\label{int13}
V_{ij}= U^{u\dagger}U^d.
\end{equation}
If one considers only two quark generations, 
the complex phases in $V$
can be absorbed by means of redefinitions of
the quark fields.
$V$ is then a orthogonal matrix parametrized by the Cabibbo angle

\begin{equation}\label{int14}
V=\begin{pmatrix}
\cos{\theta_c} & \sin{\theta_c} \\ -\sin{\theta_c}&\cos{\theta_c}
\end{pmatrix}.\end{equation}
Including the three quark generations the matrix $V$ contains 3 rotation angles and one CP-violating phase
(starting from a general $3\times 3$ unitary matrix, redefinitions of 
the quark fields do not remove of all the phases). 
This is the Cabibbo-Kobayashi-Maskawa (CKM) matrix.
The entries of the CKM matrix can be obtained experimentally from data on weak decays of
quarks and, in some cases, from neutrino deep inelastic scattering.
Assuming unitarity and the existence of only three families of quarks,
the CKM matrix in the standard parametrization, $V_{CKM}^{\text{(std)}}$, is given by

\begin{equation}\label{int15}
V_{CKM}^{\text{(std)}} = V_1 \: V_2 \: V_3,
\end{equation}
where

\begin{equation}\label{int16}
V_1=\begin{pmatrix} 1&0&0\\0&\cos{\theta_y}&\sin{\theta_y}\\
                    0&-\sin{\theta_y} &\cos{\theta_y}  \end{pmatrix},\;
V_2=\begin{pmatrix} \cos{\theta_z}&0&\sin{\theta_z}e^{-i\phi}\\
           0&1&0\\-\sin{\theta_z}e^{i\phi}&0&\cos{\theta_z}\end{pmatrix},\;
V_3=\begin{pmatrix} \cos{\theta_x}&\sin{\theta_x}&0\\
          -\sin{\theta_x}&\cos{\theta_x}&0\\0&0&1  \end{pmatrix}.
\end{equation}
The following are reference values for the three angles and the phase \cite{PDG}:

\begin{equation}\label{int17}
\theta_x \approx 0.226\;;\;\qquad\theta_y\approx 0.041\;;\;\qquad 
\theta_z\approx 0.0037\;;\;\qquad \phi\approx\pi/3\:.
\end{equation}
With these values one obtains, to three decimals 
in each matrix entry:

\begin{equation}\label{int18}
V_{CKM}^{\text{(std)}} = \begin{pmatrix}
0.973&0.233&0.002-0.003\: i\\
-0.233&0.972&0.041\\
0.008-0.003\: i&-0.040-0.001\: i&0.999
\end{pmatrix}.
\end{equation}
At present all $CP$-violating phenomena found in laboratory experiments can be explained by the phase $\phi$ of the CKM matrix.\\

For leptons the situation is somewhat different because they do not transform under $SU(3)_\text{color}$, having no contribution to the  QCD $\theta$-term.
Further, it is known by now that at least two of the three neutrinos are massive,
but we do not know whether neutrinos are Dirac or Majorana fermions.
If neutrinos are Dirac particles then the structure of the charged weak leptonic current in the mass basis is completely analogous to the quark sector,
with a leptonic mixing matrix $V_\text{lept}$ \cite{Maki:1962mu}
which has four observable parameters:
three angles and one $CP$-violating phase.
If neutrinos are Majorana particles then  $V_\text{lept}$ contains three angles and three observable $CP$-violating phases.
In the remainder of this thesis we shall ignore neutrino masses and discuss only the quark sector.

\section{Alternative Models of EWSB}\label{section:alternatives}

In this Section we discuss some unsatisfactory aspects of the SM.
In particular we concentrate on three features, or problems, related to the Higgs sector of the model.
Then, we turn to some extensions of the SM and comment on how they could solve these problems.

The SM successfully describes the strong and electroweak interactions.
It parametrizes the EWSB, but cannot be considered as an explanation of this phenomenon.
In the SM the dimensionful parameter of the Higgs potential $\mu^2$ is chosen
to be positive (see the convention in Eq. (\ref{sm15})), yielding EWSB.
This choice is possible but is not explained by any dynamics.
A more satisfactory theory would explain dynamically why this symmetry breaking takes place.

Let us mention a second aspect.
The $\beta$ function related to the Higgs self-coupling $\lambda$ is a positive quantity,
leading to a Landau pole and to triviality of the $\lambda\phi^4$ theory 
\cite{Wilson:1971dh,Wilson:1973jj}.
The theory is trivial in the sense that if one demands it to be valid up to arbitrary large energy scales, then the renormalized Higgs self-coupling $\lambda$ must go to zero, that is, the theory becomes a non-interacting one.
Triviality of the $\lambda\phi^4$ theory has been also confirmed by simulations on the lattice 
\cite{Luscher:1988uq,Kuti:1987nr,Hasenfratz:1987eh,Hasenfratz:1988kr,Bhanot:1990ai}.
Besides, it is believed that the inclusion of the SM gauge and Yukawa interactions do not affect the previous conclusions.
Triviality implies that the SM must be considered as an effective theory valid below some cutoff scale $\Lambda_\text{Triviality}$ which is associated with the Landau pole.
Imposing the condition $m_h<\Lambda_\text{Triviality}$ one gets an upper bound for the Higgs boson mass, $m_h=\mathcal{O}(1\; TeV)$ 
\cite{Cabibbo:1979ay,Dashen:1983ts,Lindner:1985uk}.
An important point is that the Landau pole is connected with the appearance of new physics.
After fixing the Higgs mass, and therefore the position of the Landau pole,
new physics must appear at energies of the order $\Lambda_\text{Triviality}$ or below
(it is possible that new physics appears before the Landau pole is reached).

Another unpleasant feature of the SM is the hierarchy problem.
In the SM there is no explanation of why the EWSB scale is $246\; GeV$
(and not 20 orders of magnitude higher or lower).
This is of course closely related to the first unsatisfactory aspect of the SM we mentioned.
Specifically, the  hierarchy problem is the lack of an answer to the question
why the EWSB scale is much smaller than the 
GUT    ($\Lambda_\text{GUT}   \sim10^{15}\; GeV$) or 
Planck ($\Lambda_\text{Planck}\sim10^{19}\; GeV$) scale.
Another important physical scale is $\Lambda_\text{QCD}\sim 100\; MeV$.
The generation of this scale is however understood from dimensional transmutation.
QCD is an asymptotically free theory where the running coupling $\alpha_3$ increases logarithmically with decreasing energy scale.
Therefore starting at the energy scale $\Lambda_\text{GUT}$ with a value of   $\alpha_3$ smaller but of order 1, this coupling constant becomes naturally non-perturbative at the scale  $\Lambda_\text{QCD}\ll\Lambda_\text{GUT}$.
In this way a dimensionless parameter of the model ($\alpha_3$) can be expressed in terms of a dimensionful one ($\Lambda_\text{QCD}$).
Many of the extensions of the SM include new asymptotically free gauge interactions in order to provide a solution to the EW hierarchy problem.
Closely related to this problem is the fine-tuning problem.
Radiative corrections to the squared Higgs mass are proportional to $\Lambda^2$, which is the upper limit of the validity range of the theory used as cutoff regulator.
These huge corrections occur because there is no (approximate)
symmetry which protects the Higgs mass.
Radiative corrections to other parameters of the SM diverge only logarithmically.
In order to have a prediction for the Higgs mass of the order of the EW scale
(the order of the tree level value)
huge fine-tunings between the divergent quantum corrections and 
their counterterms 
must be made order by order in perturbation theory.
One can, however, adopt a pragmatic point of view and make the cancellation of the divergent quantities without considering it as a problem.\\

Among the many extentions of the SM that have been proposed we would like to mention three types,
namely supersymmetry and two types of models of dynamical symmetry breaking.
We start with the supersymmetric extentions.
Supersymmetry is a symmetry between fermions and bosons. 
Supersymmetric theories have a certain number of superpairs which are made of one fermion and one boson, both having the same mass if supersymmetry is not broken.
The most studied supersymmetric extension of the SM is the
minimal supersymmetric SM (MSSM).
In this model for each SM degree of freedom a new one is introduced.
Besides the Higgs sector must be extended to a
 two-Higgs doublet model in order that the currents that couple to gauge bosons are anomaly-free.
In this way the particle content of the theory is more than doubled as compared with the SM.
However the experimentally observed particle spectrum does not include any superpartner.
Thus supersymmetry, if is realized in this way in nature, must be spontaneously broken.
From the non-observation of supersymmetry so far,
one concludes that the scale $\Lambda_\text{SUSY}$ at which this symmetry breaking takes place must be $\Lambda_\text{SUSY}>\mathcal{O}(1\; TeV)$.

It can be said that the MSSM solves almost all of the previous problems in a very nice manner.
However, important questions related to the supersymmetry breaking mechanism appear.
The MSSM provides a dynamical explanation for EWSB.
Starting at very high scales with a Higgs potential having a $SU(2)_L\times U(1)_Y$ invariant minimum (i.e. $<\Omega|\, H_i\, |\Omega>= 0$),
the evolution down to scales of the order of the EW scale, which is dictated by the perturbative renormalization group equations, leads to a low-energy Higgs potential with a  $SU(2)_L\times U(1)_Y$ non-symmetric minimum,
i.e. to radiative EWSB. 
The fine-tuning problem is also solved.
This occurs because the Higgs masses are protected by supersymmetry.
In order to have an acceptable level of tuning in the calculation of these masses,
the scale $\Lambda_\text{SUSY}$ at which supersymmetry is restored cannot be much higher than the scale of EWSB.
Finally to the hierarchy  problem:
In the MSSM, as we said, the EW scale is explained in terms of $\Lambda_\text{SUSY}$.
However, it is not clear how the last scale is generated.
The hierarchy problem of the SM has its analogy in the MSSM.
Namely, how can one understand the hierarchy between  $\Lambda_\text{SUSY}$ and  $\Lambda_\text{GUT}$.
The key, unsolved question is:
Which mechanism causes supersymmetry breaking?\\

Concerning EWSB there is one common feature between the SM and the MSSM:
EWSB is caused by the condensation of elementary Higgs bosons.
On the other hand models of EW dynamical symmetry breaking are based on the hypothesis that EW symmetry breaking is caused by the condensation of fermion-antifermion pairs
(for a review see \cite{Hill:2002ap}).
The
idea is taken from the theory of superconductivity:
On the one side there is the phenomenological Ginzburg-Landau model of superconductivity,
which corresponds to the Higgs model.
However, superconductivity is dynamically explained in the Bardeen-Cooper-Schrieffer theory by the Bose condensation of $e^-e^-$ pairs.

In particle physics two type of models of this class have been considered.
Namely models where known (SM) fermion pairs condense (top-condensation),
and models where pairs made of new type of fermions condense (technicolor).
In both cases a new strong interaction is postulated.
The elementary particle content of these models includes only fermions and gauge bosons.
The absence of fundamental scalar fields is supported by the fact that so far
no particle of this type has been found in nature.
Due to the fact that these models do not include fundamental scalar particles,
the first and second problems of the SM we mentioned above do not arise.
The third one can potentially be understood from dimensional transmutation.

The most prominent example of this class of models is
technicolor (TC) \cite{Weinberg:1979bn,Susskind:1978ms}
which is directly inspired by QCD.
In these models a new  non-abelian technicolor $SU(N_\text{TC})$ gauge group is added to the SM $SU(3)_\text{color}\times SU(2)_L\times U(1)_Y$ one.
Besides, new fermions, technifermions, which carry technicolor charges are postulated.
The $SU(2)_L\times U(1)_Y$ charge assignment of technifermions is similar to the one of the SM fermions.
Thus, the condensation of technifermion bilinears breaks the $SU(2)_L\times U(1)_Y$ symmetry to the electromagnetic $U(1)$.
The logarithmic evolution of the technicolor running coupling constant provides a dynamical explanation for the origin of the EW scale.
In this way technicolor is able to break the EW symmetry and,
through the Higgs mechanism,
to provide the weak gauge bosons with masses.
However, it fails in generating fermion masses.
A new sector, e.g. extended technicolor \cite{Dimopoulos:1979es,Eichten:1979ah},
must be included in order to explain them.

Finally we consider top-condensation models
\cite{NAMBU89,Miransky:1988xi,Miransky:1989ds,Bardeen:1989ds,Cvetic:1997eb}.
The main motivation for this type of models is given by the fact that the top mass and the scale of EW symmetry breaking are of the same order.
This suggests that the top quark plays an active r\^{o}le in the EWSB mechanism.
In top-condensation models a four-fermion interaction term involving  SM fermions is postulated at some high energy scale $\Lambda$.
The strong four-fermion interaction induces the formation of fermion-antifermion 
(mainly from the third quark generation) bound states,
which correspond to composite Higgs bosons, and the condensation of the corresponding composite fields.
These condensates transform non-trivially under $SU(2)_L\times U(1)_Y$ transformations and therefore break the EW symmetry.
As a consequence gauge bosons and fermions acquire masses.
In the minimal models no new particles are  postulated.
On the other hand these models are generally considered as effective theories valid only for energies scales below $\Lambda$ (see, however, second scenario below).

Two possible scenarios concerning the origin of the four-fermion interaction term can be considered.
In the traditional one, a non-abelian gauge theory becomes strongly-coupled at a scale $\sim\Lambda$.
For energies below $\Lambda$ the new interaction is effectively described by operators constructed with the fields corresponding to the light ($<\Lambda$) degrees of freedom of the theory.
The new interaction must violate the flavor symmetry,
i.e. must be non-universal, in order to generate the observed fermion mass pattern.
Topcolor \cite{Hill:1991at}, topcolor assisted technicolor \cite{Hill:1994hp},
and top-quark see-saw \cite{Dobrescu:1997nm,Chivukula:1998wd} are  examples of  theories of this type. 

At low energies the most important non-renormalizable operators are the ones having the lowest dimension.
Therefore, only dimension 6 four-fermion operators are considered
(however, higher dimensional operators could also be important).
Besides, only four-fermion operators made of (pseudo)scalar fermion bilinears are generally considered 
(they lead to (pseudo)scalar composite fields),
ignoring the ones made of (axial)vector fermion bilinears.
Note that the distinction between (pseudo)scalar and (axial)vector bilinears is ambiguous due to Fierz identities.

In the second scenario the four-fermion interaction term acquires a more fundamental status.
It is assumed that 
the SM with the Higgs sector being replaced by a general dimension 6 four-fermion interaction is a (non-perturbatively) renormalizable theory \cite{Gies:2003dp}.
This is the case if one or more of the non-Gaussian ultraviolet stable fixed points found in  \cite{Gies:2003dp} using the point-like approximation are established.

\chapter{Spontaneous $CP$ Symmetry Breaking}\label{chapter:spontaneous_cpv}

In this Chapter we consider models in which the $CP$ symmetry is spontaneously broken  (for related reviews see \cite{Branco:2003xh,Bernreuther:1998ju,Peccei:1990zx}).
The Lagrangian of such models is invariant under a $CP$ transformation, but the related vacuum is not.
In this way the breaking of the $CP$ symmetry is put on the same footing as the breaking of the EW symmetry.
This sort of models attempts to explain $CP$ violation by the same mechanism which breaks the EW symmetry, while in models with explicit $CP$ violation the Lagrangian does not possess that symmetry from the beginning.
Concerning the strong $CP$ problem these models potentially offer a solution because the parameter $\bar{\theta}$ could become a small calculable quantity.
In the following Sections we consider models with spontaneous $CP$ violation, first due to condensation of fundamental scalar fields, and then due to condensation of fermion-antifermion composite fields.
We discuss also a potential problem of spontaneous $CP$ violation  coming from cosmology.
Finally we comment on flavor-changing neutral currents (FCNCs).

\section{Spontaneous $CP$ Symmetry Breaking with Fundamental Scalars}
\label{section:lee_model}

The VEV of one Higgs field can always be made real by means of a global $SU(2)_L\times U(1)_Y$ transformation.
In order to have spontaneous $CP$ violation complex VEVs of scalar fields, and hence a non-minimal Higgs sector, are needed.
In this Section we briefly describe the models of Lee \cite{Lee:iz} and Weinberg \cite{Weinberg:1976hu}  
(or their generalizations to the case of 3 quark families) 
which are based on the SM gauge group.
These are two representative examples of models which incorporate spontaneous $CP$ violation by condensation of fundamental Higgs fields.

The model of Lee \cite{Lee:iz} contains, in addition to the SM fermion and gauge boson fields, 2 Higgs doublets, $H^{(1)}$ and $H^{(2)}$. 
The VEVs of the electrically neutral components can always be written in the following way

\begin{equation}\label{scpv1}
<\phi^{0(1)}>\: =\frac{v}{\sqrt{2}}\:,\qquad
<\phi^{0(2)}>\: =\frac{w\: e^{i\eta}}{\sqrt{2}}\: ,
\end{equation} 
where $H^{(i)}=(\phi^{0(i)},\phi^{-(i)})^T$ and $v,w>0$.
A value of $\eta\neq 0,\pm\pi$ leads in general to $CP$ non-conservation.
The VEVs, Eq. (\ref{scpv1}), minimize the Higgs potential which is sensitive to the phase $\eta$.
The most general gauge-invariant renormalizable potential for 2 Higgs doublets is given by

\begin{equation}\label{scpv2}
\begin{split}
V(H^{(1)},H^{(2)})&=V_0(H^{(1)},H^{(2)})+[-\mu_{12}^2\: H^{(1)\dagger}H^{(2)}
+\lambda_5 (H^{(1)\dagger}H^{(2)})^2     \\
&\;\: +\lambda_6 (H^{(1)\dagger}H^{(2)})(H^{(1)\dagger}H^{(1)})
      +\lambda_7 (H^{(1)\dagger}H^{(2)})(H^{(2)\dagger}H^{(2)}) + h.c.],
\end{split}
\end{equation} 
where $V_0$ is the part of the potential which is independent from the relative phase between the two electrically neutral Higgs fields:

\begin{equation}\label{scpv3}
\begin{split}
V_0(H^{(1)},H^{(2)})&= -\mu_1^2 \: H^{(1)\dagger}H^{(1)} -\mu_2^2\: H^{(2)\dagger}H^{(2)}
+\lambda_1 (H^{(1)\dagger}H^{(1)})^2
+\lambda_2 (H^{(2)\dagger}H^{(2)})^2     \\
&\;\: +\lambda_3 (H^{(1)\dagger}H^{(1)})(H^{(2)\dagger}H^{(2)})
      +\lambda_4 (H^{(1)\dagger}H^{(2)})(H^{(2)\dagger}H^{(1)}).
\end{split}
\end{equation}
By definition the parameters of the Higgs potential are real and thus 
$\int d^3x \; V(t,\vec{x})$ invariant under a $CP$ transformation

\begin{equation}\label{scpv3_b}
H^{(1,2)}(t,\vec{x}) \overset{CP}{\longrightarrow} H^{*(1,2)}(t,-\vec{x}).
\end{equation} 
The potential has its minimum located at a non-trivial value of $\eta$ given by 

\begin{equation}\label{scpv4}
\cos\eta=\frac{2\,\mu_{12}^2-\lambda_6\, v^2-\lambda_7\, w^2}
		    {4\lambda_5\, v w},
\end{equation}
if 
\begin{equation}\label{scpv5}
\lambda_5>0,\quad\text{and}\quad
\bigg|\frac{2\,\mu_{12}^2-\lambda_6\, v^2-\lambda_7\, w^2}{4\lambda_5\, v w}\bigg|< 1.
\end{equation}
In the following we assume that these conditions are fulfilled.
Because we start with a $CP$-invariant theory the Yukawa couplings in the weak basis are  all real.
However due to the phase $\eta$, complex mass matrices are generated:

\begin{equation}\label{scpv6}
\begin{split}
(\tilde{M}_u)_{ij}=(g_u^{(1)})_{ij}\:\frac{v}{\sqrt{2}}+(g_u^{(2)})_{ij}\:
                                              \frac{w\:e^{i\eta}}{\sqrt{2}},\\
(\tilde{M}_d)_{ij}=(g_d^{(1)})_{ij}\:\frac{v}{\sqrt{2}}+(g_d^{(2)})_{ij}\:
                                              \frac{w\:e^{i\eta}}{\sqrt{2}},
\end{split}
\end{equation}
where $\tilde{M}_u$,  $\tilde{M}_d$ are the quark mass matrices and $g_u^{(i)}$, $g_d^{(i)}$ are the Yukawa coupling matrices in the weak basis.
Diagonalizing the quark mass matrices one gets in general  a complex CKM matrix.
Thus the model contains a number of $CP$-violating interactions, namely $W^\pm$ boson exchange (as in the SM), charged and (flavor violating and flavor conserving) neutral Higgs boson exchange.
Nevertheless $CP$ violation in this model is determined by a single parameter, to wit $\eta$.
Furthermore FCNCs are present in this model already at tree level (see Section \ref{section:fcnc}).
This requires a mechanism for their suppression in order to avoid conflict with experimental data.

Let us also consider the specific case $\mu_{12}^2=\lambda_6=\lambda_7=0$,
$\lambda_5\neq 0$ \cite{Branco:tn}
which can be motivated by imposing a discrete $Z_2$ symmetry on the Higgs potential,
e.g. by demanding the Higgs potential to be invariant under

\begin{equation}\label{scpv7}
H^{(1)}\longrightarrow  H^{(1)}\: , \qquad\qquad
H^{(2)}\longrightarrow -H^{(2)}.
\end{equation}
In this case we get from Eq. (\ref{scpv4}):

\begin{equation}\label{scpv8}
<\phi^{0(1)}>\: =   \frac{v} {\sqrt{2}}\:,\qquad
<\phi^{0(2)}>\: =\pm\frac{iw}{\sqrt{2}}\: ,
\end{equation} 
where the two minima in Eq. (\ref{scpv8}) are related to each other by the transformation Eq. (\ref{scpv7}).
If we define the field $\tilde{H}^{(2)}\equiv i\, H^{(2)}$ we see that the Higgs potential as well as the vacuum are invariant under the conventional $CP$ transformation related to the fields ${H}^{(1)}$ and $\tilde{H}^{(2)}$.
We have to inspect also the Yukawa sector.
If the quark fields transform under the $Z_2$ transformation according to one of the following two cases:

\begin{equation}\label{scpv9}
\begin{array}{rl}
&\text{(a)}\qquad u_{iR} \longrightarrow + u_{iR} \:,\qquad
			  d_{iR} \longrightarrow + d_{iR} \:,\qquad  \:\text{(type I)},\\ 
\text{or}&\text{(b)}\qquad u_{iR} \longrightarrow + u_{iR} \:,\qquad
			  d_{iR} \longrightarrow - d_{iR} \:,\qquad  \text{(type II)},
\end{array} 
\end{equation} 
where $u_{iR}$ and $d_{iR}$, with $i=1,2,3$, are the up- and down-type right-handed quark fields respectively,
and we require that the Yukawa interactions respect the discrete symmetry,
then the factor $i$ introduced in the Yukawa terms due to the replacement $H^{(2)}\longrightarrow -i\tilde{H}^{(2)}$ can be eliminated.
In the case (a) no factor $i$ is actually introduced in the Yukawa terms because the Higgs field $H^{(2)}$ does not couple directly to quarks.
In case (b) the factor $i$ can be absorbed by redefining the right-handed quark fields.
One ends up with a $CP$-invariant 2HD model type I (case (a)) or type II  (case (b)).

If on the other hand the discrete $Z_2$ symmetry is not imposed on the Yukawa sector the resulting theory is very different as discussed above.
In particular the motivation to put the tree level parameters $\mu_{12}^2=\lambda_6=\lambda_7=0$ is lost because
these terms are generated radiatively through fermion-loop contributions.\\

In the Weinberg model \cite{Weinberg:1976hu} some discrete symmetry \cite{Glashow:1976nt} 
(as the one given in Eqs. (\ref{scpv7}) and (\ref{scpv9}))
is imposed in such a way that FCNCs are avoided at tree level.
Right-handed quark fields of the same electrical charge couple only to one Higgs doublet with non-vanishing VEV.
In this way each quark mass matrix is proportional to a Yukawa coupling matrix. A diagonalization of the quark mass matrix corresponds to a diagonalization of the associated Yukawa coupling matrix.
This implies that there are no FCNCs at tree level.
Minimization of  the model's tree level potential $V(\{H^{(i)}\})$, with $i=1,2,3$, being invariant under the discrete symmetry, yields the VEVs and in particular their phases.
Only if the model possesses 3 (or more) Higgs doublets exists a parameter subspace for which the $CP$ symmetry is spontaneously broken by non-trivial phases of the Higgs VEVs.
The resulting theory has a real CKM matrix and has $CP$-non-invariant interactions only in the Higgs sector, that is,
charged Higgs boson exchange and flavor-diagonal neutral Higgs boson exchange.
However, these models are ruled out because they cannot explain, on the one hand, the observed $CP$ phenomena in $K$ and $B$ meson decays and, on the other hand, why an electric dipole moment $d_n$ of the neutron has not been observed
($|d_n|_{\text{exp}}<6.3\times10^{-26} e$ cm  \cite{Harris:jx}).

\section{Dynamical $CP$ Symmetry Breaking}

By dynamical $CP$ violation we denote spontaneous $CP$ violation caused by fermion-antifermion condensation rather than by condensation of fundamental scalar fields.

The dynamics in the two cases are very different.
If fundamental scalar fields condense, weak coupling is normally assumed.
This justifies a tree level analysis.
One considers renormalizable gauge-invariant operators built of Higgs fields only and constructs with them the Higgs potential.
The Higgs VEVs and their possible $CP$-violating phases are obtained by minimizing it.
On the contrary, fermion-antifermion condensation is a strong-coupling phenomenon.
In order to make predictions drastic approximations are necessary.

In the case of technicolor theories the assumed picture \cite{Peccei:1990zx} is as follows.
Technicolor interactions become strong at the EW scale and break the chiral EW symmetries through the formation of techniquark condensates, in analogy to chiral symmetry breaking in QCD.
After EWSB the theory and in particular its vacuum possesses a residual global symmetry $H$.
On the other hand, in order to provide fermions with mass terms, additional interactions, the so-called extended technicolor (ETC) interactions, are postulated.
The ETC interactions break the residual global symmetry $H$ explicitly, and thereby lift the degeneracy of the vacuum. 
In this way the ETC interactions select one direction in the degenerated vacuum.
It is possible that the selected vacuum is not invariant under a $CP$ transformation, leading to $CP$ violation by the charged weak current interaction ($W$ exchange), and by the ETC interactions.

In this thesis we are interested in dynamical symmetry breaking caused by the condensation of quark-antiquark pairs, dominated by $\bar{t}t$ pairs.
Differently from technicolor theories, in this model the same interactions are responsible for EWSB, fermion mass generation, and for spontaneous $CP$ symmetry breaking.
In Chapter \ref{chapter:3fam} we consider this model for EWSB involving the 3 quark families in detail and show that dynamical $CP$ symmetry breaking,
including generation of a $CP$-violating CKM matrix, is possible.\\

Dynamical $CP$ violation in top-condensation models has been investigated in \cite{Andrianov:1996pf} starting from a different type of interactions as the ones considered in this thesis, namely non-local four-fermion interaction terms 
\cite{Andrianov:1999xd,Andrianov:1996ag}.
In \cite{Andrianov:1996pf} four-fermion interaction terms involving space-time derivatives (thus, dimension $>6$ operators) 
and quarks belonging to the third family were considered.
For a special configuration of the four-fermion coupling constants dynamical $CP$ violation occurs leading to a $CP$-violating composite two-Higgs model.
The question of dynamical generation of the CKM matrix is not studied in this paper.

\section{Domain Walls}

Here we briefly mention a potential cosmological problem which appears if spontaneous $CP$ symmetry breaking occurs at the EW scale \cite{Zeldovich:1974uw}.
The effective potential $V_{\text{eff}}$ related to a $CP$ invariant Lagrangian obeys in general the following relation

\begin{equation}\label{scpv10}
V_{\text{eff}}(\{H^{(i)}\})=V_{\text{eff}}(\{H^{(i)*}\}),
\end{equation} 
where $H^{(i)}$ denotes fundamental or composite Higgs fields and $H^{(i)*}$ their complex conjugate fields.
For the case of weak-coupled fundamental fields Eq. (\ref{scpv10}) can be easily checked. 
The effective potential is just a gauge-invariant function of the scalar fields and their complex conjugate fields with real parameters
(for the 2HD model at tree level see Eqs. (\ref{scpv2}) and (\ref{scpv3})).
In technicolor theories Eq. (\ref{scpv10}) is replaced \cite{Peccei:1990zx} by

\begin{equation}\label{scpv11}
V_{\text{eff}}(W)=V_{\text{eff}}(W^*),
\end{equation} 
where $W$ is a function of the transformation matrices between the weak and mass basis of the quark fields and $W^*$ denotes its complex conjugate.

Due to Eq. (\ref{scpv10}) domains with different signs of the phases of the VEVs are formed at the EW phase transition.
These domains are separated by walls with energy density much bigger than the closure energy of the universe 
(after taking into account the effect of the universe expansion) \cite{Zeldovich:1974uw}.
If one considers this problem to be a serious one, some solution must be found in order that spontaneous $CP$ symmetry breaking at the EW scale is viable.

\section{Flavor Changing Neutral Currents}\label{section:fcnc}

It can be shown \cite{Gronau:1987xz,Branco:1979pv} that the requirements of spontaneous $CP$ symmetry breaking at the EW scale, absence of FCNCs at tree level\footnote{Here we refer to the general case without assuming any discrete symmetry.}, and a realistic CKM matrix cannot be simultaneously satisfied.
For this reason, in models with spontaneous $CP$ symmetry breaking the Yukawa coupling matrices in the mass basis cannot be completely diagonal.

To be more precise let us consider $n$ fundamental or composite Higgs fields $H^{(i)}$ which couple  to the 6 quark fields.
The Yukawa interactions of the neutral Higgs fields are given by

\begin{equation}\label{scpv12}
\mathcal{L}_{\text{Yukawa-neutral}} = -\sum_{i=1}^n\;\Big(\:
 \bar{u}_L\,g^{(i)}\, u_R\:\phi^{0(i)}  
+\bar{d}_L\,h^{(i)}\, d_R\:\phi^{0(i)*}
\;+\; h.c.\;\Big),
\end{equation}
where the $3\times 3$ matrices $g^{(i)}$, $h^{(i)}$ are the Yukawa couplings,
the fields $u_{R/L}$, $d_{R/L}$ denote the up- and down-type chiral fermion fields,
and $\phi^{0(i)}$ denote the neutral components of the Higgs fields $H^{(i)}$.
The dynamics of the theory determine that some or all the Higgs fields acquire non-vanishing VEVs.
Inserting these VEVs $<\phi^{0(i)}>$ in Eq. (\ref{scpv12}) one obtains the quark mass matrices.
The up- and down-type mass matrices are in general not diagonal.
Bringing them to a diagonal form one introduces the CKM matrix in the charged current.
In the mass basis the neutral Yukawa terms are given by

\begin{equation}\label{scpv13}
\mathcal{L}_{\text{Yukawa-neutral}} = -\sum_{i=1}^n\;\Big(\:
 \bar{u}_L^{'}\,\lambda_u^{(i)}\, u_R^{'}\:\phi^{0(i)}  
+\bar{d}_L^{'}\,\lambda_d^{(i)\dagger}\, d_R^{'}\:\phi^{0(i)*}
\;+\; h.c.\;\Big),
\end{equation}
where the $3\times 3$ matrices $\lambda_u^{(i)}$, $\lambda_d^{(i)}$ are the Yukawa couplings in the  mass basis and the primed fields
$u_{R/L}^{'}$, $d_{R/L}^{'}$ denote the chiral fermion fields also in the mass basis.

Now we can state more precisely what was proven in \cite{Gronau:1987xz}.
If the Yukawa couplings in the weak basis $g^{(i)}$, $h^{(i)}$ are real, and thus $\mathcal{L}_{\text{Yukawa}}$ is $CP$-invariant, and the CKM matrix has the experimentally observed form
(small mixing between generations and a non-vanishing $CP$-violation phase),
then the Yukawa coupling matrices in the mass basis $\lambda_u^{(i)}$, $\lambda_d^{(i)}$,
with $i=1,\dots,n$, cannot be all diagonal, i.e. FCNCs at tree level cannot be avoided.
We shall come back to this issue in Chapter \ref{chapter:3fam}.

\chapter{Nambu-Jona-Lasino Approach for the EWSB: the Minimal Scheme}
\label{chapter:min}

\section{The Lagrangian}

Before the discovery of the top quark a new type of model of EWSB was proposed.
Experiments had indicated that the top quark is very heavy,
$(m_t)_\text{exp} > 80 \: GeV$ at that time.
This motivated the possibility of
a quark-antiquark (mainly from the third generation) bound state
playing the r\^ole of the Higgs boson,
being in this case a composite Higgs boson \cite{NAMBU89,Miransky:1988xi,Miransky:1989ds,Bardeen:1989ds}
(for reviews, see \cite{Cvetic:1997eb,Hill:2002ap}).

In this chapter we consider the simplest model of this type.
Its Lagrangian consists of the usual SM-Lagrangian but
without the elementary Higgs field.
In its place, a four-fermion interaction term is considered.
The Lagrangian is of the form

\begin{equation}\label{min1}
\mathcal{L}= \sum_k \;\bar{\Psi}_k \:i \gamma^\mu D_\mu \,\Psi_k
    -\frac{1}{4} \sum_i \;(F_{\mu\nu}^{(i)\,a})^2
    +\mathcal{L}_{\text{4f}},
\end{equation}
where the first sum is over all left- and right-handed fermions of the theory and
the second contains the 3 Yang-Mills terms of the SM-symmetry group,
$SU(3)_c\times SU(2)_L\times U(1)_Y$.
The Lagrangian $\mathcal{L}$ is locally invariant under this symmetry group.
In the simplest model \cite{Bardeen:1989ds}
the four-fermion interaction term $\mathcal{L}_{\text{4f}}$,
is given by

\begin{equation}\label{min2}
\mathcal{L}_{\text{4f}} =
G_t\, (\bar{\psi}_L t_R) (\bar{t}_R\psi_L),
\end{equation}
where $\psi_L=(t_L,\, b_L)^T$, and
$t$ and $b$ are the top and bottom quark fields.
The $SU(3)_c$ and $SU(2)_L$ indices are suppressed.
A color-index contraction in each parenthesis and a $SU(2)_L$-index
contraction between $\bar{\psi}_L$ and $\psi_L$ are understood.
The term $\mathcal{L}_{\text{4f}}$ should be considered
as an effective interaction
produced by some underlying physics.
In this Section we use $\mathcal{L}_{\text{4f}}$ as a given starting point and do not postulate any possible origin.
The whole theory is defined for energies $E \lesssim\Lambda$.
The scale $\Lambda$ is a parameter of the theory identified with the scale of
the four-fermion term $\mathcal{L}_{\text{4f}}$
(remember that $G_t$ has mass dimension $m^{-2}$).
All momentum integrals of the theory
are regularized using $\Lambda$ as a spherical cutoff.

\section{The Gap Equation}

In the following we investigate the consequences of the
four-fermion interaction Eq. (\ref{min2}) for EWSB.
We make several approximations which, however,
are assumed to preserve those features of the full model, Eq. (\ref{min1}),
which are relevant to EWSB.

The calculation is made at first order in the $1/N$ expansion,
where $N=3$ is the number of colors.
Besides, the $SU(3)_c\times SU(2)_L\times U(1)_Y$ gauge interactions are neglected.
With these approximations only the quarks of the third family interact.
The relevant part of the Lagrangian Eq. (\ref{min1}) is given by

\begin{equation}\label{min3}
\mathcal{L}=\mathcal{L}_0 + \mathcal{L}_{\text{I}},
\end{equation}
where $\mathcal{L}_0$ is the free Lagrangian and
$\mathcal{L}_{\text{I}}$ the interaction term

\begin{eqnarray}
\mathcal{L}_0 &=&  \bar{t}\:i \gamma^\mu \partial_\mu \,t     \label{min4}
               +    \bar{b}\:i \gamma^\mu \partial_\mu \,b,  \\
\mathcal{L}_{\text{I}}&=& \mathcal{L}_{\text{4f}}\;
=\; G_t\, (\bar{\psi}_L t_R) (\bar{t}_R\psi_L).                  \label{min5}
\end{eqnarray}
The Lagrangian Eq. (\ref{min3}) is a version of the Nambu-Jona-Lasinio model
\cite{Nambu:tp}.

There are no fermion mass terms in Eqs. (\ref{min4}), (\ref{min5}),
which would violate the $SU(2)_L\times U(1)_Y$ gauge symmetry.
Below we see how this symmetry is broken by the vacuum.
The top quark acquires a dynamical mass
(the right-handed component of the bottom quark field
does not interact and in consequence the bottom field cannot get a Dirac mass term).
Let us start redefining conventionally the free and interaction terms

\begin{equation}\label{min6}
\mathcal{L}=\mathcal{L}_0^{'} + \mathcal{L}_{\text{I}}^{'},
\end{equation}
with

\begin{figure}[tbp]
\begin{center}
\psfig{file=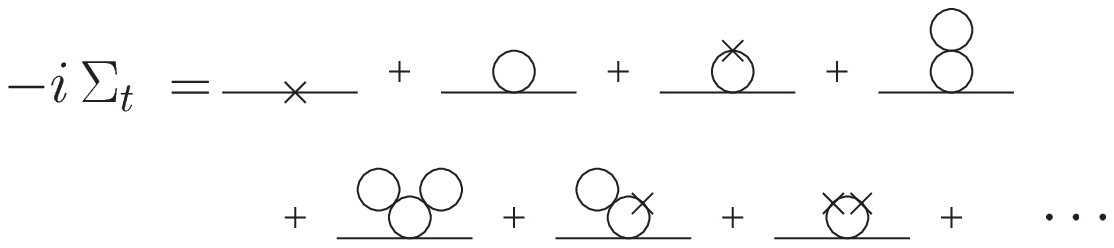,width = 12.0cm}
\end{center}
\caption{Feynman diagrams contributing to the top quark self-energy.
The crosses represent $im_t$ insertions.}
\label{fig:gap_eq1}
\end{figure}
\begin{figure}[tbp]
\begin{center}
\psfig{file=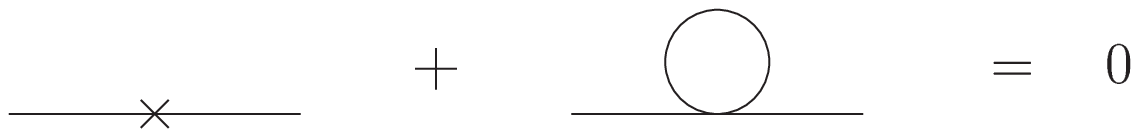,width = 10.0cm}
\end{center}
\caption{Diagrammatic representation of the gap equation. The crosses represent $im_t$ insertions.}
\label{fig:gap_eq2}
\end{figure}


\begin{eqnarray}
\mathcal{L}_0^{'} &=&  \bar{t}\:i \gamma^\mu \partial_\mu \,t     \label{min7}
               +    \bar{b}\:i \gamma^\mu \partial_\mu \,b
      -\: m_t \:\bar{t}t,\\
\mathcal{L}_{\text{I}}^{'}&=& m_t \:\bar{t}t +
                  G_t\, (\bar{\psi}_L t_R) (\bar{t}_R\psi_L),  \nonumber \\
 &=& m_t\: \bar{t}t + \frac{G_t}{4} \left[
(\bar{t}t)(\bar{t}t)+(\bar{t}\: i\gamma_5 t)(\bar{t}\: i\gamma_5 t)+
(\bar{b}(1+\gamma_5) t)(\bar{t}(1-\gamma_5) b) \right].      \label{min8}
\end{eqnarray}
Using the convention given by Eqs. (\ref{min4}) and (\ref{min5})
one would obtain the same result we arrive below.
In Eqs. (\ref{min7}) and (\ref{min8}) $m_t$ denotes the physical top mass of the interacting theory.
Thus, the top self-energy $\Sigma_t$ must vanish for $p^2=m_t^2$:

\begin{equation}\label{min10}
\Sigma_t\big|_{p^2=m_t^2} =0,
\end{equation}
where the diagrams contributing to $-i\Sigma_t$ are shown in Fig. \ref{fig:gap_eq1}.
Remember that we perform calculations to leading order in the $1/N$ expansion.
The limit $N\longrightarrow \infty$ is taken keeping $G N$ fixed.
In this approximation the top self-energy $\Sigma_t$ is momentum independent.
The condition Eq. (\ref{min10}) can be expressed in a simplified manner: If the sum of the two first diagrams of Fig. \ref{fig:gap_eq1}
is zero, then Eq. (\ref{min10}) is fulfilled\footnote{This can be shown by introducing an auxiliary field and demanding that the associated 1PI one-point function of the shifted theory vanishes.}.
Eq. (\ref{min10}) is therefore equivalent to

\begin{equation}\label{min11}
i m_t+\frac{i\, G_t}{4}(-1)\, 2N\int\frac{d^4l}{(2\pi)^4}
\tr\left(\frac{i}{l\contract -m_t}\right)      =0.
\end{equation}
This equation is expressed diagrammatically in Fig. \ref{fig:gap_eq2}.
Taking the trace one obtains

\begin{equation}\label{min11a}
m_t \left[ 1-2i N G_t \int \frac{d^4l}{(2\pi)^4}
\;\frac{1}{l^2 -m_t^2}\right] =0 .
\end{equation}
After doing the Wick rotation and performing the momentum integral using $\Lambda$ as spherical cutoff, we obtain

\begin{equation}\label{min12}
m_t = m_t\,G_t \:\frac{N\Lambda^2}{8\pi^2}
\left( 1-\frac{m_t^2}{\Lambda^2} \, \log(\Lambda^2/m_t^2 -1)\right).
\end{equation}
This is the gap equation for the top mass.
We shall derive this equation again using auxiliary fields in Chapter \ref{chapter:eff_potential}.
Note that the factor inside the outer parenthesis is smaller than 1.
For $G_t$ smaller than a critical coupling,
$G_t < 8\pi^2/(N\Lambda^2) \equiv G_{\text{crit}}$,
only $m_t=0$ is a solution of Eq. (\ref{min12}).
If $G_t$ is bigger than $G_{\text{crit}}$,
the gap equation possesses 2 solutions,
the symmetrical one $m_t=0$ and a non-symmetrical solution $m_t\neq 0$.
In Chapter \ref{chapter:eff_potential} we shall see that for $G_t >  G_{\text{crit}}$
the solution $m_t\neq 0$ corresponds to the state of lowest energy of the model, the vacuum.
The top mass is given in this case by the equation

\begin{equation}\label{min13}
1 = G_t \:\frac{N\Lambda^2}{8\pi^2}
\left( 1-\frac{m_t^2}{\Lambda^2} \, \log(\Lambda^2/m_t^2 -1)\right).
\end{equation}
Eq. (\ref{min13}) reveals that in order to obtain a top mass much smaller than the cutoff, $m_t\ll\Lambda$, a fine-tuning is needed.
The parameters of the theory, $G_t$ and $\Lambda$, must be chosen
in such a way that $G_t/G_{\text{crit}}=1+\delta$, with $\delta\ll 1$.

The dynamically created top mass breaks the $SU(2)_L\times U(1)_Y$ symmetry
of the Lagrangian Eq. (\ref{min1}).
The theory remains, however, invariant under a subgroup of the original symmetry,
namely under electromagnetic $U(1)_{\text{em}}$ transformations.
Due to the Goldstone theorem we expect, as in the SM, 3 Goldstone bosons.
In the next Section we find these massless modes and, in addition,
a neutral scalar boson explicitly.
All of them are fermion-antifermion bound states. \\

In order to motivate extensions of this minimal scheme,
we give some comments about the possible mass terms which can be dynamically generated.
The term $m_t \:(\bar{t}_L t_R+\bar{t}_R t_L)$ is the only mass term that can be generated by the interaction Eq. (\ref{min5}).\footnote{
One could say that more general terms are in this case also possible,
e.g. $\bar{b}_L t_R + \text{h.c.}$ or $\bar{t}\, i\gamma_5\,t$.
However these terms are related to the term $\bar{t}_L t_R+\bar{t}_R t_L$ by $SU(2)_L$ or $t_R$-$U(1)$-chiral transformations, respectively.
The Lagrangian Eq. (\ref{min1}) is invariant under these transformations.}
If instead of Eq. (\ref{min2}) we consider gauge invariant four-quark interaction terms involving all quarks,
other dynamical mass terms could also be induced.
In particular, it could happen that the electroweak symmetry gets broken completely
if in addition to the usual electrically neutral mass terms a term such as $\bar{b}t+\bar{t}b$ appears
(this term violates the $U(1)_{\text{em}}$ symmetry).
We refer to this possibility as the vacuum alignment problem.
Furthermore complex mass terms which mix quarks of the same electric charge could also appear.
They can lead to a non-trivial CKM-matrix and,
if the Lagrangian of the model is $CP$-invariant,
to spontaneous $CP$-breaking.
We study this possibility in Chapter \ref{chapter:3fam}.

\section{Scalar and Goldstone Modes}\label{section:min_poles}

Quark-antiquark bound states show up as poles
in two-point correlation functions of quark-antiquark composite fields

\begin{equation}\label{min16a}
i\mathrm{G}(p^2)= \int d^4x \: e^{ipx}
<\Omega|T\{ \mathcal{O}(x)\,\mathcal{O}^\dagger(0)\}|\Omega>_{\text{amputated}},
\end{equation}
where the field $\mathcal{O}(x)$ is a Lorentz scalar, quark-bilinear composite field
and $\mathcal{O}^\dagger(x)$ its hermitian adjoint.

In this Section we calculate such correlation functions in the
scalar, pseudoscalar, and charged channels.
We shall find 3 massless modes corresponding to the 3 Goldstone bosons,
and a fourth pole in the scalar channel corresponding to a composite Higgs boson.
In order to obtain the poles,  it is necessary to use
the gap equation (Eq. (\ref{min13})) in the calculations.

The composite fields we deal with are

\begin{equation}\label{min16b}
\begin{split}
\mathcal{O}_s (x)  &=\bar{t}(x)t(x), \\
\mathcal{O}_p (x)  &=\bar{t}(x) \,i\gamma_5\, t(x), \\
\mathcal{O}_c^- (x)&=\bar{t}(x)(1- \gamma_5)b(x), \\
\mathcal{O}_c^+ (x)&=\bar{b}(x)(1+ \gamma_5)t(x),
\end{split}
\end{equation}
where the indices $s$, $p$, and $c$ stand for scalar, pseudoscalar, and charged, respectively.

\begin{figure}[tbp]
\begin{center}
\psfig{file=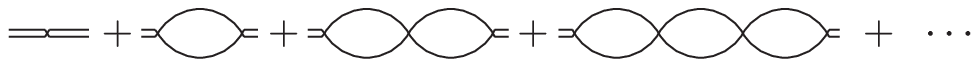,width = 13.8cm}
\end{center}
\caption{Feynman diagrams contributing to two-point correlation functions
of fermion-antifermion composite fields.}
\label{fig:four-point}
\end{figure}

The Feynman diagrams contributing to the two-point correlation functions
are shown in Fig. \ref{fig:four-point}.
In the scalar channel the amplitude is given by

\begin{eqnarray}
i\mathrm{G}_s(p^2) &=&
\frac{i G_t}{4}\,2+\left(\frac{iG_t}{4}\right)^2 4(-1) N
\int\frac{d^4l}{(2\pi)^4}\tr\left( \frac{i}{l\contract -m_t}
\;\;\frac{i}{p\contract +l\contract -m_t}\right) +\dots      \nonumber    \\
&=& \frac{i G_t}{2} \left(                                        \label{min15}
1-\frac{i G_t N I_s}{2}+\left( -\frac{i G_t N I_s}{2}\right)^2+\dots\right),
\end{eqnarray}
where $I_s$ is given by

\begin{equation}\label{min16}
I_s=\int\frac{d^4l}{(2\pi)^4}\tr\left( \frac{i}{l\contract -m_t}
\;\;\frac{i}{l\contract +p\contract -m_t}\right).
\end{equation}

If one uses, as we do, a spherical cutoff as regulator,
the Feynman integrals are
in general not invariant under a shift of the integration variables.
Convergent or logarithmically divergent integrals are invariant under this operation,
while linearly or more than linearly divergent integrals are not.
In the last case additional surface terms appear \cite{JAUCH-ROHRLICH55}.
A consequence of the appearance of surface terms is that ambiguities show up in some
correlation functions.
In particular, the mass of the Higgs particle becomes ambiguous.
On the other hand the position of the Goldstone boson poles is not affected
\cite{Willey:1993cp}.
These surface terms are treated in more detail in Appendix \ref{appendix:surface_terms}.
In what follows we neglect surface terms,
assuming that their ambiguous contributions are not physically relevant.

It is convenient to write $I_s$ as a quadratically divergent plus
a logarithmically divergent contribution.
Neglecting surface terms we obtain for $I_s$:

\begin{equation}\label{min17}
I_s= -4 \int \frac{d^4l}{(2\pi)^4} \;\frac{1}{l^2 -m_t^2}
+2(p^2-4m_t^2)\int \frac{d^4l}{(2\pi)^4} \;
                   \frac{1}{(l^2 -m_t^2)((l+p)^2 -m_t^2)}.
\end{equation}
Summing the geometric series in Eq. (\ref{min15}) we get

\begin{equation}\label{min18}
i\mathrm{G}_s(p^2)= \frac{i G_t}{2} \left(1+\frac{i G_t N I_s}{2}\right)^{-1}.
\end{equation}
Using Eq. (\ref{min17}) and the gap equation we finally obtain

\begin{equation}\label{min19}
i\mathrm{G}_s(p^2)= \frac{1}{2N(p^2-4m_t^2)} \left(
\int \frac{d^4l}{(2\pi)^4}\;\frac{1}{(l^2 -m_t^2)((l+p)^2 -m_t^2)}
                \right)^{-1} .
\end{equation}
Thus the amplitude of the scalar channel has a pole at $p^2=4m_t^2$.
That means that there is a composite scalar particle, a composite Higgs boson, with mass equal to $2 m_t$.
This quantitative prediction of the model for the mass of the composite Higgs particle should not be considered as an exact one, mainly because of the crude approximation we are doing.
For example in a renormalization group analysis including gauge interactions this result receives important corrections (see Chapter \ref{chapter:composite_condition}).
Concerning the regularization procedure and its possible influence on this result, some comments are made in the next Section.

In a similar way one finds the amplitudes in the pseudoscalar and charged channels.
One gets

\begin{eqnarray}
i\mathrm{G}_p(p^2) &=&  \frac{1}{2N p^2} \left(         \label{min20}
\int \frac{d^4l}{(2\pi)^4}\;\frac{1}{(l^2 -m_t^2)((l+p)^2 -m_t^2)}\right)^{-1}, \\
i\mathrm{G}_c(p^2) &=&  \frac{1}{8N} \left(         \label{min21}
\int \frac{d^4l}{(2\pi)^4}\;\frac{p(l+p)}{(l^2 -m_t^2)(l+p)^2}\right)^{-1}.
\end{eqnarray}
Note that none of the three four-point correlation functions depends directly
on the coupling constant $G_t$.
From Eqs. (\ref{min20}) and (\ref{min21}) one can see that
the pseudoscalar and charged amplitudes have a pole at $p^2=0$.
These massless modes are the Goldstone bosons.
In the next Section we see how these modes manifest themselves.

\section{Gauge boson masses}\label{section:gauge_boson_masses}

Finally we study in this minimal scheme the masses of the gauge bosons,
which appear in the theory according to the Higgs mechanism.
Recall that this denotes the mechanism where Goldstone bosons related to local symmetries are eaten by the corresponding gauge fields which, in turn, get massive.
Let us remark that the Higgs mechanism does not require  {\it elementary} scalar bosons \cite{Jackiw:1973tr,Cornwall:1973ts}.

As a consequence of the gauge invariance, the gauge-boson self-energies
$\Pi^{\mu \nu}_A$, with $A=\gamma$, $Z^0$, $W^{\pm}$, must fulfill the Ward identity $p_\mu\Pi^{\mu \nu}_A=0$
(where $p_\mu$ is the external momentum).
The self-energies can then be written as

\begin{table}[tbp]
\begin{equation*}\Large
\begin{array}{|!{\;}c!{\;}|!{\;}c!{\;}|!{\;}c!{\;}|}
\hline
& & \\
\int\frac{d^4l}{(2\pi)^4}\;\frac{1}{(l^2-M)^2} &
\frac{i}{(4\pi)^2}\int_0^{\Lambda^2} \frac{dq^2\, q^2}{(q^2+M)^2} &
\frac{i}{(4\pi)^{d/2}}\frac{\Gamma(2-d/2)}{M^{2-d/2}} \\
& & \\
\hline  \hline
& & \\
\int\frac{d^4l}{(2\pi)^4}\;\frac{l^2}{(l^2-M)^2} &
\frac{-i}{(4\pi)^2}\int_0^{\Lambda^2} \frac{dq^2\, q^4}{(q^2+M)^2} &
\frac{-i}{(4\pi)^{d/2}}\frac{d}{2}\,\frac{\Gamma(1-d/2)}{M^{1-d/2}} \\
& & \\
\hline
\end{array}
\end{equation*}
\caption{Table showing momentum integrals using different regularizations.
In the first column the generic integrals are shown.
In the second column these integrals are regularized using a cutoff $\Lambda$.
In the third column the same integrals are shown using dimensional regularization,
where $d=4-\epsilon$, the space-time dimension, is used as regulator.
The first and second rows show logarithmically and quadratically divergent integrals, respectively.}
\label{table:regularizations}
\end{table}

\begin{equation}\label{min22}
\Pi^{\mu \nu}_A=(g^{\mu \nu}p^2-p^\mu p^\nu)\,\Pi_A(p^2).
\end{equation}
It is known that a momentum cutoff is not a good regulator,
in the sense that it does not respect the Ward identities.
In order to obtain self-energies of the form given in Eq. (\ref{min22})
we use as an intermediate stage dimensional regularization.
We proceed as follows:
first we calculate the self-energies using dimensional regularization
(which is a good regulator).
Then we express the results as a function of the cutoff $\Lambda$ using Table \ref{table:regularizations}.

By using dispersion relations \cite{Gherghetta:1994cr}
it is possible to obtain directly transverse gauge bosons self-energies.
In this approach one imposes the appearance of the Goldstone bosons
related to the $SU(2)_L\times U(1)_Y\longrightarrow U(1)_{em}$
symmetry breaking.
As a consequence, one obtains the gap equation and the mass pole at $p^2=4m_t^2$
in the scalar channel (the same we obtained using a spherical cutoff as regulator).
Still another possibility is to use proper time regularization.
For a comparison between these different schemes
and the spherical cutoff regularization
in the context of a next-to-leading order calculation in $1/N$, see \cite{Cvetic:1996uq}.
For simplicity we stick to the spherical cutoff regularization.

\begin{figure}[tbp]
\begin{center}
\psfig{file=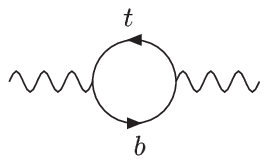,width = 4.0cm}
\end{center}
\caption{Feynman diagram associated with
the self-energy contribution $\left[ i\Pi^{\mu \nu}_W\right]_1$.}
\label{fig:self_energy1}
\end{figure}

\begin{figure}[tbp]
\begin{center}
\psfig{file=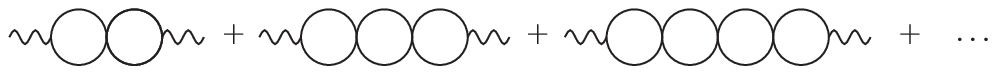,width = 15.8cm}
\end{center}
\caption{Feynman diagrams contributing to
the self-energy $\left[ i\Pi^{\mu \nu}_W\right]_2$,
which contain four-fermion interaction insertions.}
\label{fig:self_energy2}
\end{figure}

We calculate the self-energies to all orders in $G_t$ in the $N\longrightarrow\infty$ limit.
In this approximation the contribution of gauge boson loops is suppressed by a factor $N^{-1}$ as compared with the quark-loop contribution shown in Fig.  \ref{fig:self_energy1} and are therefore omitted.
Consequently a gauge fixing is not required.
Let us first calculate the self-energy of the $W^\pm$ boson.
We consider two types of contributions

\begin{equation}\label{min24}
i\Pi^{\mu \nu}_W=\left[ i\Pi^{\mu \nu}_W\right]_1+\left[ i\Pi^{\mu \nu}_W\right]_2,
\end{equation}
where $\left[ i\Pi^{\mu \nu}_W\right]_1$ is the contribution we obtain without considering
the four-fermion interaction $\mathcal{L}_{\text{4f}}$ (Fig. \ref{fig:self_energy1}),
and $\left[ i\Pi^{\mu \nu}_W\right]_2$ contains all diagrams with $G_t$ insertions
(Fig. \ref{fig:self_energy2}).
The first term is given by

\begin{equation}\label{min25}
\left[i\Pi^{\mu \nu}_W\right]_1=
\left( \frac{i g_2}{\sqrt{2}}\right)^2
\int\frac{d^4l}{(2\pi)^4} (-1)N \tr\left(
\projectorl \, \frac{i}{l\contract +p\contract} \,\gamma^\mu\,
\projectorl \, \frac{i}{l\contract - m_t     } \,\gamma^\nu \right) ,
\end{equation}
where $\projectorl=(1-\gamma_5)/2$ and $g_2$ denotes the $SU(2)_L$ gauge coupling.
After integration we obtain

\begin{equation}\label{min26}
\begin{split}
\left[i\Pi^{\mu \nu}_W\right]_1 &= (g^{\mu\nu}p^2-p^\mu p^\nu)
\left( \frac{i g_2}{\sqrt{2}}\right)^2  2N
\int_0^1 dx \: 2x(1-x) \:\int\frac{d^4l}{(2\pi)^4}\;\frac{1}{(l^2-D)^2} \\
&\qquad\quad -m_t^2 \,g^{\mu\nu}\left( \frac{i g_2}{\sqrt{2}}\right)^2  2N
\int_0^1 dx \: (1-x) \:\int\frac{d^4l}{(2\pi)^4}\;\frac{1}{(l^2-D)^2} ,
\end{split}
\end{equation}
with $D=-x(1-x)p^2+(1-x)m_t^2$.
For $m_t=0$, $\left[ i\Pi^{\mu \nu}_W\right]_1$ fulfills the Ward identity.
In order to obtain a transversal self-energy for $m_t \neq 0$
one must also consider the contributions coming from the interaction $\mathcal{L}_{\text{4f}}$ which is responsible for
the generation of $m_t$.
Because of the tensor structure of $\left[ i\Pi^{\mu \nu}_W\right]_2$
(see Fig. \ref{fig:self_energy2})
this term must be proportional to $p^\mu p^\nu$.
Therefore, the only way for $i\Pi^{\mu \nu}_W$ to respect the Ward identity is
that the second term of Eq. (\ref{min24}) is given by

\begin{equation}\label{min27}
\left[i\Pi^{\mu \nu}_W\right]_2=
m_t^2 \,\frac{p^\mu p^\nu}{p^2}\left( \frac{i g_2}{\sqrt{2}}\right)^2  2N
\int_0^1 dx \: (1-x) \:\int\frac{d^4l}{(2\pi)^4}\;\frac{1}{(l^2-D)^2}.
\end{equation}
Note that this expression does not depend on $G_t$.
Now we show that this is indeed what one obtains for $\left[i\Pi^{\mu \nu}_W\right]_2$ from direct calculation.
We use the fact that the sum of the diagrams of Fig. \ref{fig:self_energy2}
is proportional to the sum of the diagrams given in Fig. \ref{fig:four-point}
(which are calculated in Eq. (\ref{min21})).
The contribution is given by

\begin{equation}\label{min28}
\left[i\Pi^{\mu \nu}_W\right]_2=
\left( \frac{i g_2}{\sqrt{2}}\right)^2  N^2
\left[\int \frac{d^4l}{(2\pi)^4}
      \frac{m_t \tr(\gamma^\mu(l\contract+p\contract))}{(l^2-m_t^2)^2 (l+p)^2}
\right]
i\mathrm{G}_c(p^2)
\left[\int \frac{d^4l}{(2\pi)^4}
      \frac{m_t \tr(\gamma^\nu(l\contract+p\contract))}{(l^2-m_t^2)^2 (l+p)^2}
\right] .
\end{equation}
Using Eq. (\ref{min21}) one obtains Eq. (\ref{min27}). The self-energy
fulfills the Ward identity as it should.
Extracting the tensor structure from $\Pi^{\mu \nu}_W$ (using Eq. (\ref{min22}))
one obtains finally

\begin{equation}\label{min29}
\Pi_W(p^2)=
\left( \frac{i g_2}{\sqrt{2}}\right)^2
\frac{2N}{(4\pi)^2}
\int_0^1 dx
\left(  2x(1-x) -(1-x)\frac{m_t^2}{p^2} \right)
\left[  \log(\Lambda^2/D) -1            \right],
\end{equation}
with $D=-x(1-x)p^2+(1-x)m_t^2$.
The Goldstone boson contribution makes $\Pi^{\mu \nu}_W(p^2)$ singular at $p^2=0$.
This shifts the $W^\pm$ mass away from zero.
The dressed propagator is then given by

\begin{equation}\label{min30}
\int d^4x \: e^{ipx} <\Omega| \text{T}\:\{ W_\mu^+(x) W_\nu^-(0)\} |\Omega>\:=
\frac{-i (g_{\mu\nu}-p_\mu p_\nu/p^2)}{p^2}
\left( \frac{1}{1-\Pi_W(p^2)} \right) ,
\end{equation}
where $W_\mu^\pm (x)$ are the $W^\pm$ fields.
We define $\bar{g}_2(p^2)$ and $\bar{f}(p^2)$ as

\begin{eqnarray}
\frac{1}{\bar{g}_2^2(p^2)} &=& \frac{1}{g_2^2}
+\frac{N}{(4\pi)^2}
\int_0^1 dx\: 2x(1-x)
\left[  \log(\Lambda^2/D) -1    \right], \label{min31}    \\
\bar{f}^2(p^2)  &=&
m_t^2\,\frac{ N}{(4\pi)^2}
\int_0^1 dx\: (1-x)
\left[  \log(\Lambda^2/D) -1    \right], \label{min32}
\end{eqnarray}
such that

\begin{equation}\label{min33}
\frac{1-\Pi_W(p^2)}{g_2^2}  \:=\:
\frac{1}{\bar{g}_2^2(p^2)}
-\frac{\bar{f}^2(p^2)}{p^2}.
\end{equation}
The $W^\pm$ mass is given by the pole of its dressed propagator (Eq. (\ref{min30}))
through the condition

\begin{equation}\label{min34}
\begin{split}
M_W^2 &= \bar{f}^2(M_W^2)\:\bar{g}_2^2(M_W^2)  \\
&= m_t^2\:\:
\frac{  g_2^2 N/(4\pi)^2 \int_0^1 dx\,  (1-x)[\log(\Lambda^2/D) -1]}
     {1+g_2^2 N/(4\pi)^2 \int_0^1 dx\,2x(1-x)[\log(\Lambda^2/D) -1]},
\end{split}
\end{equation}
with $D$ evaluated at $p^2=M_W^2$.\\
On the other hand the Fermi constant is given by

\begin{equation}\label{min35}
G_F=\frac{1}{4\sqrt{2} \:\bar{f}^2(0)}.
\end{equation}
The experimental value of the Fermi constant is
$(\sqrt{2}\,G_F)^{-1/2}=246\:GeV$.
Using the experimental value of the top mass $m_t=178\: GeV$ and
the number of colors $N=3$ Eq. (\ref{min35}) gives the condition

\begin{equation}\label{min36}
\left( \frac{246}{178}\right)^2 \approx
\frac{12}{(4\pi)^2}\int_0^1 dx\,(1-x)
\left[ \log(\Lambda^2/((1-x)m_t^2))-1\right].
\end{equation}
The last equation demands a value of $\Lambda\approx 3\times 10^{13} GeV$.

Next we perform a similar calculation for the neutral gauge bosons.
We consider the fields $A_3^\mu(x)$ and $B^\mu(x)$ from the $SU(2)_L$ and $U(1)_Y$ gauge groups, respectively.
The difference in this case is that the gauge bosons mix.
Instead of calculating a single self-energy we have to calculate a $2\times 2$
matrix.
As in the case of the charged-boson self-energy
we distinguish two contributions

\begin{equation}\label{min37}
i\Pi^{\mu \nu}_{jk} = \left[ i\Pi^{\mu \nu}_{jk} \right]_1
                    + \left[ i\Pi^{\mu \nu}_{jk} \right]_2,
\end{equation}
where $j,k=1,2$ refers to the fields $B^\mu(x)$ ($j,k=1$) and $A_3^\mu(x)$ ($j,k=2$).
Similar as in Eq. (\ref{min22}) we define

\begin{equation}\label{min38}
\Pi^{\mu \nu}_{ij}=(g^{\mu \nu}p^2-p^\mu p^\nu)\,\Pi_{ij}(p^2).
\end{equation}
The dressed propagator is given in this case by

\begin{equation}\label{min39}
\int dx\: e^{ipx}
\begin{pmatrix}
<   B^\mu(x)\, B^\nu(0)>  & <   B^\mu(x)\, A_3^\nu(0)>  \\
< A_3^\mu(x)\, B^\nu(0)>  & < A_3^\mu(x)\, A_3^\nu(0)>
\end{pmatrix}
= \frac{-i (g^{\mu\nu}-p^\mu p^\nu/p^2)}{p^2}
\left( 1-\Pi(p^2) \right)^{-1},
\end{equation}
where $< A^\mu(x)\, B^\nu(0)>$ stands for
$<\Omega|T\{ A^\mu(x)\, B^\nu(0)\}|\Omega>$
and $( 1-\Pi(p^2))$ is a $2\times 2$ matrix.
The first contribution to the self-energy,
$\left[ i\Pi^{\mu \nu}_{jk}\right]_1$, is given by

\begin{equation}\label{min40}
\begin{split}
\left[ i\Pi^{\mu \nu}_{jk}\right]_1 =
-2N\sum_{q=t,b}\: &\int \frac{d^4l}{(2\pi)^4}\int dx\,\frac{1}{(l^2-D_q)^2}\times \\
& \Big[  \, (g^{\mu \nu}p^2-p^\mu p^\nu) \,2x(1-x)
\begin{pmatrix}
g_1^2\,(Y_{q\text{L}}^2 + Y_{q\text{R}}^2) &  g_1 g_2 \,T_q^3 Y_{q\text{L}} \\
g_1 g_2\, T_q^3 Y_{q\text{L}}              &  g_2^2  \,(T_q^3)^2
\end{pmatrix}   \\
& \qquad -m_q^2 \,g^{\mu \nu}
\begin{pmatrix}
g_1^2\,     (Y_{q\text{L}}-Y_{q\text{R}})^2  & g_1 g_2\, T_q^3(Y_{q\text{L}}-Y_{q\text{R}})\\
g_1 g_2\,T_q^3(Y_{q\text{L}}-Y_{q\text{R}})  & g_2^2  \, (T_q^3)^2
\end{pmatrix}
\Bigg] ,
\end{split}
\end{equation}
with $D_q=-x(1-x)p^2+m_q^2$.
The weak hypercharges are $Y_{t\text{L}}=Y_{b\text{L}}=-1/6$, $Y_{t\text{R}}=-2/3$,
$Y_{b\text{R}}=1/3$ and the weak isospin charges are
$T_t^3=-T_b^3=1/2$.
The $U(1)_Y$ and $SU(2)_L$ gauge couplings are denoted by $g_1$ and $g_2$, respectively.
Again $\left[ i\Pi^{\mu \nu}_{jk}\right]_1$ is not transversal for non-vanishing quark masses.
Adding the contributions with $G_t$ insertions
(only the pseudoscalar channel, i.e. the Goldstone boson, contributes)
one obtains

\begin{equation}\label{min41}
\begin{split}
\Pi(p^2) &=
-\frac{2N}{(4\pi)^2}\int_0^1 dx\,2x(1-x) \left[\log(\Lambda^2/D_t) -1\right]
\begin{pmatrix}
17/36 \:g_1^2   & -1/12\: g_1 g_2 \\
-1/12\: g_1 g_2 &  1/4 \: g_2^2
\end{pmatrix}    \\
&\quad-\frac{2N}{(4\pi)^2}\int_0^1 dx\,2x(1-x) \left[\log(\Lambda^2/D_b) -1\right]
\begin{pmatrix}
5/36\: g_1^2   & 1/12\: g_1 g_2 \\
1/12\: g_1 g_2 &  1/4\:  g_2^2
\end{pmatrix}    \\
&\quad+\frac{2m_t^2 N}{p^2(4\pi)^2}\int_0^1 dx \left[\log(\Lambda^2/D_t) -1\right]
\frac{1}{4}
\begin{pmatrix}
g_1^2   &  g_1 g_2 \\
g_1 g_2 &  g_2^2
\end{pmatrix} ,
\end{split}
\end{equation}
where Eq. (\ref{min38}) was used.
After a rearrangement we obtain

\begin{equation}\label{min42}
\begin{pmatrix}
1/g_1 &   0   \\
0     & 1/g_2
\end{pmatrix}
\Big( 1-\Pi(p^2) \Big)
\begin{pmatrix}
1/g_1 &   0   \\
0     & 1/g_2
\end{pmatrix}
=
\begin{pmatrix}
1/g_1^2(p^2) &   0   \\
0          & 1/g_2^2(p^2)
\end{pmatrix}
-\frac{f^2(p^2)}{p^2}
\begin{pmatrix}
1 & 1 \\
1 & 1
\end{pmatrix} ,
\end{equation}
with $g_1(p^2)$, $g_2(p^2)$, and $f(p^2)$ given by

\begin{equation}\label{min43}
\frac{1}{g_1^2(p^2)} = \frac{1}{g_1^2} + \frac{2N}{(4\pi)^2}\int_0^1 dx\,2x(1-x)
\left( \frac{5}{9}  \left[\log(\Lambda^2/D_t) -1\right]
      +\frac{1}{18} \left[\log(\Lambda^2/D_b) -1\right] \right),
\end{equation}

\begin{equation}\label{min44}
\frac{1}{g_2^2(p^2)} = \frac{1}{g_2^2} + \frac{2N}{(4\pi)^2}\int_0^1 dx\,2x(1-x)
\left( \frac{1}{3}  \left[\log(\Lambda^2/D_t) -1\right]
      +\frac{1}{6}  \left[\log(\Lambda^2/D_b) -1\right] \right),
\end{equation}

\begin{equation}\label{min45}
f^2(p^2)= \frac{p^2}{6}\frac{N}{(4\pi)^2}\int_0^1 dx\,2x(1-x)\,\log(D_b/D_t)
    +\frac{m_t^2}{2}\frac{N}{(4\pi)^2}
     \int_0^1 dx\left[\log(\Lambda^2/D_t) -1\right].
\end{equation}
The masses of the $\gamma$ and $Z^0$ particles are given by the poles of
the propagator Eq. (\ref{min39})
(the values of $p^2$ for which $1-\Pi(p^2)$ has vanishing eigenvalues).
For  $p^2\longrightarrow 0$ one can see in Eq. (\ref{min42}) that the last term dominates the sum. This matrix has all entries equal and hence possesses a vanishing eigenvalue.
As expected, a pole in the propagator located at $p^2=0$ is found.
The second pole related to the $Z^0$ particle is located at $p^2=M_Z^2$,
with $M_Z^2$ given by the condition

\begin{equation}\label{min46}
M_Z^2= f^2(M_Z^2)\left[ g_1^2(M_Z^2)+g_2^2(M_Z^2)\right].
\end{equation}
From Eqs. (\ref{min34}) and (\ref{min46}) one obtains

\begin{equation}
\frac{M_W^2}{M_Z^2}  \left(
\frac{g_1^2(M_Z^2)+g_2^2(M_Z^2)}{\bar{g}_2^2(M_W^2)} \right)
=\frac{\bar{f}^2(M_W^2)}{f^2(M_Z^2)}.
\end{equation}
The difference between $f^2$ and $\bar{f}^2$ is essentially the usual correction
to the $\rho$ parameter due to the weak isospin breaking effects.\\


We can in general say that the self-energies $\Pi^{\mu\nu}$
do not depend on the details of the four-fermion interactions we consider,
as long as each of these interactions is a product of two Lorentz (pseudo)scalar bilinears.
By separating the self-energies in the contributions $[\Pi^{\mu\nu}]_1$ and $[\Pi^{\mu\nu}]_2$ we saw that the part containing four-fermion interaction insertions $[\Pi^{\mu\nu}]_2$ is completely determined by its Lorentz structure and the requirement that $\Pi^{\mu\nu}$ must satisfy $p_\mu\Pi^{\mu\nu}=0$.
For this reason the gauge boson propagators,
and in consequence the gauge boson masses,
do not change if we consider more a general four-fermion interaction $\mathcal{L}_{\text{4f}}$,
as we do in Chapters \ref{chapter:1fam} and \ref{chapter:3fam}.

\chapter{Auxiliary Fields and the Effective Potential}\label{chapter:eff_potential}

\section{Auxiliary Fields}

For the study of EWSB we have used until now a formalism
involving only fermionic fields.
This is mainly because we considered a model without elementary scalar fields,
as for instance a fundamental Higgs field.
The particle content of this model includes only gauge vector bosons and chiral fermions.
However, as we saw in Chapter \ref{chapter:min},
fermion-antifermion bound states can appear in the scalar, pseudoscalar, and charged channels.
These spin-0 channels correspond to relevant degrees of freedom of the model.
In this Section we introduce new fields, auxiliary fields,
with the quantum numbers of the resonant channels.
This allows us to work directly with these degrees of freedom.

The auxiliary field formalism \cite{Gross:1974jv,Kugo:1976tq} is very convenient, especially if one goes beyond the minimal scheme studied in Chapter \ref{chapter:min},
in particular for the more realistic case where the six quarks are considered.
The formalism is also useful for studying next-to-leading order corrections in the
$1/N$ expansion
\cite{Cvetic:1994cc,Cvetic:1995qn,Cvetic:1996uq}.\footnote{To see the connection between the formalisms with and without auxiliary fields in the case of one auxiliary field, see \cite{Cvetic:1992nh}.
In this paper the effective potential is calculated diagrammatically.}

In general there is no prescription which tells us which and how many auxiliary fields should be introduced.
That would imply the knowledge of the relevant degrees of freedom at each scale,
something we normally do not know.
An example of this difficulty are Fierz rearrangements which often lead to ambiguities \cite{Dudas:1993cj,Jaeckel:2002rm}.

Motivated by the results of Section \ref{chapter:min} we consider in this Section
a generalized four-fermion effective interaction defined at the scale $\Lambda$ which involves the 3 quark generations.
Next, we introduce auxiliary fields with the same quantum numbers of the resonant channels
found in Section \ref{chapter:min},
i.e., the quantum numbers of the SM Higgs doublet field.
The auxiliary fields are, in the case of EWSB,
relevant degrees of freedom at scales comparable with the bound state masses.

The effective interaction term we work with is given by

\begin{equation}\label{int1}
\mathcal{L}_{\text{4f}} =
G_{ijkl}\,(\bar{\psi}_{iL} u_{jR}) (\bar{u}_{lR}\psi_{kL})+
G_{ijkl}^{'}\,(\bar{\psi}_{iL} d_{jR}) (\bar{d}_{lR}\psi_{kL})+
[G_{ijkl}^{''}\,\epsilon^{ab}(\bar{\psi}_{iL}^a d_{jR})
    (\bar{\psi}_{kL}^b u_{lR})+ h.c. ],
\end{equation}
where the coupling constants $G$ and the quark fields $u_R$, $d_R$, $\psi_L=(u_L,d_L)^T$ have indices $i$, $j$, $k$, $l$,
which go from the first to the third quark generation,
and the antisymmetric matrix $\epsilon^{ab}$ is given by
\begin{equation}\label{int2}
\epsilon^{ab} =
\begin{pmatrix} 0 & -1\; \\ 1 & \;0 \end{pmatrix}.
\end{equation}

We introduce now the auxiliary fields $H^{(i)}$
by replacing the four-fermion interaction $\mathcal{L}_{\text{4f}}$
by the equivalent term  $\mathcal{L}_{\text{aux}}$:

\begin{equation}\label{int3}
\mathcal{L}_{\text{4f}} \longrightarrow
\mathcal{L}_{\text{aux}}=
- \sum_{i=1}^n m_{H_i}^2 H^{(i)\dagger}H^{(i)}+\mathcal{L}_{\text{Yukawa}},
\end{equation}
where
\begin{equation}\label{int4}
\mathcal{L}_{\text{Yukawa}} = -\sum_{i=1}^n\;
(\;g_{kl}^{(i)}\:\bar{\psi}_{kL}u_{lR}\:H^{(i)}+
 h_{kl}^{(i)}\:\epsilon^{ab}\:\bar{\psi}_{kL}^a d_{lR}\:H^{b(i)*}
\;+\; h.c.\;),
\end{equation}
with complex parameters $g^{(i)}_{kl}$, $h^{(i)}_{kl}$,
and real mass parameters $m_{H_i}^2$.
The $n$ auxiliary fields $H^{(i)}$ are $SU(2)_L$ doublets, $SU(3)$ singlets,
and possess weak hypercharge +1/2.
The auxiliary fields have mass terms and couple to fermions through Yukawa terms.
A main feature is that at the scale $\Lambda$ no kinetic term
and no quartic boson interaction are present.
This is the so-called compositeness condition.
In Chapter \ref{chapter:composite_condition} we shall see how these terms are induced at lower energies.

Starting from $\mathcal{L}_{\text{aux}}$ we recover the effective interaction term
$\mathcal{L}_{\text{4f}}$, showing that both formulations of the model are equivalent.
This can be achieved either
in the path integral formalism or using the Euler-Lagrange equations for  the auxiliary fields.
In the path integral formalism the auxiliary fields $H^{(i)}$ can be easily integrated out
from the generating functional
because they appear only quadratically in the Lagrangian.
Alternatively, one can impose the constrains, i.e. Euler-Lagrange equations,
for the auxiliary fields:

\begin{equation}\label{int5}\begin{split}
H^{a(i)}&=\frac{1}{m_{H_i}^2}
    \:(-g_{kl}^{(i)*}\:\bar{u}_{lR}\psi_{kL}^a
        +\epsilon^{ab}h_{kl}^{(i)} \:\bar{\psi}_{kL}^b d_{lR}),\\
H^{a(i)*}&=\frac{1}{m_{H_i}^2}
    \:(-g_{kl}^{(i)}\:\bar{\psi}_{kL}^a u_{lR}
    +\epsilon^{ab}h_{kl}^{(i)*}\bar{d}_{lR}\psi_{kL}^b).
\end{split}\end{equation}
Replacing Eqs. (\ref{int5}) in Eq. (\ref{int3}),
the auxiliary fields are eliminated from $\mathcal{L}_{\text{aux}}$ and the four-quark interaction term $\mathcal{L}_{\text{4f}}$ is recovered.
The relations between the Yukawa couplings $g^{(i)}$, $h^{(i)}$
and the couplings $G$ are then given by

\begin{equation}\label{int6}\begin{split}
G_{ijkl}&=\sum_{r=1}^n\; \frac{g_{ij}^{(r)}g_{kl}^{(r)*}}{m_{H_r}^2},\\
G_{ijkl}^{'}&=\sum_{r=1}^n\;\frac{h_{ij}^{(r)}h_{kl}^{(r)*}}{m_{H_r}^2},\\
G_{ijkl}^{''}&=\sum_{r=1}^n\;\frac{h_{ij}^{(r)}g_{kl}^{(r)}}{m_{H_r}^2},
\end{split}\end{equation}

In the following we call the fields $H^{(i)}$ (composite) Higgs fields.
The idea behind this is that, as we said before,
at scales below $\Lambda$ these fields become
dynamical.
The fields $H^{(i)}$ play a similar r\^ole as the Higgs field in
the SM, the difference being, however, that in the present consideration they are composite fields.

\section{The Effective Potential}

We calculate the effective potential associated with the auxiliary fields $H^{(i)}$.
The effective potential allows us to find the vacuum of the theory.
We make the same approximations we did in Chapter \ref{chapter:min}.
We consider only the leading order contributions in the $1/N$ expansion,
i.e. in the $N\longrightarrow \infty$ limit keeping $G N$ fixed,
which is equivalent to the fermionic determinant approximation.
The effective potential \cite{Coleman:1973jx,Jackiw:1974cv} associated with the Lagrangian Eq. (\ref{min1}) with $\mathcal{L}_{\text{4f}}$ given by Eq. (\ref{int1}) but effectively replaced by $\mathcal{L}_{\text{aux}}$ in Eqs. (\ref{int3}) and  (\ref{int4}), is given by

\begin{equation}\label{eff1}
V_{\text{eff}}(\{H^{(i)}\})=m_{H_i}^2 H^{(i)\dagger}H^{(i)}
-i\int\frac{d^4k}{(2\pi)^4} \;
\log\det(\mathcal{D}^{-1}\{H^{(i)},k\}),
\end{equation}
where $\mathcal{D}^{-1}$ is the fermionic propagator in momentum
space. It is a function of the scalar fields $H^{(i)}$ and of the momentum $k$.
In Eq. (\ref{eff1}) and in the following a sum over repeated indices is understood.
As usual we extract the propagator from the fermionic
quadratic terms of the Lagrangian,
which in our model correspond to all the fermionic terms:

\begin{equation}\label{eff2}
\mathcal{L}_{\text{kin}}+\mathcal{L}_{\text{aux}}\supset
\bar{u}_{jR}\,i \partial\contract u_{jR}+
\bar{u}_{jL}\,i \partial\contract u_{jL}+
\bar{d}_{jR}\,i \partial\contract d_{jR}+
\bar{d}_{jL}\,i \partial\contract d_{jL}+\mathcal{L}_{\text{Yukawa}},
\end{equation}
where $\mathcal{L}_{\text{kin}}$ contains the fermionic kinetic terms.
We calculate now $\det(\mathcal{D})$.
All terms in Eq. (\ref{eff2}) are diagonal in the color
indices, therefore

\begin{equation}\label{eff3}
\det(\mathcal{D})=(\det D)^N,
\end{equation}
where $D$ contains the momentum space quadratic terms
associated with the 12 Weyl spinors of the three generations of quarks.
With respect to these degrees of freedom $D$ is a $12 \times 12$ matrix and is given
in the basis
$(u_{1L},u_{2L},u_{3L},d_{1L},d_{2L},d_{3L},u_{1R},u_{2R},u_{3R},d_{1R},d_{2R},d_{3R})$
by

\begin{equation}\label{eff4}
D= k\contract -M,
\end{equation}
where the term $k\contract$ is proportional to the $12 \times 12$ identity matrix, and
$M$ is given by

\begin{equation}\label{eff5}
M=
\begin{pmatrix}
0 & \tilde{M} \\ \tilde{M}^\dagger & 0
\end{pmatrix},
\end{equation}
with

\begin{equation}\label{eff6}
\tilde{M}=
\begin{pmatrix}
g^{(i)}H^{1(i)} & \;  -h^{(i)}H^{2(i)*}  \\
g^{(i)}H^{2(i)} &     h^{(i)}H^{1(i)*}
\end{pmatrix}.
\end{equation}
In the last expression $g^{(i)}$ and $h^{(i)}$ are $3 \times 3$ matrices.
Now we use the identity

\begin{equation}\label{eff7}
\det(k\contract -M)=\sqrt{\det(-k^2+M^2)},
\end{equation}
perform the Wick rotation $k_0=i k_4$, and consider an extra factor 2 in the exponent of
the last expression coming from the 2 spinorial degrees of
freedom  of each of the 12 Weyl spinors
($k^2+M^2$ is diagonal in the spinor indices).
We obtain

\begin{equation}\label{eff8}
\det(k\contract -M)\longrightarrow\det(k^2+M^2),
\end{equation}
with

\begin{equation}\label{eff9}
M^2=
\begin{pmatrix}
\tilde{M}\tilde{M}^\dagger&0\\
0&\tilde{M}^\dagger\tilde{M}
\end{pmatrix}.
\end{equation}
Finally, we obtain the fermionic determinant

\begin{equation}\label{eff10}
\det(\mathcal{D})=[\det(k^2+A)]^{2N},
\end{equation}
with

\begin{equation}\label{eff11}
A=\tilde{M}^\dagger\tilde{M}.
\end{equation}

Now we return to the effective potential.
After performing the Wick rotation and the spherical momentum integration,
Eq. (\ref{eff1}) takes the following form

\begin{equation}\label{eff12}
V_{\text{eff}}(\{H^{(i)}\})=m_{H_i}^2 H^{(i)\dagger}H^{(i)}
-\frac{1}{16\pi^2}\int_0^{\Lambda^2} k^2\,dk^2\:
\log\det(\mathcal{D}),
\end{equation}
where the momentum cutoff $\Lambda$ is identified
with the energy scale of the new interaction $\mathcal{L}_{4f}$.
Now inserting the determinant calculated in Eq. (\ref{eff10}) we get

\begin{equation}\label{eff13}
V_{\text{eff}}(\{H^{(i)}\})=m_{H_i}^2 H^{(i)\dagger}H^{(i)}
-\frac{N}{8\pi^2}\int_0^{\Lambda^2} k^2\,dk^2\:
\log\det(k^2+A),
\end{equation}
with

\begin{equation}\label{eff14}
A=
\begin{pmatrix}
 g^{(i)\dagger}g^{(j)} H^{(i)\dagger}H^{(j)} &
 g^{(i)\dagger}h^{(j)} \epsilon^{ab} H^{a(i)*}H^{b(j)*} \\
-h^{(i)\dagger}g^{(j)} \epsilon^{ab} H^{a(i)}H^{b(j)} &
 h^{(i)\dagger}h^{(j)} H^{(j)\dagger}H^{(i)}
\end{pmatrix}.
\end{equation}
The effective potential is of course gauge invariant.

We shall use the effective potential Eq. (\ref{eff13}) in different models.
In the next Section we apply this method to the case of having only one auxiliary field and an arbitrary number of quarks.
Considering more auxiliary fields increases the complexity in the dependence of the effective potential on the Higgs fields.
In Chapter \ref{chapter:1fam} we study a model with two auxiliary fields involving only one quark generation.
Finally, in Chapter \ref{chapter:3fam} the case with two auxiliary fields and 3 quark generations is analyzed.

\section{The Case of One Auxiliary Field}\label{section:1hd}

We consider here models with four-fermion interactions which can be rewritten by means  of only one auxiliary field.
We shall consider 3 quark generations. This includes the minimal scheme treated in Section \ref{chapter:min}.
The Eqs.(\ref{int6}) reduce to

\begin{equation}\label{eff15}\begin{split}
m_H^2 \: G_{ijkl}      &= g_{ij}\: g_{kl}^{*},\\
m_H^2 \: G_{ijkl}^{'}  &= h_{ij}\: h_{kl}^{*},\\
m_H^2 \: G_{ijkl}^{''} &= h_{ij}\: g_{kl},
\end{split}\end{equation}
with $i,j,k,l=1,2,3$.
Thus, the couplings $G$ in Eq. (\ref{int1}) must fulfill the following conditions

\begin{equation}\label{eff16}
G_{ijkl} G_{mnrs}^{'}=
G_{mnij}^{''} G_{rskl}^{''*}\: ,
\end{equation}
for all $i,\dots,s$.
In the case of one family it reduces to $G\:G^{'}=|G^{''}|^2$, or,
in the notation of Chapter \ref{chapter:1fam} to

\begin{equation}\label{eff17}
G_t\,G_b= |G_{tb}|^2 .
\end{equation}
By means of gauge transformations any constant configuration of the auxiliary field $H$ can be brought into the form

\begin{equation}\label{eff18}
H=\bigg(
            \begin{matrix}
              \frac{\phi}{\sqrt{2}}\\0
              \end{matrix}\bigg),
\end{equation}
with the classical field $\phi>0$.
The effective potential, which is gauge-invariant, is in this case a function of one variable, namely $\phi$.
The matrix $A$ given in Eq. (\ref{eff14}) is now

\begin{equation}\label{eff19}
A=
\begin{pmatrix}
g^\dagger g \frac{\phi^{2}}{2} & 0 \\
0 & h^\dagger h \frac{\phi^{2}}{2}
\end{pmatrix}.
\end{equation}
It is convenient to express $A$ in the quark mass basis where it is a diagonal matrix
($\det(k^2+A)$ is invariant under the replacement $A\longrightarrow U\, A\, U^\dagger$,
with $U$ being a unitary matrix):

\begin{equation}\label{eff20}
A\longrightarrow
\begin{pmatrix}
\lambda_{q_1}^2 \frac{\phi^{2}}{2} & & 0 \\
 & \ddots & \\
0 & & \lambda_{q_6}^2 \frac{\phi^{2}}{2}
\end{pmatrix}  \equiv
\begin{pmatrix}
m_{q_1}^2  & & 0 \\
 & \ddots & \\
0 & & m_{q_6}^2
\end{pmatrix},
\end{equation}
where the parameters $\lambda_{q_i}$, with $i=1,\dots,6$, are the Yukawa couplings in the quark mass basis,
and $m_{q_i }$, with $i=1,\dots,6$, are the quark masses.
Replacing Eq. (\ref{eff20}) in Eq.(\ref{eff13}) one obtains

\begin{equation}\label{eff21}
V_{\text{eff}}(\phi)=m_H^2 \frac{\phi^{2}}{2}
-\frac{N}{8\pi^2}\sum_{i=1}^6 \int_0^{\Lambda^2} k^2\,dk^2\:
\log\left( k^2+\lambda_{q_i}^2 \frac{\phi^{2}}{2}\right) .
\end{equation}
The VEV $<\phi>$ of $\phi$ is given by the value of $\phi$ which minimizes
the effective potential.
The first and the second derivatives of $V_{\text{eff}}$ are given by

\begin{equation}\label{eff22}
\frac{\partial V_{\text{eff}}}{\partial \phi}=
\phi \left[
m_H^2-\frac{N}{8\pi^2}\sum_{i=1}^6 \:
\lambda_{q_i}^2 \int_0^{\Lambda^2} \frac{k^2\,dk^2}{k^2+m_{q_i}^2}
\right],
\end{equation}
and
\begin{equation}\label{eff23}
\begin{split}
\frac{\partial^2 V_{\text{eff}}}{\partial \phi^2}&=
    m_H^2-\frac{N}{8\pi^2}\sum_{i=1}^6 \:
    \lambda_{q_i}^2 \int_0^{\Lambda^2} \frac{k^2\,dk^2}{k^2+m_{q_i}^2} \\
&\qquad + \phi^{2}\frac{N}{8\pi^2}\sum_{i=1}^6 \:
    \lambda_{q_i}^4 \int_0^{\Lambda^2} \frac{k^2\,dk^2}{(k^2+m_{q_i}^2)^2}.
\end{split}
\end{equation}
The first derivative condition $\frac{\partial V_{\text{eff}}}{\partial \phi}=0$ has two solutions. A gauge symmetrical one at $\phi=0$,
and a non-symmetrical solution, i.e. with $\phi\neq 0$, determined by

\begin{equation}\label{eff24}
1=\frac{N\Lambda^2}{8\pi^2}\sum_{i=1}^6 \:
\frac{\lambda_{q_i}^2}{m_H^2}
\left( 1-\frac{m_{q_i}^2}{\Lambda^2} \, \log(\Lambda^2/m_{q_i}^2)\right),
\end{equation}
with

\begin{equation}\label{eff25}
m_{q_i}=\lambda_{q_i} \frac{\phi}{\sqrt{2}}.
\end{equation}
Eq. (\ref{eff24}) must be understood as an equation for $\phi$ with the parameters
$N$, $\Lambda$, $\lambda_{q_i}$, $m_H^2$ kept fixed.
Again a fine-tuning is necessary in order to get $m_{q_i}\ll \Lambda$.
Eq. (\ref{eff24}) has a solution only if the following inequality is satisfied:

\begin{equation}\label{eff26}
\sum_{i=1}^6 \:\frac{\lambda_{q_i}^2}{m_H^2}
> G_{\text{crit}} \equiv 8\pi^2/(N\Lambda^2).
\end{equation}
If the condition Eq. (\ref{eff26}) is not fulfilled one gets $<\phi>=0$,
and the EW symmetry is not broken.
If on the other hand the coupling constants do fulfill Eq. (\ref{eff26})
one can see from Eq. (\ref{eff23}) that the second derivative of $V_{\text{eff}}(\phi)$ is negative at $\phi=0$ (the quadratically divergent terms domain) and positive at the non-zero value of $\phi$ determined by Eq. (\ref{eff24}).
Thus, in this case the minimum of $V_{\text{eff}}(\phi)$ is located at $<\phi>\neq 0$,
and the fermions acquire masses given by Eq. (\ref{eff25}).

The condition Eq. (\ref{eff26}) can be rewritten as a function of the original coupling constants $G$
(which fulfill as well Eq. (\ref{eff16})). One gets

\begin{equation}\label{eff27}
\sum_{i,j=1}^3 \left( G_{ijij}+G_{ijij}^{'} \right)
> G_{\text{crit}}.
\end{equation}

Besides, taking the scalar neutral component of $H$ from Eq. (\ref{int5}), one gets

\begin{equation}\label{eff28}
\phi(x)=\frac{-1}{\sqrt{2}\: m_H^2} \sum_{i=1}^6 \:\lambda_{q_i}\:
\bar{q}_i^{'} q_i^{'}(x),
\end{equation}
where $q_i^{'}$ denotes the quark fields in the mass basis and $\phi(x)$ the field operator.
The composite Higgs particle is a quark-antiquark bound state.
It is mainly made of top-antitop, but also contains the other quark flavors.

For completeness we also calculate the mass of the composite Higgs particle in this framework.
The two-point proper vertex (amputated 1PI correlation function) associated with the field $\phi(x)$,
$i\Gamma_{\phi,\phi}(p^2)$,
which corresponds to the inverse $\phi$ propagator is given by

\begin{equation}\label{eff30}
i\Gamma_{\phi,\phi}(p^2)=\frac{i N}{16\pi^2} \sum_{i=1}^6 \:
(p^2-4m_{q_i}^2)\: \lambda_{q_i}^2 \int_0^{\Lambda^2}
\frac{k^2\,dk^2}{[(p+k)^2+m_{q_i}^2](k^2+m_{q_i}^2)},
\end{equation}
where $i\Gamma_{\phi,\varphi}(p^2)$ for arbitrary fields $\phi$ and $\varphi$ is defined as

\begin{equation}\label{eff29}
i\Gamma_{\phi,\varphi}(p^2)= \int d^4x \: e^{ipx}
<\Omega|T\{\phi(x) \,\varphi(0)\}|\Omega>_{\text{amputated,1PI}}.
\end{equation}
The physical mass $m_\text{pole}$ of the composite Higgs boson is obtained from the condition $\Gamma_{\phi,\phi}(p^2=m_\text{pole}^2)=0$.
It follows from Eq. (\ref{eff30}) that, neglecting the quark mass dependence of the momentum integrals, the position of the composite Higgs mass is given by \cite{Suzuki:1989nv}

\begin{equation}\label{eff31}
\sum_{i=1}^6 \:
(m_\text{pole}^2-4m_{q_i}^2)\: m_{q_i}^2 =0.
\end{equation}
Due to the factors $m_{q_i}^2$ the term related to the top-quark dominates the sum.
The mass of the composite Higgs is $m_\text{pole}\approx 2m_t$. We see that the result obtained for the minimal scheme is stable under the inclusion of further fermions with masses much smaller than $m_t$.

The coupling constants between the composite Higgs and the fermions are also of interest. At the energy scale $\mu=\Lambda$ the Yukawa term in the mass basis is given by

\begin{equation}\label{eff32}
\mathcal{L}_\text{Yukawa}=- \sum_{i=1}^6 \:\lambda_{q_i} \:\frac{\phi}{\sqrt{2}}\:
\bar{q}_i^{'} q_i^{'}.
\end{equation}
The relevant couplings are however the ones defined at scales much lower than $\Lambda$ where the Higgs field $\phi$ possesses a kinetic term.
In order to have an induced kinetic term for $\phi$ with the conventional prefactor at scales $\mu\ll\Lambda$ - for definiteness, we put here and in the following $\mu=0$ - the field $\phi$ must be scaled.
The coefficient of $p^2$ in the expression for $\Gamma_{\phi,\phi}(p^2)$ calculated in Eq. (\ref{eff30}) gives the correct factor

\begin{equation}\label{eff33}
\phi\longrightarrow \Big(
\frac{N}{16\pi^2}\sum_{i=1}^6 \: \lambda_{q_i}^2 \log\Lambda^2/m_{q_i}^2
\Big)^{-1/2}    \phi.
\end{equation}
On the other hand the fermionic kinetic terms do not receive corrections in this approximation and hence the fermion fields need not be scaled.
Replacing Eq. (\ref{eff33}) in Eq. (\ref{eff32}) one gets

\begin{equation}\label{eff34}
\mathcal{L}_\text{Yukawa}=
-\sum_{i=1}^6 \: f_i\:\frac{\phi}{\sqrt{2}}\: \bar{q}_i^{'} q_i^{'},
\end{equation}
with

\begin{equation}\label{eff35}
f_j= m_{q_j}\Big(\frac{N}{16\pi^2}\sum_{i=1}^6 \: m_{q_i}^2
           \log\Lambda^2/m_{q_i}^2
     \Big)^{-1/2}.
\end{equation}
We see that the Yukawa couplings at the scale $\mu=0$, $f_j$, do not depend on the four-fermion couplings explicitly. They depend on them only indirectly through the quark masses.
Besides, the top Yukawa coupling for the case in which only the top-quark gets a mass, $f_t=[N/(16\pi^2) \log\Lambda^2/m_t^2 ]^{-1/2}$,
is also stable under the inclusion of further fermions that are much lighter than the top quark.

By considering only one flavor in Eqs. (\ref{eff24}), (\ref{eff27}), and (\ref{eff30}) we recover the gap equation, the critical coupling condition, and the mass of the composite Higgs found in Chapter \ref{chapter:min}.

\chapter{One Generation of Quarks}\label{chapter:1fam}

\section{Four-fermion interaction term}

In this Section we assume the existence of a more general four-fermion effective interaction 
involving both quarks of the third family.
We determine the main properties of this model in the $N\longrightarrow\infty$ limit.

The most general (dimension 6) gauge-invariant four-fermion interaction term
which can be written as a sum of products of fermion bilinears 
with the quantum numbers and Lorentz structure of the SM Higgs boson are given by

\begin{equation}\label{h1} 
\mathcal{L}_{\text{4f}} = 
G_t\,   (\bar{\psi}_L t_R) (\bar{t}_R\psi_L)+
G_b\,   (\bar{\psi}_L b_R) (\bar{b}_R\psi_L)+
\left[ G_{tb} \,\epsilon^{ab}
        (\bar{\psi}_L^a b_R)(\bar{\psi}_L^b t_R)
+ h.c. \right],  
\end{equation} 
where
$\psi_L=
\begin{pmatrix}
 t_L\\ 
 b_L
\end{pmatrix}$ and
$ \epsilon^{ab}=
\begin{pmatrix}
0 & -1 \\ 
1 &  0
\end{pmatrix} $.
\\
Due to the hermiticity of the Lagrangian $G_t$ and $G_b$ are real.
One can set $G_{tb}$ also real (or positive) by redefining one of the right-handed fermion fields. 
In this way the interaction term $\mathcal{L}_{\text{4f}}$
possesses only real coupling constants. 
In any case the Lagrangian -- more precisely $\int d^3x\; \mathcal{L}(x)$ -- is invariant under a $CP$ 
transformation\footnote{We ignore here the QCD $\theta$-term.} of the fields.

\section{Auxiliary fields and the effective potential}

In order to study the ground state of the theory 
it is convenient to introduce $n$ spin-zero auxiliary fields $H^{(i)}$. 
Following Chapter \ref{chapter:eff_potential} we replace $\mathcal{L}_{\text{4f}}$ by

\begin{equation}\label{h2} 
\mathcal{L}_{\text{aux}}=
- \sum_{i=1}^n m_{H_i}^2 H^{(i)\dagger}H^{(i)}+\mathcal{L}_{\text{Yukawa}},
\end{equation} 
with
\begin{equation}\label{h3} 
\mathcal{L}_{\text{Yukawa}} = -\sum_{i=1}^n\;
(\;g_t^{(i)}\:\bar{\psi}_L t_R\:H^{(i)}+
   g_b^{(i)}\:\epsilon^{ab}\:\bar{\psi}_L^a b_R\:H^{b(i)*}
\;+\; h.c.\;),
\end{equation} 
where $g_t^{(i)}$ and $g_b^{(i)}$ are the Yukawa coupling constants
and $m_{H_i}^2$ are mass parameters associated with the auxiliary fields.
The relations between the coupling constants in the two formulations of the model are given by

\begin{equation}\label{h4}
\begin{split}
G_t   &=\sum_{r=1}^n\;\frac{g_t^{(r)}g_t^{(r)*}}{m_{H_r}^2},\\
G_b   &=\sum_{r=1}^n\;\frac{g_b^{(r)}g_b^{(r)*}}{m_{H_r}^2},\\
G_{tb}&=\sum_{r=1}^n\;\frac{g_t^{(r)}g_b^{(r)}}{m_{H_r}^2}.
\end{split}
\end{equation}
In order to parameterize the space of couplings $G$,
it is enough to consider $n=2$ and real coupling constants $g$
(for $G_t$ and/or $G_b$ negative or if $G_{tb}^2> G_t G_b$,
negative parameters $m_{H_r}^2$ are needed).
Therefore we restrict ourselves to $n=2$ and
real coupling constants $g$ in the following.

The self-interaction $\mathcal{L}_{\text{4f}}$ given in Eq. (\ref{h1}) possesses 3 parameters $G_t$, $G_b$, and $G_{tb}$. 
This term is replaced by $\mathcal{L}_{\text{aux}}$ with $n=2$ which has 4 parameters $g_t^{(1)}$, $g_b^{(1)}$, $g_t^{(2)}$, and $g_b^{(2)}$,
one more than the original Lagrangian $\mathcal{L}_{\text{4f}}$
(the mass parameters $m_{H_i}^2$ can be set equal to 1 by scaling the auxiliary fields $H^{(i)}$).
At this stage it seems that there is an inconsistency.
We clarify this point at the end of this Chapter.

We consider now the effective potential of the model
in the auxiliary field formulation.
In Section \ref{chapter:eff_potential}, 
the effective potential was calculated for a general four-fermion interaction $\mathcal{L}_{\text{4f}}$. 
Using this result we obtain for the one family case

\begin{equation}\label{h5}
V_{\text{eff}}=
\sum_{i=1,2} m_{H_i}^2 H^{(i)\dagger}H^{(i)}
-\frac{N}{8\pi^2}\int_0^{\Lambda^2} k^2\,dk^2\:
\log\det(k^2+A),
\end{equation}
with the $2\times 2$ matrix $A$ given by

\begin{equation}\label{h6}
A=
\begin{pmatrix}
 g_t^{(i)}g_t^{(j)} H^{(i)\dagger}H^{(j)} &
-g_t^{(i)}g_b^{(j)} \epsilon^{ab} H^{a(i)*}H^{b(j)*} \\
-g_t^{(i)}g_b^{(j)} \epsilon^{ab} H^{a(i)} H^{b(j)} &
 g_b^{(i)}g_b^{(j)} H^{(i)\dagger}H^{(j)}
\end{pmatrix},
\end{equation}
where summation over the indices $i$ and $j$ is understood.

\section{Minimum of the effective potential}
\subsection{First derivatives of the effective potential}

The ground state of the theory is found by minimizing the effective potential
with respect to the auxiliary fields $H^{(1)}$ and $H^{(2)}$.
Due to the gauge invariance of the effective potential it is possible to gauge any field
configuration into the following form:

\begin{equation}\label{h7}
H^{(1)}=\bigg(\begin{matrix} 
              \frac{v^{'}}{\sqrt{2}}\\0
              \end{matrix}\bigg),\qquad\qquad
H^{(2)}=\bigg(\begin{matrix} 
              \frac{w^{'}e^{i\eta^{'}}}{\sqrt{2}}\\z{'}
              \end{matrix}\bigg),
\end{equation}
with $v^{'}$, $w^{'}$ , $z^{'}\geq 0$.
Note that in the 1HD case the effective potential depends only on one variable, while in the 2HD case it depends on four, making the task of finding its minimum more laborious.
In the following $v^{'}$, $w^{'}$, $z^{'}$, $\eta^{'}$ denote the classical fields
and the corresponding non-primed symbols denote their VEVs,

\begin{equation}\label{hGb2}
<H^{(1)}>\,=\bigg(\begin{matrix} 
              \frac{v}{\sqrt{2}}\\0
              \end{matrix}\bigg),\qquad\qquad
<H^{(2)}>\,=\bigg(\begin{matrix} 
              \frac{w\;e^{i\eta}}{\sqrt{2}}\\z
              \end{matrix}\bigg).
\end{equation}
Using Eq. (\ref{h7}) in Eq. (\ref{h5}) the effective potential becomes

\begin{equation}\label{h8}
V_{\text{eff}}=V_{\text{eff}}(v^{'},w^{'},\eta^{'},z^{'2}).
\end{equation} 
(As we see later it is convenient to use  $z^{'2}$ instead of $z^{'}$).
In order to preserve the electromagnetic $U(1)$ symmetry, the VEV $z^2$ must be zero.

Next we inspect the effective potential as a function of these 4 variables.
We shall restrict ourselves to the parameter subspace with $z^{'2}=0$
and search for local minima in this region.
It is possible to show \cite{Harada:1990wg} that for $z^{'2}\neq 0$ there is no local minimum
(at least for $m_t\neq m_b$).
The following conditions are sufficient in order to have a local minimum 
at a point with $z^{'2}= 0$:
\begin{equation}\label{h8a}
\begin{split}
&\text{a)}\; \frac{\partial V_{\text{eff}}}{\partial\theta} = 0\;,
         \text{ for } \theta=v^{'},w^{'},\eta^{'}, \\
&\text{b)}\; \frac{\partial V_{\text{eff}}}{\partial z^{'2}} > 0,\\
&\text{c)}\; \text{The $3\times 3$ Hessian matrix associated with the variables} \\
&\quad \text{$v^{'}$, $w^{'}$ and $\eta^{'}$ is positive definite.}
\end{split}
\end{equation}

The conditions a) evaluated at the point 
$v^{'}=v$, $w^{'}=w$, $\eta^{'}=\eta$, and $z^{'}=0$ 
are given by

\begin{equation}\label{h10}
v \bigg[ m_{H_1}^2-
\frac{N\Lambda^2}{8\pi^2}\sum_{q=t,b}
\bigg(1-\frac{m_q^2}{\Lambda^2}\log\bigg(\frac{\Lambda^2}{m_q^2}+1\bigg)\bigg)
\Big( (g_q^{(1)})^2+g_q^{(1)}g_q^{(2)} \frac{w}{v}\cos\eta \Big) \bigg]
=0,
\end{equation}

\begin{equation}\label{h11}
w \bigg[ m_{H_2}^2-
\frac{N\Lambda^2}{8\pi^2}\sum_{q=t,b}
\bigg(1-\frac{m_q^2}{\Lambda^2}\log\bigg(\frac{\Lambda^2}{m_q^2}+1\bigg)\bigg)
\Big( (g_q^{(2)})^2+g_q^{(1)}g_q^{(2)} \frac{v}{w}\cos\eta \Big) \bigg]
=0,
\end{equation}

\begin{equation}\label{h12}
v w \sum_{q=t,b}
\bigg(1-\frac{m_q^2}{\Lambda^2}\log\bigg(\frac{\Lambda^2}{m_q^2}+1\bigg)\bigg)
g_q^{(1)}g_q^{(2)} \sin\eta 
=0,
\end{equation}
where
\begin{equation}\label{h13}
m_q=\bigg| g_q^{(1)}\frac{v}{\sqrt{2}}+g_q^{(2)}
\frac{w\, e^{i\eta}}{\sqrt{2}}
\bigg|.
\end{equation} 
The first derivative of the effective potential with respect to $z^{'2}$ is given by

\begin{equation}\label{h14}
\begin{split}
\frac{\partial V_{\text{eff}}}{\partial z^{'2}}= 
& \;m_{H_2}^2-
\frac{N\Lambda^2}{8\pi^2}\sum_{q=t,b}
\bigg(1-\frac{m_q^2}{\Lambda^2}\log\bigg(\frac{\Lambda^2}{m_q^2}+1\bigg)\bigg)
(g_q^{(2)})^2 \\
&+\frac{N}{8\pi^2} 
\int_0^{\Lambda^2}
\frac{k^2 \, dk^2}{(k^2+m_t^2)(k^2+m_b^2)}\frac{v^2}{2}
(g_t^{(1)}g_b^{(2)}-g_t^{(2)}g_b^{(1)})^2.
\end{split}
\end{equation} 
Using Eq. (\ref{h11}), with $w\neq 0$, the last expression can be written as

\begin{equation}\label{h15}
\begin{split}
\frac{\partial V_{\text{eff}}}{\partial z^{'2}}= 
& \;
\frac{N\Lambda^2}{8\pi^2}\sum_{q=t,b}
\bigg(1-\frac{m_q^2}{\Lambda^2}\log\bigg(\frac{\Lambda^2}{m_q^2}+1\bigg)\bigg)
g_q^{(1)}g_q^{(2)} \frac{v}{w}\cos\eta \\
&+\frac{N}{8\pi^2} 
\int_0^{\Lambda^2}
\frac{k^2 \, dk^2}{(k^2+m_t^2)(k^2+m_b^2)}\frac{v^2}{2}
(g_t^{(1)}g_b^{(2)}-g_t^{(2)}g_b^{(1)})^2.
\end{split}
\end{equation} 

\subsection{The case $G_{tb}=0$}\label{section:pq_sym}

If $G_{tb}$ in Eq. (\ref{h1}) is equal to zero, the Lagrangian has,
in addition to the local $SU(2)_L\times U(1)_Y$ and 
global baryon number symmetries,
an extra global symmetry, namely

\begin{equation}\label{h16}
\begin{split}
\psi_L  &\longrightarrow e^{-i\gamma}\psi_L, \\
t_R,\,b_R &\longrightarrow e^{i\gamma}t_R,\,e^{i\gamma}b_R,
\end{split}
\end{equation} 
which is known in the context of fundamental Higgs fields as
Peccei-Quinn (PQ) symmetry \cite{Peccei:1977hh,Peccei:1977ur}.\\

It is convenient to introduce the two auxiliary fields 
in a way that $H^{(1)}$ couples only to $t_R$ and $H^{(2)}$ only to $b_R$
(with $g_t^{(2)}=g_b^{(1)}=0$).\footnote{This model and its generalizations for more quark families ($H^{(1)}$ couples only to up-type right-handed quarks and $H^{(2)}$ couples only to down-type right-handed quarks) are called  type-II 2 Higgs doublet (2HD) models. In these models FCNCs are naturally suppressed.}
These two composite scalar fields transform under the PQ symmetry as
\begin{equation}\label{h17}
\begin{split}
H^{(1)}  &\longrightarrow e^{-2i\gamma}\,H^{(1)}, \\
H^{(2)}  &\longrightarrow e^{ 2i\gamma}\,H^{(2)}.
\end{split}
\end{equation} 
Due to this extra symmetry of the Lagrangian and of the effective potential 
one can eliminate the dependence of $V_{\text{eff}}$ on the phase $\eta^{'}$ 
(we choose $\eta^{'}=0$).
That is, the effective potential is a function of $v^{'}$, $w^{'}$, and $z^{'2}$ only.

We can choose $g_t^{(1)},\, g_b^{(2)}>0$ (see Eqs. (\ref{h4})).
In this way we get automatically positive quark mass parameters:

\begin{equation}\label{}
m_t= g_t^{(1)}\frac{v}{\sqrt{2}}, \qquad
m_b= g_b^{(2)}\frac{w}{\sqrt{2}}.
\end{equation}

Now we turn to the conditions (\ref{h10})-(\ref{h12}).
In the case in which both auxiliary fields condense
($v,w\neq 0$), Eqs. (\ref{h10}) and (\ref{h11}) become

\begin{equation}\label{h19}
m_{H_1}^2=
\frac{N\Lambda^2}{8\pi^2}
\bigg(1-\frac{m_t^2}{\Lambda^2}\log\bigg(\frac{\Lambda^2}{m_t^2}+1\bigg)\bigg)
 (g_t^{(1)})^2,
\end{equation}

\begin{equation}\label{h20}
m_{H_2}^2=
\frac{N\Lambda^2}{8\pi^2}
\bigg(1-\frac{m_b^2}{\Lambda^2}\log\bigg(\frac{\Lambda^2}{m_b^2}+1\bigg)\bigg)
 (g_b^{(2)})^2,
\end{equation}
These two conditions have exactly the same form as the condition obtained in the 
minimal scheme
(compare with Eq. (\ref{eff24})).
As in that case, they can be fulfilled only if $G_t, G_b>G_{\text{crit}}$.

We check now that the electromagnetic $U(1)$ symmetry is conserved, i.e. that $z>0$. 
As can be seen from Eq. (\ref{h15})
the first derivative of the effective potential with respect to $z^{'2}$ is always bigger than zero:

\begin{equation}\label{h21}
\frac{\partial V_{\text{eff}}}{\partial z^{'2}}= 
\frac{v^2 N}{16 \pi^2}(g_t^{(1)}g_b^{(2)})^2
\int_0^{\Lambda^2}
\frac{k^2 \, dk^2}{(k^2+m_t^2)(k^2+m_b^2)}>0.
\end{equation} 

Finally, the $2\times 2$ Hessian matrix is positive definite.
The non-vanishing second derivatives of the effective potential,
$\partial^2V_\text{eff}/\partial v^{'2}$ and $\partial^2V_\text{eff}/\partial w^{'2}$,
are bigger than zero at the critical point.

In summary we have seen that in the case we have $G_{tb}=0$ and $G_t, G_b>G_{\text{crit}}$ the EW symmetry is broken. Both auxiliary fields $H^{(1)}$ and $H^{(2)}$ condense in such a way that the electromagnetic $U(1)$ symmetry remains unbroken.

\subsection{The case $G_{tb} \neq 0$}

Now we consider the case $G_{tb} \neq 0$ for a non-symmetrical stationary point with $v$, $w\neq 0$.
The following equations must hold (from Eqs. (\ref{h10})-(\ref{h12})):

\begin{center}
\begin{eqnarray}
m_{H_1}^2&=&      \label{h22}
\frac{N\Lambda^2}{8\pi^2}\sum_{q=t,b}
\bigg(1-\frac{m_q^2}{\Lambda^2}\log\bigg(\frac{\Lambda^2}{m_q^2}+1\bigg)\bigg)
\Big( (g_q^{(1)})^2+g_q^{(1)}g_q^{(2)} \frac{w}{v}\cos\eta \Big),  \\
m_{H_2}^2&=&      \label{h23}
\frac{N\Lambda^2}{8\pi^2}\sum_{q=t,b}
\bigg(1-\frac{m_q^2}{\Lambda^2}\log\bigg(\frac{\Lambda^2}{m_q^2}+1\bigg)\bigg)
\Big( (g_q^{(2)})^2+g_q^{(1)}g_q^{(2)} \frac{v}{w}\cos\eta \Big),  \\
& & \sum_{q=t,b}     
\bigg(1  -\frac{m_q^2}{\Lambda^2}\log\bigg(\frac{\Lambda^2}{m_q^2}+1\bigg)\bigg)
g_q^{(1)}g_q^{(2)} \sin\eta 
=0.\qquad \qquad   \label{h24}
\end{eqnarray} 
\end{center}
The last equation can be fulfilled only if 

\begin{equation}\label{h24a}
\sin\eta=0.
\end{equation} 
To see this, assume that $\sin\eta\neq0$. In that case the following equality must hold:

\begin{equation*}
\sum_{q=t,b}     
\bigg(1-\frac{m_q^2}{\Lambda^2}\log\bigg(\frac{\Lambda^2}{m_q^2}+1\bigg)\bigg)
g_q^{(1)}g_q^{(2)}=0.
\end{equation*} 
But this implies $G_{tb}=0$.
The case $G_{tb}=0$ was treated separately in the previous Subsection. 

We also have to check the condition b) of (\ref{h8a}). 
Taking only the quadratically divergent terms, Eq. (\ref{h15}) becomes

\begin{equation}\label{h25}
\frac{\partial V_{\text{eff}}}{\partial z^{'2}}\approx
\frac{N\Lambda^2}{8\pi^2}
(g_t^{(1)}g_t^{(2)}+g_b^{(1)}g_b^{(2)}) \frac{v}{w}\cos\eta .
\end{equation} 
In order to fulfill condition b) the relation
\begin{equation}\label{h25a}
(g_t^{(1)}g_t^{(2)}+g_b^{(1)}g_b^{(2)})\cos\eta >0,
\end{equation} 
must hold.
This can be achieved by choosing the sign of $\cos\eta$ equal to the 
sign of $(g_t^{(1)}g_t^{(2)}+g_b^{(1)}g_b^{(2)})$.

For the condition c) we need the Hessian of $V_{\text{eff}}$.
For $\sin\eta=0$, using Eqs. (\ref{h22}) and  (\ref{h23}) in the calculation we obtain

\begin{equation}\label{hhessian1}
\begin{split}
\frac{\partial^2 V_{\text{eff}}}{\partial\theta_a\partial\theta_b}&=
\frac{N\Lambda^2}{8\pi^2} \sum_{q=t,b}     
\bigg(1-\frac{m_q^2}{\Lambda^2}\log\bigg(\frac{\Lambda^2}{m_q^2}+1\bigg)\bigg)
g_q^{(1)}g_q^{(2)} \cos\eta  
\begin{pmatrix}
w/v & -1  & 0  \\ 
-1  & v/w & 0  \\ 
0   &  0  & v w
\end{pmatrix}    \\
&\quad +\frac{N}{8\pi^2} \sum_{q=t,b} 
\:\int_0^{\Lambda^2} 2 m_q^2 \,\frac{dx\, x}{(x+m_q^2)^2}
\begin{pmatrix}
(g_q^{(1)})^2               & g_q^{(1)}g_q^{(2)} \cos\eta & 0 \\ 
g_q^{(1)}g_q^{(2)} \cos\eta & (g_q^{(2)})^2               & 0 \\ 
0 & 0 & 0
\end{pmatrix} ,
\end{split}
\end{equation} 
with $\theta_a=v^{'},w^{'},\eta^{'}$.
Considering only terms of order $\Lambda^2$ we have

\begin{equation}\label{hhessian2}
\frac{\partial^2 V_{\text{eff}}}{\partial\theta_a\partial\theta_b}\approx
\frac{N\Lambda^2}{8\pi^2}
(g_t^{(1)}g_t^{(2)}+g_b^{(1)}g_b^{(2)}) \cos\eta
\begin{pmatrix}
w/v & -1  & 0  \\ 
-1  & v/w & 0  \\ 
0   &  0  & v w
\end{pmatrix} .
\end{equation} 
Thus, using Eq. (\ref{h25a}), 
we see that the Hessian matrix possesses two positive eigenvalues and one equal to zero.
Expanding in $m_q^2/\Lambda^2$ the next contribution to the eigenvalue,
which is zero at leading order, is given by

\begin{equation}\label{hhessian3}
\frac{N}{8\pi^2} \sum_{q=t,b} 
\:\int_0^{\Lambda^2}\frac{dx\, x}{(x+m_q^2)^2}\:\frac{4m_q^2}{v^2+w^2}\: > 0.
\end{equation} 
Therefore the Hessian evaluated at the point given by Eqs. (\ref{h22}) and  (\ref{h23})
is positive definite.

We compare this local minimum with the symmetrical point $v=w=0$.
To do that we do not calculate the Hessian at $v^{'}=w^{'}=0$,
we just compare the value of $V_{\text{eff}}$ at the two points.
Using the following two equations

\begin{equation}\label{hhessian4}
\frac{m_{H_1}^2 v^2}{2}+\frac{m_{H_2}^2 w^2}{2}=
\frac{N\Lambda^2}{8\pi^2} \sum_{q=t,b}     
\bigg(1-\frac{m_q^2}{\Lambda^2}\log\bigg(\frac{\Lambda^2}{m_q^2}+1\bigg)\bigg)\, m_q^2,
\end{equation} 

\begin{equation}\label{hhessian5}
\int_0^{\Lambda^2} dx\, x \left[ \log(x+m^2)-\log(x)\right] =
m^2\Lambda^2-\frac{m^4}{2}\left( \log(\Lambda^2/m^2)+1/2\right) 
+\mathcal{O}(m^6/\Lambda^2),
\end{equation} 
one can see that the asymmetric point has a lower energy than the symmetric one.\\

We consider now the Lagrangian (\ref{h1})
in the approach used in Section \ref{chapter:min},
i.e. without introducing auxiliary fields.
Doing that, 
one should investigate the possibility of having further symmetry breaking mass terms
besides the mass terms $m_q \,\bar{\psi}_q\psi_q$, with $q=t,\:b$.
Additional terms like $\tilde{m}_q \,\bar{\psi}_q i \gamma_5 \psi_q$ would in general violate the 
$CP$ symmetry in the four-fermion interaction term
(by going to the fermion mass basis, the coupling $G_{tb}$ can become complex)
while a term like $m_{tb} \,\bar{\psi}_t\psi_b\;+\: h.c.$
would break the electromagnetic $U(1)$ symmetry.
The auxiliary field method offers, as we saw,
a convenient framework to treat these phenomena.
We saw that the VEVs of the auxiliary fields are real and that $z=0$.
This is equivalent to show that these additional mass terms ($\tilde{m}_q$ and $m_{tb}$) 
are not dynamically generated.
Considering only the usual mass terms we obtain the following gap 
equations\footnote{A numerical analysis of these equations including also gluon exchange is made in \cite{King:bx} for $\Lambda=10^{15}\, GeV$.} 
for $m_t$ and  $m_b$

\begin{figure}[tbp]
\begin{center}
\psfig{file=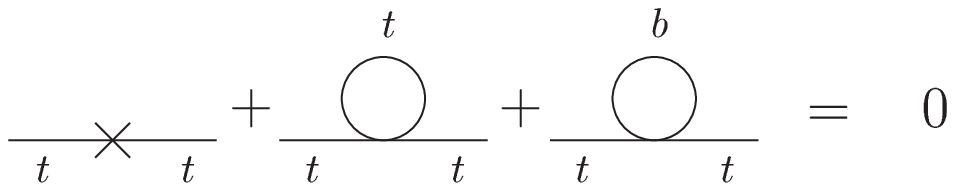,width = 9.0cm}
\end{center}
\caption{Diagrammatic representation of the gap equation for $m_t$.}
\label{fig:gap_1fam}
\end{figure}

\begin{eqnarray}
m_t&=&2i N \int \frac{d^4l}{(2\pi)^4} \left\lbrace    \label{h26}
	\frac{m_t\: G_t}{l^2 -m_t^2}+\frac{m_b\: G_{tb}}{l^2 -m_b^2}\right\rbrace ,  \\
m_b&=&2i N \int \frac{d^4l}{(2\pi)^4} \left\lbrace    \label{h27}
	\frac{m_t\: G_{tb}}{l^2 -m_t^2}+\frac{m_b\: G_b}{l^2 -m_b^2}\right\rbrace , 
\end{eqnarray} 
where negative mass parameters are allowed.
Eq. (\ref{h26}) is shown diagrammatically in Fig. \ref{fig:gap_1fam}.
For  Eq. (\ref{h27}) there is an analogous representation.
One can also see that Eqs. (\ref{h26}), (\ref{h27}) may be obtained from Eqs. (\ref{h22}), (\ref{h23}).

Now we discuss Eqs. (\ref{h26}) and (\ref{h27}) assuming $G_{tb}\neq 0$.
These equations can only be fulfilled if $m_t$ and  $m_b$ 
are both zero or both non-zero.
We consider $m_q^2/\Lambda^2 \ll 1$ and estimate the values of
$G_t$, $G_b$ and $G_{tb}$ which are needed in order to have non-vanishing
quark masses.
From these equations we obtain

\begin{equation}\label{h28}
\begin{split}
m_t (G_t-G_\text{crit})+m_b\: G_{tb}   \approx 0,  \\
m_t\: G_{tb} + m_b (G_b-G_\text{crit}) \approx 0,
\end{split} 
\end{equation} 
with $G_\text{crit}=8\pi^2/(N\Lambda^2)$.
For the top-bottom mass ratio this implies

\begin{equation}\label{h28a}
\Big| \frac{m_t}{m_b} \Big| \approx
\Big| \frac{G_{tb}}{G_\text{crit}-G_t} \Big| \approx
\Big| \frac{G_\text{crit}-G_b}{G_{tb}} \Big| .
\end{equation} 
In order to obtain non-zero quark masses 
(remember we are assuming here $G_{tb}\neq 0$)
the couplings $G$ must then obey the equation:

\begin{equation}\label{h29}
\frac{1}{G_\text{crit}}  (G_t+G_b)-
\frac{1}{G_\text{crit}^2}(G_t G_b - G_{tb}^2)
=1+\mathcal{O}(m_q^2/\Lambda^2).
\end{equation} 
If the condition for having only one Higgs doublet,
$G_t\,G_b= G_{tb}^2$, holds, one obtains
$(G_t+G_b)/G_\text{crit}=1+\mathcal{O}(m_q^2/\Lambda^2)$ as expected.

For a given $G_{tb}$ ($\neq 0$), the values of $G_t$ and $G_b$ which fulfill Eq. (\ref{h29}) describe a hyperbola. 
$G_t$, $G_b$ need not necessarily be positive.
As in the previous cases, in order to obtain quark masses much smaller than $\Lambda$,
it is necessary to consider the terms suppressed by a factor $m_q^2/\Lambda^2$
and do fine-tuning.

There is a question for which we do not present a direct answer because it involves fine-tuning of many parameters:
Given the parameters of the theory, $G_t$, $G_b$, and $G_{tb}$,
is there symmetry breaking or not?
It is enough to say that if Eqs. (\ref{h26}) and (\ref{h27})
can be fulfilled (with $m_q\neq 0$), EWSB occurs.
We do not give here an analogon to the symmetry breaking condition $G_t>G_\text{crit}$ obtained for the minimal scheme.\\

In summary we have seen in this Subsection that for a $\mathcal{L}_{\text{4f}}$ interaction term of the form given in Eq. (\ref{h1}) with $G_{tb} \neq 0$ EWSB occurs if Eqs. (\ref{h26}) and (\ref{h27}) are fulfilled (with $m_q\neq 0$).
Besides, the $CP$ symmetry is not spontaneously broken. 
For the auxiliary field analysis we restricted ourselves to the case where both auxiliary Higgs doublets condense,
because only in this case spontaneous electromagnetic or $CP$ symmetry breaking are possible.
In Section \ref{chapter:3fam} we shall see that for a four-fermion interaction term
involving the 3 families, spontaneous $CP$ violation can occur.

\section{Composite Higgs boson masses}\label{section:masses_1fam}

\subsection{Goldstone bosons}\label{ss41}

In order to calculate the mass spectrum of the composite Higgs bosons,
it is helpful to identify the Higgs boson degrees of freedom which correspond to Goldstone bosons.
In this Subsection we express the neutral and charged Goldstone bosons
in the auxiliary field basis.
We shall use this information for choosing a convenient basis for the calculation of the Higgs boson self-energies.

The Goldstone theorem states that for every spontaneously broken generator
of a continuous global symmetry
a massless boson, i.e., a Goldstone boson, appears. 
If the broken symmetry is a local one,
the Goldstone boson does not appear in the physical particle spectrum of theory.
In this case the Higgs mechanism takes place: 
The Goldstone boson is absorbed by the gauge boson associated with the broken local symmetry, which in turn gets massive. 
A massive gauge boson possesses 3 degrees of freedom, one more than a massless one. 

If a global symmetry is broken by the VEVs of some scalar fields,
the Goldstone theorem says also how to express the associated Goldstone boson field
as a function of the scalar fields.
The Goldstone bosons are given by the infinitesimal displacements of the scalar field VEVs under the transformations generated by the broken generators

\begin{equation}\label{hGb1}
G^{(k)}  \propto\delta^{(k)} <\phi> = i\; T^{(k)}<\phi>,\qquad \text{for } k=1,2,3,
\end{equation}
where $G^{(k)}$ are the Goldstone boson fields, $T^{(k)}$ the broken generators, 
and $<\phi>$ represents the VEVs of the scalar fields.
Let us write the VEVs of $H^{(1)}$ and $H^{(2)}$, given in Eq. (\ref{hGb2}),
using a real matrix (only to avoid $i$ factors)

\begin{equation}\label{hGb3}
<\phi>\,=
\begin{pmatrix}
\frac{v}{\sqrt{2}} & \frac{w}{\sqrt{2}} \cos\eta \\ 
0 & \frac{w}{\sqrt{2}} \sin\eta \\ 
0 & 0 \\ 
0 & 0
\end{pmatrix} ,
\end{equation}
where the two columns refer to the two Higgs doublets and 
the four rows refer from above to below to
the real part of the neutral component,
the imaginary part of the neutral component,
the real part of the charged component, and
the imaginary part of the charged component, respectively.
The broken generators of $SU(2)_L\times U(1)_Y$ are given by

\begin{eqnarray}
i \label{hGb4}
\begin{pmatrix} 0 & -i \\ i & 0\end{pmatrix} 
&\longrightarrow
i\, T^{(1)}=
\begin{pmatrix}
0 & 0  & 1 & 0 \\ 
0 & 0  & 0 & 1 \\ 
-1 & 0 & 0 & 0 \\ 
0 & -1 & 0 & 0
\end{pmatrix},  \\
i \label{hGb5}
\begin{pmatrix} 0 & 1 \\ 1 & 0\end{pmatrix} 
&\longrightarrow
i\, T^{(2)}=
\begin{pmatrix}
0 & 0 & 0 & -1 \\ 
0 & 0 & 1 & 0 \\ 
0 & -1 & 0 & 0 \\ 
1 & 0 & 0 & 0
\end{pmatrix},  \\
i \label{hGb6}
\begin{pmatrix} 1 & 0 \\ 0 & 0\end{pmatrix} 
&\longrightarrow
i\, T^{(3)}=
\begin{pmatrix}
0 & -1 & 0 & 0 \\ 
1 & 0 & 0 & 0 \\ 
0 & 0 & 0 & 0 \\ 
0 & 0 & 0 & 0
\end{pmatrix}.
\end{eqnarray} 
Inserting into Eq. (\ref{hGb1}) we obtain the following 
expressions for the charged and neutral Goldstone boson fields

\begin{eqnarray}
G^{\pm}  \propto & v\:\phi^{\pm (1)}+w\: e^{-i\eta}\:\phi^{\pm (2)},\label{hGb7}\\
G \:\: \propto & \mathcal{I}m(\,v\:\phi^{0(1)}+w \:e^{-i\eta}\:\phi^{0(2)}),\label{hGb8}
\end{eqnarray} 
where the fields $\phi^{0(i)}$, $\phi^{\pm(i)}$ are components of the Higgs fields $H^{(i)}=(\phi^{0(i)},\phi^{-(i)})^T$.

\subsection{Change of Basis of Auxiliary Fields}
\label{subsection:new_bases}

We define new bases for the auxiliary fields in the neutral and charged sectors.
We choose the new bases in such a way that the Goldstone bosons, identified in the previous Subsection, become basis vectors.
Later we calculate the two-point proper-vertex matrices in these bases.
The advantage is that in each sector, neutral and charged, one of the basis vectors of the proper-vertex matrix is already an eigenvector with associated eigenvalue equal to zero, i.e. a pole of the propagator.

In the four-dimensional neutral sector we define the new basis by

\begin{equation}\label{h58}
\begin{pmatrix} \varphi^1 \\ \varphi^2 \\ \varphi^3 \\ G
\end{pmatrix}
= R\;
\begin{pmatrix}
\mathcal{R}e\,\phi^{0(1)}\\
\mathcal{I}m\,\phi^{0(1)}\\
\mathcal{R}e\,\phi^{0(2)}\\
\mathcal{I}m\,\phi^{0(2)}
\end{pmatrix},
\end{equation}
where the orthogonal transformation matrix $R$ is given by

\begin{equation}\label{h59}
R=\frac{1}{\sqrt{v^2+w^2}}
\begin{pmatrix}
w & 0 & -v\,\cos\eta & -v\,\sin\eta \\
0 & w &  v\,\sin\eta & -v\,\cos\eta \\
v & 0 &  w\,\cos\eta &  w\,\sin\eta\\
0 & v & -w\,\sin\eta &  w\,\cos\eta
\end{pmatrix}.
\end{equation}
The field $G$ is the normalized Goldstone boson found in Eq. (\ref{hGb8}).
In this new basis the mass term of the neutral bosonic fields is given by

\begin{equation}\label{h60}
- \sum_{i=1,2} m_{H_i}^2\; H^{(i)\dagger}H^{(i)}\supset
-\frac{1}{2}
\begin{pmatrix}   \varphi^1, \varphi^2 , \varphi^3 , G
\end{pmatrix}
\mathcal{M}
\begin{pmatrix} \varphi^1\\ \varphi^2 \\ \varphi^3 \\ G
\end{pmatrix},
\end{equation}
with

\begin{equation}\label{h61}
\mathcal{M}=\frac{2}{v^2+w^2}
\begin{pmatrix} 
w^2 m_{H_1}^2 + v^2 m_{H_2}^2 &0& v w (m_{H_1}^2-m_{H_2}^2) &0\\
0& w^2 m_{H_1}^2 + v^2 m_{H_2}^2 &0& v w (m_{H_1}^2-m_{H_2}^2)\\
v w (m_{H_1}^2-m_{H_2}^2) &0& v^2 m_{H_1}^2 + w^2 m_{H_2}^2 &0\\
0& v w (m_{H_1}^2-m_{H_2}^2) &0& v^2 m_{H_1}^2 + w^2 m_{H_2}^2
\end{pmatrix}.
\end{equation}

Now we turn to the charged sector which is composed of two charged fields.
The new basis is defined by

\begin{equation}\label{h70}
\begin{pmatrix}
\varphi^\pm \\ G^\pm 
\end{pmatrix} 
=\frac{1}{\sqrt{v^2+w^2}}
\begin{pmatrix}
w\, e^{i\eta} & -v \\ 
v & w\, e^{-i\eta}
\end{pmatrix} 
\begin{pmatrix}
\phi^{\pm(1)} \\ \phi^{\pm(2)}
\end{pmatrix} ,
\end{equation} 
where the charged field $G^\pm$ is the normalized charged Goldstone boson found in Eq. (\ref{hGb7}). 
In this new basis the mass term for the charged scalar fields is given by

\begin{eqnarray}\label{h71}
\lefteqn{- \sum_{i=1,2} m_{H_i}^2\; H^{(i)\dagger}H^{(i)}\supset}\\
& &-\frac{1}{v^2+w^2}
\begin{pmatrix} \varphi^+ & G^+ \end{pmatrix} 
\begin{pmatrix}
w^2 m_{H_1}^2+v^2 m_{H_2}^2        &  vw\, e^{i\eta} (m_{H_1}^2-m_{H_2}^2) \\ 
vw\, e^{-i\eta} (m_{H_1}^2-m_{H_2}^2) &  v^2 m_{H_1}^2 + w^2 m_{H_2}^2 
\end{pmatrix} 
\begin{pmatrix} \varphi^- \\ G^- \end{pmatrix}. \nonumber
\end{eqnarray} 

We shall use these auxiliary field bases in the Subsection \ref{subsection:Gtb_no_0} and in Chapter \ref{chapter:3fam}.

\subsection{The case $G_{tb}=0$}

For $G_{tb}=0$ the neutral sector is approximately twice as large as the neutral sector
of the minimal scheme (see Section \ref{section:1hd}).
In the neutral sector of the minimal scheme one continuous symmetry is spontaneously broken.
A Goldstone boson appears.
Besides there is a neutral Higgs boson with a mass equal to $2m_t$.
In the present case there are, at least classically, two broken symmetries.
The one with the quantum numbers of the $Z$ boson and the PQ symmetry.
We obtain here a Goldstone boson and an axion \cite{STRONGCP}.
The PQ symmetry is actually anomalous which implies that the 
axion\footnote{An axion at the electroweak scale is experimentally ruled out.}
gets a mass from instanton contributions.
Furthermore in analogy with the minimal scheme there are two Higgs bosons with masses $2m_t$ and $2m_b$.\\

Let us now consider the charged sector. 
The relevant part of the Lagrangian is given in the quark mass basis by

\begin{equation}\label{h51}
\begin{split}
\mathcal{L}_{\text{aux}}\supset
&-\left[ m_{H_1}^2 \phi^{+(1)}\phi^{-(1)}+
            m_{H_2}^2 \phi^{+(2)}\phi^{-(2)}\right] \\
&\quad -\left[ g_t^{(1)}\phi^{+(1)}\:\bar{t}_R^{'} b_L^{'} -
            g_b^{(2)}\phi^{+(2)}\:\bar{t}_L^{'} b_R^{'}\;+\; h.c.
  \right].
\end{split}
\end{equation}
The $2\times 2$ proper vertex matrix of the charged fields,
shown diagrammatically in Fig. \ref{fig:2hd_1fam_charged}, is given by

\begin{equation}\label{h52}
i\Gamma_{\phi^{+(i)},\phi^{-(j)}}(p^2)=\frac{iN}{16 \pi^2}
\begin{pmatrix}
g_t^{(1)} & 0  \\ 
0 & g_b^{(2)}
\end{pmatrix} 
U(p^2)
\begin{pmatrix}
g_t^{(1)} & 0  \\ 
0 & g_b^{(2)}
\end{pmatrix} ,
\end{equation}
(see Eq. (\ref{eff29}) for the definition of $i\Gamma_{\phi,\varphi}(p^2)$) with

\begin{equation}\label{h53}
\begin{split}
U(p^2)=&\begin{pmatrix}
p^2-m_t^2-m_b^2 & 2 m_t m_b \\ 
2 m_t m_b & p^2-m_t^2-m_b^2
\end{pmatrix} 
I(m_t^2,m_b^2;p^2)\\
&\quad+
\begin{pmatrix} 1 & 0 \\ 0 & -1 \end{pmatrix} 
\left( 
m_t^2 \log(\Lambda^2/m_t^2)- m_b^2 \log(\Lambda^2/m_b^2)
\right),
\end{split}
\end{equation}
where $I(m_t^2,m_b^2;p^2)$ is defined as

\begin{equation}\label{h53b}
I(m_t^2,m_b^2;p^2)=
\frac{16 \pi^2}{i} \int \frac{d^4l}{(2\pi)^4}\:
\frac{1}{(l^2-m_t^2)[(l+p)^2-m_b^2]}.
\end{equation} 
First we evaluate the matrix $U$ for a 
vanishing external momentum squared:

\begin{equation}\label{h54}
U(p^2=0)=
\begin{pmatrix}
-2m_b^2 &  2 m_t m_b\\ 
2 m_t m_b & -2m_t^2
\end{pmatrix} 
I(m_t^2,m_b^2;0).
\end{equation} 
The determinant of $U(p^2=0)$ is zero.
This reflects the
appearance of the expected charged Goldstone boson. 
To find the mass of the charged composite Higgs, we need the value of $p^2$ at which the second eigenvalue of $U$ becomes 0.
If we make the following approximation

\begin{equation}\label{h55}
\log(\Lambda^2/m_t^2),\,\log(\Lambda^2/m_b^2),\,
I(m_t^2,m_b^2;p^2)\longrightarrow \log\Lambda^2,
\end{equation}
where $\log\Lambda^2$ states, e.g., for $\log(\Lambda/1\,GeV)^2$,
the matrix $U$ takes the form

\begin{equation}\label{h56}
U(p^2)\approx
\begin{pmatrix}
p^2-2m_b^2 &  2 m_t m_b\\ 
2 m_t m_b & p^2-2m_t^2
\end{pmatrix}
\log\Lambda^2.
\end{equation}
The last matrix has vanishing eigenvalue at $p^2=2(m_t^2+m_b^2)$.
If one calculates the second eigenvalue from $U(p^2)$ without the last approximation,
one finds that the charged Higgs mass, for example for $\Lambda=10^{10}\: GeV$, 
differs only by 1\%.

\subsection{The case $G_{tb}\neq 0$}\label{subsection:Gtb_no_0}

Now we consider the case when $G_{tb}\neq 0$ and both auxiliary fields condense.
Let us first calculate the composite Higgs masses in the neutral sector.
It is convenient to adopt the auxiliary field basis defined in Eqs. (\ref{h58})-(\ref{h61}) in which the neutral Goldstone boson field is one of the basis vectors.
The orthogonal matrix $R$ is given, for $\sin\eta=0$, by

\begin{equation}\label{h59b}
R=\frac{1}{\sqrt{v^2+w^2}}
\begin{pmatrix}
w & 0 & -v\,\cos\eta & 0 \\
0 & w &  0           & -v\,\cos\eta \\
v & 0 &  w\,\cos\eta &  0 \\
0 & v &  0           &  w\,\cos\eta
\end{pmatrix}.
\end{equation}
Note that only the VEV of $\varphi^3$ is different from zero, $<\varphi^3>=\sqrt{(v^2+w^2)/2}$.
The Yukawa coupling term involving the neutral fields $\phi^{0(i)}$ is given by

\begin{equation}\label{h62} 
\mathcal{L}_{\text{Yukawa-neutral}} = -\sum_{i=1,2}\;
(\;\lambda_t^{(i)}\:\bar{t}^{'}_L t^{'}_R\:\phi^{0(i)}+
   \lambda_b^{(i)}\:\bar{b}^{'}_L b^{'}_R\:\phi^{0(i)*}
\;+\; h.c.\;),
\end{equation} 
where the couplings $\lambda$ are the Yukawa couplings in the quark mass basis.
Due to the condition (\ref{h24a}) these are real coupling constants which can differ from the respective Yukawa couplings in the weak basis $g$ only by a sign.
Using Eq. (\ref{h58}) we obtain the Yukawa term in the new basis:

\begin{equation}\label{h63} 
\begin{split}
\mathcal{L}_{\text{Yukawa-neutral}} &=
-\frac{(w\,\lambda_t^{(1)}-v\cos\eta\;\lambda_t^{(2)})}{\sqrt{v^2+w^2}}  
(\varphi^1 \,\bar{t}^{'} t^{'}+\varphi^2 \,\bar{t}^{'} i\gamma_5\, t^{'})  \\
&\;\quad -\frac{\sqrt{2}\, m_t}{\sqrt{v^2+w^2}}
(\varphi^3 \,\bar{t}^{'} t^{'}+G \,        \bar{t}^{'} i\gamma_5\, t^{'}) 
+\;\dots,
\end{split}
\end{equation} 
where the dots represent analogous terms for the bottom field.
The only difference which appears in the $b$-quark terms is that in this case it is necessary to replace $\gamma_5$ by $-\gamma_5$, as we can see from Eq. (\ref{h62}).
There are top-loop and bottom-loop contributions to the two-point functions
we want to calculate. 
These two contributions have however the same form.
A change of the sign in front of $\gamma_5$ has no consequences for the neutral two-point functions. 

As we saw at the beginning of this Chapter the Lagrangian we consider is $CP$-invariant.
This symmetry is also not broken by the vacuum of the theory because $\sin\eta=0$.
As a consequence the $CP$-even ($\varphi^1$ and $\varphi^3$) and  
the $CP$-odd ($\varphi^2$ and $G$) fields do not mix and the $4\times 4$ two-point proper vertex matrix is a block diagonal one.

\begin{figure}[tbp]
\begin{center}
\psfig{file=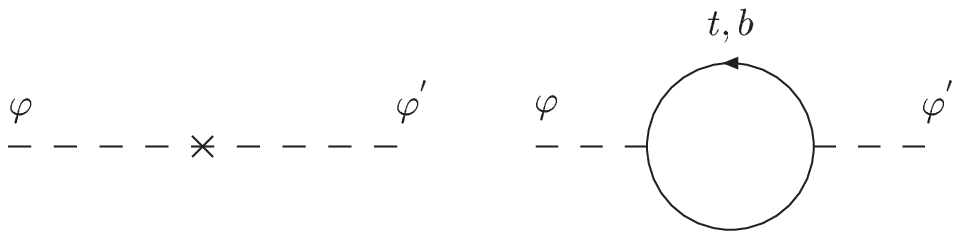,width = 8.0cm}
\end{center}
\caption{Feynman diagrams contributing to the neutral two-point proper vertices of auxiliary fields in the case of one quark family. The fields $\varphi$ and $\varphi^{'}$ stand for the four bosonic fields $\varphi^1$, $\varphi^2$, $\varphi^3$ and $G$.}
\label{fig:2hd_1fam_neutral}
\end{figure}

The Feynman diagrams we have to calculate are shown in Fig. \ref{fig:2hd_1fam_neutral}.
Using Eqs. (\ref{h22}) and (\ref{h23}), 
the non-vanishing two-point functions are 

\begin{eqnarray}
i\Gamma_{\varphi^2,\varphi^2}(p^2)&=&\frac{i N}{8\pi^2(v^2+w^2)} \Bigg\{  
		\;p^2 \sum_{q=t,b} K_q^2\,I(m_q^2;p^2)  \nonumber\\
 & &	-2\Lambda^2\sum_{q=t,b}\left(1-\frac{m_q^2}{\Lambda^2}\log(\Lambda^2/m_q^2)\right) 
		\left(\frac{w^3}{v}+\frac{v^3}{w} +2vw\right) 
		\lambda_q^{(1)}\lambda_q^{(2)}\cos\eta  \Bigg\}, \nonumber\\ \label{h64}  \\
i\Gamma_{G,G}(p^2)          &=& \frac{i N}{8\pi^2(v^2+w^2)}      
      \; p^2 \sum_{q=t,b}\: 2 \: m_q^2\: I(m_q^2;p^2),    \label{h65}     \\
i\Gamma_{\varphi^2,G}(p^2)  &=& \frac{i N}{8\pi^2(v^2+w^2)}     
      \; p^2 \sum_{q=t,b} \sqrt{2}\: m_q\: 
     K_q\, I(m_q^2;p^2),   \label{h66}           
\end{eqnarray} 
in the $CP$-odd sector, and

\begin{eqnarray}
i\Gamma_{\varphi^1,\varphi^1}(p^2)&=&\frac{i N}{8\pi^2(v^2+w^2)} \Bigg\{ 
	\;\sum_{q=t,b}\: (p^2-4 m_q^2)\: K_q^2\: I(m_q^2;p^2)  \nonumber \\
 & &	-2\Lambda^2\sum_{q=t,b}\left(1-\frac{m_q^2}{\Lambda^2}\log(\Lambda^2/m_q^2)\right) 
		\left(\frac{w^3}{v}+\frac{v^3}{w} +2vw\right) 
		\lambda_q^{(1)}\lambda_q^{(2)}\cos\eta \Bigg\}, \nonumber \\
       \label{h67}   \\
i\Gamma_{\varphi^3,\varphi^3}(p^2) &=& \frac{i N}{8\pi^2(v^2+w^2)}      
       \sum_{q=t,b}\: 2 \: m_q^2\:(p^2-4 m_q^2)\: I(m_q^2;p^2),    \label{h68}     \\
i\Gamma_{\varphi^1,\varphi^3}(p^2) &=& \frac{i N}{8\pi^2(v^2+w^2)}  
    \:\sum_{q=t,b} \sqrt{2}\: m_q\:(p^2-4 m_q^2)   
     K_q\, I(m_q^2;p^2),  \label{h69}            
\end{eqnarray} 
in the $CP$-even sector.
The factor $K_q$ is defined as

\begin{equation}\label{h73}
K_q\equiv w \cos\eta\,\lambda_q^{(1)}-v\,\lambda_q^{(2)},\qquad \text{for }q=t,b.
\end{equation}
The integral $I(m_q^2;p^2)$ is given by

\begin{equation}\label{h69b} 
I(m^2;p^2)= 
\frac{16 \pi^2}{i} \int \frac{d^4l}{(2\pi)^4}\:
\frac{1}{(l^2-m^2)[(l+p)^2-m^2]}.
\end{equation} 
The masses of the bound states are given by the values of $p^2$ at which the proper vertex matrix has vanishing eigenvalues.
From Eqs. (\ref{h65}) and (\ref{h66}) we see that at $p^2=0$ 
a zero eigenvalue  with associated eigenvector $G$, the neutral Goldstone boson, appears.
In the $CP$-even sector the $(\varphi^1,\varphi^1)$ entry of the $2\times 2$ matrix is of order\footnote{It follows from the reasoning after Eq. (\ref{h24}) that $\sum_{q=t,b}\left(1-\frac{m_q^2}{\Lambda^2}\log(\Lambda^2/m_q^2)\right)
\lambda_q^{(1)}\lambda_q^{(2)}\neq 0$.} 
$\Lambda^2$ and therefore, for $p^2\ll\Lambda^2$,
much bigger than the other matrix elements. 
In first approximation the smaller eigenvalue of this matrix is given by
$i\Gamma_{\varphi^3,\varphi^3}(p^2)$. 
From Eq. (\ref{h68}) we see that the propagator has a pole located at $p^2\approx (2 m_t)^2$.
The other two eigenvalues,
associated with a $CP$-even and a $CP$-odd field,
are of order $\Lambda^2$. \\

\begin{figure}[tbp]
\begin{center}
\psfig{file=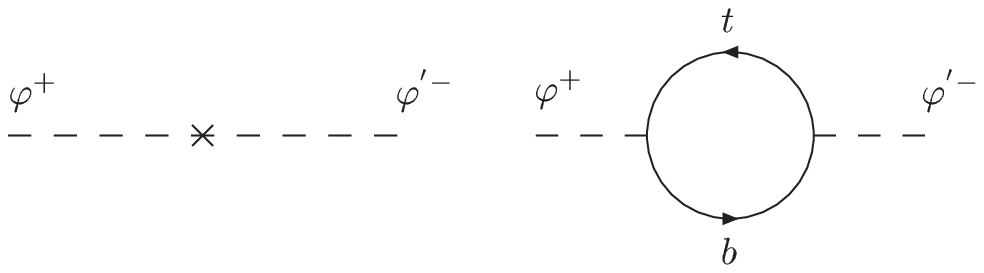,width = 8.0cm}
\end{center}
\caption{Feynman diagrams contributing to the charged two-point proper vertices of auxiliary fields in the case of one quark family. The fields $\varphi^\pm$ and $\varphi^{'\pm}$ denote the charged fields $\varphi^\pm$ and $G^\pm$.}
\label{fig:2hd_1fam_charged}
\end{figure}

Now we treat the charged sector in an analogous way.
We use the auxiliary field basis defined in Eqs. (\ref{h70}) and (\ref{h71}).
The Yukawa terms are

\begin{equation}\label{h72}
\mathcal{L}_{\text{Yukawa-charged}}=
-\frac{1}{\sqrt{v^2+w^2}}\left[ 
      \varphi^+ \,\bar{t}^{'} (K_t \,\projectorl- K_b \,\projectorr) b^{'}
+\sqrt{2}\, G^+ \,\bar{t}^{'} (m_t \,\projectorl- m_b \,\projectorr) b^{'} +h.c. \right] ,
\end{equation} 
where $\projectorl$ and $\projectorr$ are the left and right projectors.
For the two-point proper vertices,
represented by the diagrams of Fig. \ref{fig:2hd_1fam_charged},
we obtain

\begin{equation}\label{h74}
\begin{split}
i\Gamma_{\varphi^+,\varphi^-}(p^2)&=
-\frac{iN\Lambda^2}{8\pi^2 (v^2+w^2)}
\sum_{q=t,b}\left(1-\frac{m_q^2}{\Lambda^2}\log(\Lambda^2/m_q^2)\right) 
\left(\frac{w^3}{v}+\frac{v^3}{w} +2vw\right) 
\lambda_q^{(1)}\lambda_q^{(2)}\cos\eta            \\
& \quad+\frac{iN}{16\pi^2 (v^2+w^2)}  \Big\{ 
	[(p^2-m_t^2 -m_b^2)(K_t^2+K_b^2)+ 4 m_t m_b K_t K_b]\, I(m_t^2,m_b^2;p^2)      \\   
& \qquad\qquad\qquad\qquad\quad		
   +(K_t^2-K_b^2)
   \left(m_t^2 \log(\Lambda^2/m_t^2)-m_b^2 \log(\Lambda^2/m_b^2)\right) \Big\},
\end{split}
\end{equation}

\begin{equation}\label{h75}
\begin{split}
i\Gamma_{\varphi^+,G^-}(p^2)&=
\frac{\sqrt{2}\,iN}{16\pi^2 (v^2+w^2)}\Big\{  
    [p^2(m_t K_t+m_b K_b)-(m_t K_t-m_b K_b) (m_t^2-m_b^2)]\, I(m_t^2,m_b^2;p^2)\\ 
&\qquad\qquad\qquad\qquad\qquad  +(m_t K_t-m_b K_b)
	\left(m_t^2 \log(\Lambda^2/m_t^2)-m_b^2 \log(\Lambda^2/m_b^2)\right) \Big\}
\end{split}
\end{equation}

\begin{equation}\label{h76}
\begin{split}
i\Gamma_{G^+,G^-}(p^2)&=
\frac{iN}{8\pi^2 (v^2+w^2)}  \Big\{ 
	[p^2(m_t^2+m_b^2)-(m_t^2-m_b^2)^2]\, I(m_t^2,m_b^2;p^2)      \\   
& \qquad\qquad\qquad\qquad +(m_t^2-m_b^2)
   \left(m_t^2 \log(\Lambda^2/m_t^2)-m_b^2 \log(\Lambda^2/m_b^2)\right) \Big\},
\end{split}
\end{equation} 
with

\begin{equation}\label{h77} 
I(m_t^2,m_b^2;p^2)= 
\frac{16 \pi^2}{i} \int \frac{d^4l}{(2\pi)^4}\:
\frac{1}{(l^2-m_t^2)[(l+p)^2-m_b^2]}.
\end{equation} 
We have here a similar situation as in the neutral $CP$-odd sector.
The $(\varphi^+,\varphi^-)$ element is of order $\Lambda^2$ and
the other elements of the matrix vanish at $p^2=0$.
Therefore the two poles are located at $p^2=0$ and at $p^2=\mathcal{O}(\Lambda^2)$.

We turn now to the Yukawa coupling constants at low energy.
From the 8 scalar degrees of freedom which are present in this model,
3 are Goldstone bosons (they are treated in detail in Section \ref{section:gauge_boson_masses}) and 4 possess masses of order $\Lambda$,
which is the upper limit of the validity range of the model.
For energies lower than $\Lambda$ only one neutral scalar particle, with mass $m_H \approx 2m_t$, is present.
In a good approximation this particle can be described by the field $\varphi^3$.
Its Yukawa interaction at the scale $\Lambda$ is given by

\begin{equation}\label{h78} 
\mathcal{L}_{\text{Yukawa}} =
-\frac{\varphi^3}{\sqrt{v^2+w^2}}
(\sqrt{2}\, m_t \,\bar{t}^{'} t^{'}+\sqrt{2}\, m_b \,\bar{b}^{'} b^{'}).
\end{equation} 
Now we express the Yukawa coupling at zero momentum.
From the inverse of the propagator, Eq. (\ref{h68}), we obtain the appropriate scaling factor in order to get the conventional normalization of $\varphi^3$ at the scale $\mu=0$

\begin{equation}\label{h79}
\varphi^3\longrightarrow \Big(
\frac{N}{8\pi^2 (v^2+w^2)}\sum_{q=t,b} \: 2 m_q^2 \:\log\Lambda^2 /m_q^2
\Big)^{-1/2}    \varphi^3.
\end{equation}
Replacing Eq. (\ref{h79}) in Eq. (\ref{h78}) we get  the same interaction we found in Section \ref{section:1hd}, Eqs. (\ref{eff34}) and Eq. (\ref{eff35}), modified for the case of 2 quark flavors.
We see in this way that at energies much lower than $\Lambda$ this model cannot be distinguished from the one which has only one auxiliary Higgs doublet from the beginning.
In the last case the condition $G_t G_b = G_{tb}^2$ must hold, see Section \ref{section:1hd}.

\subsection{Composite Higgs boson masses: Summary}

We calculated the masses of the composite Higgs bosons when both auxiliary fields condense considering separately the cases of having $G_{tb}$ equal or different from zero.
There are qualitative differences between these two cases.
For $G_{tb}=0$ the Lagrangian possesses a PQ symmetry and the theory contains
an axion.
A further qualitative difference is related to quadratic divergences in the
two-point functions.
For $G_{tb}\neq 0$ there are quadratic divergences in these amplitudes
and as a consequence the theory predicts bound states 
with masses of order $\Lambda$, 
the upper limit of the validity range of the model.
In the case of having $G_{tb}=0$ the gap equations cancel all these divergences 
and all poles lie well below $\Lambda$. 
In the two cases there are, as expected, 3 Goldstone bosons related to 
the broken electroweak symmetry generators.
For $G_{tb}=0$ the theory predicts also an axion, 
2 neutral $CP$-even Higgs bosons with masses $2m_t$, $2m_b$ 
and a charged Higgs boson with a mass $\approx\sqrt{2(m_t^2+m_b^2)}$.
If $G_{tb}\neq 0$ one expects one neutral $CP$-even Higgs bosons
with mass $\approx 2 m_t$.
The other poles in this case are of order $\Lambda$
and cannot be interpreted a Higgs particles.

We see that both for the case of having $G_{tb}= 0$ and for $G_{tb}\neq 0$
the number of parameters at low energies is two.
Once one fixes them, e.g. by the experimental values of the quark masses, the model is completely specified.
For the case $G_{tb}\neq 0$ one would like to understand why the number of parameters is only two and not three.
The technical reason is that in order to satisfy the first derivative conditions simultaneously, the relation Eq. (\ref{h29}) must hold.
This supplies an extra constraint.

Finally it must be said that the case  $G_{tb}=0$ is experimentally ruled out because
an ``electroweak" axion has not been found.

\chapter{Three Generations of Quarks}\label{chapter:3fam}

So far we have studied models which either have only one composite Higgs doublet (Section \ref{section:1hd}), or we considered just the third family of quarks (Chapter \ref{chapter:1fam}).
In order to construct a model with three quark generations with dynamical EW and $CP$ symmetry breaking which leads to a quark mass spectrum, to family transitions, and to a $CP$ phase which are in accord with observations,
a more general four-fermion self-interaction is needed.
We investigate such a self-interaction term in this Chapter.

In order to get a realistic CKM matrix which is $CP$-violating,
one could start with a four-quark interaction term which violates the $CP$ symmetry explicitly. In that case one obtain naturally a non-trivial phase (and mixing angles) in the CKM matrix.
We are interested, however, in a different possibility.
We put the $CP$ symmetry on the same footing as the EW symmetry, that is, we consider it as a spontaneously broken symmetry.
That would give a dynamical explanation to this phenomenon (see Chapter \ref{chapter:spontaneous_cpv}). \\

The generalization of Eq. (\ref{h1}) to the case where all quarks interact is given by

\begin{equation}\label{3f1}
\mathcal{L}_{\text{4f}} =
G_{ijkl}\,(\bar{\psi}_{iL} u_{jR}) (\bar{u}_{lR}\psi_{kL})+
G_{ijkl}^{'}\,(\bar{\psi}_{iL} d_{jR}) (\bar{d}_{lR}\psi_{kL})+
[G_{ijkl}^{''}\,\epsilon^{ab}(\bar{\psi}_{iL}^a d_{jR})
    (\bar{\psi}_{kL}^b u_{lR})+ h.c. ].
\end{equation}
The notation used in this equation was introduced in Chapter \ref{chapter:eff_potential}.
The Lagrangian $\mathcal{L}_{\text{4f}}$ includes four-fermion terms which mix quark fields of different quark families.

Because we demand the Lagrangian to be $CP$-invariant,
all the coupling constants $G$ in $\mathcal{L}_{\text{4f}}$ are considered to be real.
In this case the number of parameters of Eq. (\ref{3f1}) is 171.\footnote{After requiring hermiticity of the Lagrangian there are 45 independent couplings $G$, 81  couplings $G^{'}$, and further 45 couplings $G^{''}$.}
Where we do not count the cutoff $\Lambda$ which is also a parameter of the model.

We now rewrite the interaction term, Eq. (\ref{3f1}), in terms of auxiliary fields which have the quantum numbers of the SM Higgs doublet field, as we explained in Chapter \ref{chapter:eff_potential}.
We do not consider completely arbitrary couplings $G$.
For simplicity we restrict ourselves to the subset for which the theory can be described by means of only two auxiliary fields,
$H^{(1)}$ and  $H^{(2)}$.
In terms of these auxiliary fields the four-fermion term $\mathcal{L}_{\text{4f}}$ is replaced by

\begin{equation}\label{3f2}
\mathcal{L}_{\text{4f}} \longrightarrow
\mathcal{L}_{\text{aux}}=
- \sum_{i=1}^2 m_{H_i}^2 H^{(i)\dagger}H^{(i)}+\mathcal{L}_{\text{Yukawa}},
\end{equation}
where
\begin{equation}\label{3f3}
\mathcal{L}_{\text{Yukawa}} = -\sum_{i=1}^2\;
(\;g_{kl}^{(i)}\:\bar{\psi}_{kL}u_{lR}\:H^{(i)}+
 h_{kl}^{(i)}\:\epsilon^{ab}\:\bar{\psi}_{kL}^a d_{lR}\:H^{b(i)*}
\;+\; h.c.\;).
\end{equation}
The relations between the real coupling constants in the two formulations of the model are given in Eqs. (\ref{int6}).
In consequence we restrict ourselves to a model with 36,
essentially the four $3\times 3$ Yukawa matrices, parameters
(plus the cutoff scale $\Lambda$).

We are now confronted with the following problem.
We want to find the values of the parameters of the model
such that the vacuum of the theory breaks the EW symmetry in the observed way.
Besides, the generated CKM matrix and the quark masses must correspond to their measured values.
However, the relation between the 36 parameters
of our model and the quantities to be reproduced is rather complicated.
In order to find an analytical solution we introduce in the next Section a self-consistent approach to the problem.

\section{Self-consistent Approach}

For the purpose of finding local minima of the effective potential we proceed in a similar way as we did in Chapter \ref{chapter:1fam} for a simpler Lagrangian.
Due to gauge invariance it is possible to consider the dependence of the effective potential on the auxiliary fields, $H^{(1)}$ and $H^{(2)}$ just through the four variables $v^{'}$, $w^{'}$, $\eta^{'}$, and $z^{'}$, where

\begin{equation}\label{3f4}
H^{(1)}=\bigg(\begin{matrix}
              \frac{v^{'}}{\sqrt{2}}\\0
              \end{matrix}\bigg),\qquad\qquad
H^{(2)}=\bigg(\begin{matrix}
              \frac{w^{'}e^{i\eta^{'}}}{\sqrt{2}}\\z{'}
              \end{matrix}\bigg),
\end{equation}
with $v^{'}, w^{'}, z^{'}>0$.
We examine the three conditions written in Eq. (\ref{h8a}) which are sufficient in order to have a local minimum which respects the electromagnetic $U(1)$ symmetry,
i.e. with $z=0$.
These conditions consist in having
a) vanishing first derivatives of the effective potential with respect to the variables $v^{'}$, $w^{'}$, and $\eta^{'}$,
b) a positive first derivative of $V_{\text{eff}}$ with respect to $z^{'2}$,
and c) a positive definite $3\times 3$ Hessian matrix related to the variables
$v^{'}$, $w^{'}$, and $\eta^{'}$.

We use a self-consistent method in order to find the portion of the parameter space for which the model generates realistic CKM matrix elements and quark masses.
First, we assume that the minimum of the effective potential
(denoted by non-primed symbols) is given by a field configuration with non-trivial values of $v$, $w$, and $\eta$ ($\neq 0,\pi$) and with $z=0$:

\begin{equation}\label{}
<H^{(1)}>\;=\bigg(\begin{matrix}
              \frac{v}{\sqrt{2}}\\0
              \end{matrix}\bigg),\qquad\qquad
<H^{(2)}>\;=\bigg(\begin{matrix}
              \frac{w\, e^{i\eta}}{\sqrt{2}}\\0
              \end{matrix}\bigg).
\end{equation}
VEVs of this form are necessary in order to have a theory with spontaneous $CP$ symmetry breaking and unbroken $U(1)_{em}$ symmetry.

Inserting the Higgs VEVs in the Yukawa interactions, Eq. (\ref{3f3}), one gets the quark mass term:

\begin{equation}\label{3f5}
\mathcal{L}_{\text{m}} = -\sum_{i=1}^2\;
(\;g_{kl}^{(i)}\:\bar{u}_{kL} u_{lR}\:<\phi^{0(i)}> +
 \;h_{kl}^{(i)}\:\bar{d}_{kL} d_{lR}\:<\phi^{0(i)}>^*
\;+\; h.c.\;).
\end{equation}
The quark mass matrices for the up- and down-type quarks in the last equation are in general non-diagonal.
In order to diagonalize them we perform the following chiral rotations:

\begin{equation}\label{3f6}
\begin{matrix}
u_{iR} = W_{ij}^u \;u_{jR}^{'}, & & u_{iL} = U_{ij}^u \;u_{jL}^{'},\\
&&\\
d_{iR} = W_{ij}^d \;d_{jR}^{'}, & & d_{iL} = U_{ij}^d \;d_{jL}^{'},
\end{matrix}
\end{equation}
where the primed fields denote the fermion fields in the mass basis and
$U^u$, $U^d$, $W^u$, $W^d$ are basis transformation matrices.
The CKM matrix is given by $V_{CKM}=U^{u\dagger}U^d$.
In the new basis the quark mass matrices, which are now diagonal and real are given by

\begin{equation}\label{3f7}
\begin{split}
M_u  &= \lambda_u^{(1)}\;\frac{v}{\sqrt{2}}+
         \lambda_u^{(2)}\;\frac{w\, e^{i\eta}}{\sqrt{2}}, \\
M_d  &= \lambda_d^{(1)}\;\frac{v}{\sqrt{2}}+
         \lambda_d^{(2)}\;\frac{w\, e^{i\eta}}{\sqrt{2}},
\end{split}\end{equation}
where the Yukawa couplings in the mass basis $\lambda_u^{(i)}$, $\lambda_d^{(i)}$ are defined by

\begin{equation}\label{3f8}
\begin{split}
g^{(i)} &\equiv U^u\;\lambda_u^{(i)}      \; W^{u\dagger},
                \quad\text{for }i=1,2, \\
h^{(i)} &\equiv U^d\;\lambda_d^{(i)\dagger}\;W^{d\dagger},
                \quad\text{for }i=1,2.
\end{split}\end{equation}
We emphasize that the Higgs VEVs in Eq. (\ref{3f7}) are still not determined.

Using the last definitions, the relations between the composite fields $H^{(i)}$ and their constituent quark fields are given by:

\begin{equation}\label{int11b}
\begin{split}
\phi^{0(i)}&=-\frac{1}{m_{H_i}^2} \;\Big(
\bar{u}_R^{'}\:\lambda_u^{(i)\dagger}\:u_L^{'} +
\bar{d}_L^{'}\:\lambda_d^{(i)\dagger}\:d_R^{'}  \Big),  \\
\phi^{-(i)}&=-\frac{1}{m_{H_i}^2} \;\Big(
\bar{u}_R^{'}\:\lambda_u^{(i)\dagger}\:V_{CKM}\:d_L^{'} -
\bar{u}_L^{'}\:V_{CKM}\:\lambda_d^{(i)\dagger}\:d_R^{'}  \Big),
\end{split}
\end{equation}
for $i=1,2$.
Besides, the interaction term $\mathcal{L}_{\text{Yukawa}}$ is given in this basis by

\begin{equation}\label{int11}
\begin{split}
\mathcal{L}_{\text{Yukawa}} = -\sum_{i=1}^2\;\Big(\:
& \bar{u}_L^{'}\,\lambda_u^{(i)}\, u_R^{'}\:\phi^{0(i)}
   -\bar{u}_L^{'}V_{CKM}\,\lambda_d^{(i)\dagger}\, d_R^{'}\:\phi^{+(i)}  \\
&+\:\bar{d}_L^{'}V_{CKM}^\dagger\,\lambda_u^{(i)}\, u_R^{'}\:\phi^{-(i)}
   +\bar{d}_L^{'}\,\lambda_d^{(i)\dagger}\, d_R^{'}\:\phi^{0(i)*}
\;+\; h.c.\;\Big),
\end{split}
\end{equation}
with $\phi^{+(i)}\equiv\phi^{-(i)*}$.

Combining Eqs. (\ref{3f7}) and (\ref{3f8}) and using the fact that the matrices
$g^{(i)}$ and  $h^{(i)}$ are real, it is possible to write the Yukawa
couplings in the weak basis, $g^{(i)}$, $h^{(i)}$, as a function of the Higgs VEVs, the quark masses, and the basis transformation matrices:

\begin{equation}\label{2hd5}
\begin{split}
g^{(1)} &= \frac{\sqrt{2}}{v}\;\big[\mathcal{R}e(U^u M_u W^{u\dagger})
                  -\cot\eta   \; \mathcal{I}m(U^u M_u W^{u\dagger})\big],\\
g^{(2)} &= \frac{\sqrt{2}}{w\,\sin\eta} \;
                                  \mathcal{I}m(U^u M_u W^{u\dagger}),\\
h^{(1)} &= \frac{\sqrt{2}}{v}\;\big[\mathcal{R}e(U^d M_d W^{d\dagger})
                  +\cot\eta   \; \mathcal{I}m(U^d M_d W^{d\dagger})\big],\\
h^{(2)} &= -\frac{\sqrt{2}}{w\,\sin\eta} \;
                                  \mathcal{I}m(U^d M_d W^{d\dagger}).\\
\end{split}\end{equation}

We have in this way transformed the original problem
into one which can be solved in a self-consistent way:
We must find values of $v$, $w$, and $\eta$,
to which we associate the Yukawa couplings given in Eqs. (\ref{2hd5}),
such that the resulting effective potential has its minimum at the same values $v$, $w$ and $\eta$ (besides $z=0$).
For this purpose we can vary the basis transformation
matrices $U^u$, $U^d$, $W^u$ and $W^d$.
These are arbitrary unitary matrices which must obey the condition
$U^{u\dagger}U^d=V_{CKM}$.

\section{Derivatives of the Effective Potential}

In this Section we calculate the first and second derivatives of the effective potential related to the Lagrangian Eq. (\ref{3f1}).
These expressions are needed in order to apply the minimum conditions given in Eq.  (\ref{h8a}).

The effective potential was calculated in Chapter \ref{chapter:eff_potential}.
We are considering the special case in which the Lagrangian $\mathcal{L}_{\text{4f}}$ given in Eq. (\ref{3f1}) can be rewritten with the help of only two auxiliary fields.
The function to be  minimized is given by

\begin{equation}\label{3f9}
V_{\text{eff}}(H^{(1)},H^{(2)})=m_{H_i}^2 H^{(i)\dagger}H^{(i)}
-\frac{N}{8\pi^2}\int_0^{\Lambda^2} k^2\,dk^2\:
\log\det(k^2+A),
\end{equation}
with the $6\times 6$ matrix $A$ given by

\begin{equation*}\label{}
A=
\begin{pmatrix}
 g^{(i)\dagger}g^{(j)} H^{(i)\dagger}H^{(j)} &
 g^{(i)\dagger}h^{(j)} \epsilon^{ab} H^{a(i)*}H^{b(j)*} \\
-h^{(i)\dagger}g^{(j)} \epsilon^{ab} H^{a(i)}H^{b(j)} &
 h^{(i)\dagger}h^{(j)} H^{(j)\dagger}H^{(i)}
\end{pmatrix}.
\end{equation*}
In the last equations and in the following ones a sum over repeated indices is understood.

We only consider derivatives of $V_{\text{eff}}$ at the hyperplane given by $z^{'2}=0$.
At each point of the hyperplane it is possible to diagonalize the matrix $A$.
This is equivalent to the diagonalization of the quark mass matrices that can be done performing the chiral transformations introduced in Eqs. (\ref{3f6}).
The change of fermion basis modifies the matrix $A$ of the fermion determinant in the following way:

\begin{equation}\label{2hd6}
A\longrightarrow
\begin{pmatrix}W^{u\dagger}&0\\0&W^{d\dagger}\end{pmatrix}
A
\begin{pmatrix}W^u&0\\0&W^d\end{pmatrix},
\end{equation}
where the r.h.s. of the last expression is a diagonal matrix.
The determinant $\det(k^2+A)$  does not change after the substitution Eq. (\ref{2hd6}).

We put the matrix $A$ in diagonal form because it simplifies the calculation of the derivatives of the effective potential.
In particular, the matrix $A$ at the minimum of $V_{\text{eff}}$ corresponds to a diagonal matrix formed by the squared quark masses

\begin{equation}\label{2hd7a}
A(v,w,\eta,0) = \text{diag}(m_u^2,m_c^2,m_t^2,m_d^2,m_s^2,m_b^2)=
\begin{pmatrix}
M_u^2 & 0\\0&M_d^2
\end{pmatrix}.
\end{equation}

Using Eqs. (\ref{3f6}) and (\ref{3f8}) the matrix $A$ can be written as

\begin{equation}\label{2hd7b}
A(v^{'},w^{'},\eta^{'},z^{'2}) =
\begin{pmatrix}
(\lambda_u^{(i)}H^{(i)})^\dagger (\lambda_u^{(j)}H^{(j)})&
 \epsilon^{ab}     (\lambda_u^{(i)}H^{a(i)})^\dagger
 \:V_{CKM}         (\lambda_d^{(j)}H^{b(j)})^\dagger             \\
-\epsilon^{ab}     (\lambda_d^{(i)}H^{a(i)})
 \:V_{CKM}^\dagger (\lambda_u^{(j)}H^{b(j)})               &
(\lambda_d^{(i)}H^{a(i)*}) (\lambda_d^{(j)}H^{a(j)*})^\dagger
\end{pmatrix}.
\end{equation}

\subsection{First Derivatives of the Effective Potential}
\label{subsection:1derivatives_3f}

Having expressed the effective potential in a more convenient form,
we calculate its first derivatives.
First we consider the derivatives with respect to the variables $v^{'}$,
$w^{'}$, and $\eta^{'}$.
The following relation will be useful:

\begin{equation}\label{2hd8}
\frac{\partial}{\partial\theta} \log\det(k^2+A)\Big|_{(v,w,\eta,0)} =
\sum_{i=1}^6\: \frac{1}{k^2+m_i^2}\;
\frac{\partial}{\partial\theta}A_{ii}\Big|_{(v,w,\eta,0)},
\end{equation}
for $\theta=v^{'},w^{'}$ and $\eta^{'}$.
The fact that at the evaluation point the matrix $A$ is diagonal
and that the first derivatives of its matrix elements with respect to
the variables $\theta$ are well defined
(that is not true for $z^{'2}$, see below)
leads to the previous relation.
From Eqs. (\ref{3f9}) and (\ref{2hd8}) we obtain

\begin{equation}\label{2hd9}
\begin{split}
\frac{\partial V_{\text{eff}}}{\partial
\theta}\Big|_{(v,w,\eta,0)} &=
\frac{\partial}{\partial\theta}\Big(m_{H_i}^2 H^{(i)\dagger}H^{(i)}\Big)-
\frac{N}{8\pi^2}\sum_{i=1}^6\int_0^{\Lambda^2}
\frac{k^2dk^2}{k^2+m_i^2}\;
\frac{\partial}{\partial\theta}A_{ii}\Big|_{(v,w,\eta,0)},\\
&=
\frac{\partial}{\partial\theta}\Big(m_{H_i}^2 H^{(i)\dagger}H^{(i)}\Big)-
\frac{N\Lambda^2}{8\pi^2}\sum_{i=1}^6
\bigg(1-\frac{m_i^2}{\Lambda^2}\log\bigg(\frac{\Lambda^2}{m_i^2}+1\bigg)\bigg)
\;\frac{\partial}{\partial\theta}A_{ii}\Big|_{(v,w,\eta,0)},
\end{split}
\end{equation}
for $\theta=v^{'},w^{'}$ and $\eta^{'}$.
Only the diagonal elements of $A$ are relevant to these first derivatives
of $V_{\text{eff}}$.
Writing  the matrix $A$ in block form

\begin{equation}\label{2hd10}
A=\begin{pmatrix}
A11&A12\\A21&A22
\end{pmatrix},
\end{equation}
we need to concentrate only on

\begin{equation}\label{2hd11}
\begin{split}
A11&=(\lambda_u^{(1)\dagger}\lambda_u^{(1)})\;\frac{v^{'2}}{2}
      + (\lambda_u^{(2)\dagger}\lambda_u^{(2)})\;\frac{w^{'2}}{2}
 +(\lambda_u^{(1)\dagger}\lambda_u^{(2)}e^{ i\eta^{'}}+
   \lambda_u^{(2)\dagger}\lambda_u^{(1)}e^{-i\eta^{'}})
                                                   \;\frac{v^{'}w^{'}}{2},\\
A22&=(\lambda_d^{(1)}\lambda_d^{(1)\dagger})\;\frac{v^{'2}}{2}
      + (\lambda_d^{(2)}\lambda_d^{(2)\dagger})\;\frac{w^{'2}}{2}
 +(\lambda_d^{(2)}\lambda_d^{(1)\dagger}e^{ i\eta^{'}}+
   \lambda_d^{(1)}\lambda_d^{(2)\dagger}e^{-i\eta^{'}})\;\frac{v^{'}w^{'}}{2}.
\end{split}
\end{equation}
The derivatives of the diagonal elements of $A$ with respect to $\eta^{'}$ are given by

\begin{equation}\label{2hd12}
\begin{split}
\frac{\partial}{\partial\eta^{'}} (A11)_{ii}&=
-v^{'}w^{'}\: \mathcal{I}m\Big[
  \Big(\lambda_u^{(1)\dagger}\;\lambda_u^{(2)}\Big)_{ii}
                                 \;e^{ i\eta^{'}}\Big],  \\
\frac{\partial}{\partial\eta^{'}} (A22)_{ii}&=
-v^{'}w^{'}\: \mathcal{I}m\Big[
  \Big(\lambda_d^{(2)}\;\lambda_d^{(1)\dagger}\Big)_{ii}
                                 \;e^{ i\eta^{'}}\Big].
\end{split}
\end{equation}
In this way the condition

\begin{equation}\label{2hd13}
\frac{\partial V_{eff}}{\partial\eta^{'}}\Bigg|_{(v,w,\eta,0)} = 0
\end{equation}
takes the form

\begin{equation}\label{2hd13a}
v^{'}w^{'}\:\sum_{i=1}^6
\bigg(1-\frac{m_i^2}{\Lambda^2}\log\bigg(\frac{\Lambda^2}{m_i^2}+1\bigg)\bigg)
\;\mathcal{I}m\big(
  \Sigma_{ii}^{(0)}\;e^{ i\eta^{'}}\big)\Big|_{(v,w,\eta,0)}=0,
\end{equation}
where $\Sigma^{(0)}$ is the $6\times 6$ block-diagonal matrix given by

\begin{equation}\label{2hd13b}
\Sigma^{(0)}=
\begin{pmatrix}
\lambda_u^{(1)\dagger}\lambda_u^{(2)} & 0 \\
0 & \lambda_d^{(2)}\lambda_d^{(1)\dagger}
\end{pmatrix}.
\end{equation}
In order to have spontaneous CP violation both $v^{'}$ and $w^{'}$ must be
non-zero at the minimum. Thus we obtain

\begin{equation}\label{2hd14}
\sum_{i=1}^6
\bigg(1-\frac{m_i^2}{\Lambda^2}\log\bigg(\frac{\Lambda^2}{m_i^2}+1\bigg)\bigg)
\;\mathcal{I}m\big(
  \Sigma_{ii}^{(0)}\;e^{ i\eta^{'}}\big)\Big|_{\eta^{'}=\eta}=0.
\end{equation}
Note that the last condition does not depend on $v^{'}$ or $w^{'}$ explicitly.
Now we consider the first derivative of $V_{\text{eff}}$
with respect to the variable $v^{'}$.
The derivatives of the diagonal elements of $A$ are given by

\begin{equation}\label{2hd15}
\begin{split}
\frac{\partial}{\partial v^{'}}(A11)_{ii}&=
v^{'}\:\Big(\lambda_u^{(1)\dagger}\;\lambda_u^{(1)}\Big)_{ii}+
w^{'}\:\mathcal{R}e\Big[
   \Big(\lambda_u^{(1)\dagger}\;\lambda_u^{(2)}\Big)_{ii}
                                 \;e^{i\eta^{'}}\Big],       \\
\frac{\partial}{\partial v^{'}}(A22)_{ii}&=
v^{'}\:\Big(\lambda_d^{(1)}\;\lambda_d^{(1)\dagger}\Big)_{ii}+
w^{'}\:\mathcal{R}e\Big[
   \Big(\lambda_d^{(2)}\;\lambda_d^{(1)\dagger}\Big)_{ii}
                                 \;e^{i\eta^{'}}\Big].
\end{split}
\end{equation}
Therefore the condition

\begin{equation}\label{2hd15a}
\frac{\partial V_{\text{eff}}}{\partial v^{'}}\Bigg|_{(v,w,\eta,0)} = 0
\end{equation}
takes the form

\begin{equation}\label{2hd16}
m_{H_1}^2=\frac{N\Lambda^2}{8\pi^2}\sum_{i=1}^6
\bigg(1-\frac{m_i^2}{\Lambda^2}\log\bigg(\frac{\Lambda^2}{m_i^2}+1\bigg)\bigg)
\Big(\Sigma^{(1)}_{ii} + \frac{w^{'}}{v^{'}}\:
       \mathcal{R}e(\Sigma^{(0)}_{ii}e^{i\eta^{'}})\Big)\Big|_{(v,w,\eta,0)},
\end{equation}
where we used the following notation

\begin{equation}\label{2hd16a}
\Sigma^{(i)}=
\begin{pmatrix}
\lambda_u^{(i)\dagger}\lambda_u^{(i)} & 0 \\
0 & \lambda_d^{(i)}\lambda_d^{(i)\dagger}
\end{pmatrix}\:,
\;\qquad\text{for }i=1,2.
\end{equation}
In a similar way we obtain the first derivative condition related to the variable $w^{'}$:

\begin{equation}\label{2hd17}
m_{H_2}^2=\frac{N\Lambda^2}{8\pi^2}\sum_{i=1}^6
\bigg(1-\frac{m_i^2}{\Lambda^2}\log\bigg(\frac{\Lambda^2}{m_i^2}+1\bigg)\bigg)
\Big(\Sigma^{(2)}_{ii} + \frac{v^{'}}{w^{'}}\:
       \mathcal{R}e(\Sigma^{(0)}_{ii}e^{i\eta^{'}})\Big)\Big|_{(v,w,\eta,0)}.
\end{equation}

Now we consider the dependence of the effective potential on $z^{'2}$.
The matrix $A$ at a given point $(v^{'},w^{'},\eta^{'},z^{'2})$ can be
written as a sum of the matrix $A$ evaluated at the point
$(v^{'},w^{'},\eta^{'},0)$ in its diagonal basis, and terms
proportional to $z^{'}$ and $z^{'2}$:

\begin{equation}\label{2hd18}
\begin{split}
A11&= M_u^2+ \lambda_u^{(2)\dagger}\lambda_u^{(2)}\:z^{'2}, \\
A22&= M_d^2+ \lambda_d^{(2)}\lambda_d^{(2)\dagger}\:z^{'2}, \\
A12&= \frac{z^{'}v^{'}}{\sqrt{2}}\;
 \Big(\lambda_u^{(2)\dagger}\,V_{CKM}\,\lambda_d^{(1)\dagger}-
      \lambda_u^{(1)\dagger}\,V_{CKM}\,\lambda_d^{(2)\dagger}\Big),\\
A21&= (A12)^\dagger,
\end{split}
\end{equation}
that is,

\begin{equation}\label{2hd19}
A= \begin{pmatrix}
M_u^2+ \lambda_u^{(2)\dagger}\lambda_u^{(2)}\:z^{'2} & z^{'}\,T\\
 z^{'}\,T^\dagger & M_d^2+\lambda_d^{(2)}\lambda_d^{(2)\dagger}\:z^{'2}
                 \end{pmatrix},
\end{equation}
with

\begin{equation}\label{2hd20}
T=\frac{v^{'}}{\sqrt{2}}\;
 \Big(\lambda_u^{(2)\dagger}\,V_{CKM}\,\lambda_d^{(1)\dagger}-
      \lambda_u^{(1)\dagger}\,V_{CKM}\,\lambda_d^{(2)\dagger}\Big).
\end{equation}
The determinant of $(k^2+A)$ contains only even powers of $z^{'}$.
This is because all terms have the same number of factors $z^{'}$ coming
from the submatrix $A12$ and from the submatrix $A21$.
Thus, the effective potential written as a function of $z^{'2}$ is given by

\begin{equation}\label{2hd22}
V_{\text{eff}}= \text{const}+m_{H_2}^2\, z^{'2}-\frac{N}{8\pi^2}\int_0^{\Lambda^2}
k^2dk^2 \log\det(k^2+A),
\end{equation}
where the determinant $\det(k^2+A)$ is a
sixth degree polynomial of $z^{'2}$:

\begin{equation}\label{2hd21}
\det(k^2+A)=c_0+c_2\, z^{'2}+c_4 \,z^{'4}+\ldots + c_{12}\, z^{'12}.
\end{equation}
The first derivative of the effective potential with respect to $z^{'2}$
at the point $z^{'2}=0$ is given by

\begin{equation}\label{2hd23}
\frac{\partial V_{\text{eff}}}{\partial z^{'2}}\bigg|_{z^{'2}=0}=
\:m_{H_2}^2-\frac{N}{8\pi^2}\int_0^{\Lambda^2}
\frac{k^2dk^2}{\overset{6}{\underset{i=1}{\Pi}} (k^2+m_i^2)}\;\:c_2\, ,
\end{equation}
with $c_2$ defined in Eq. (\ref{2hd21}).
There are two types of terms which contribute to $c_2$,

\begin{equation}\label{2hd24}
\frac{c_2}{\overset{6}{\underset{i=1}{\Pi}} (k^2+m_i^2)}=\;
\sum_{i=1}^6
\frac{\Sigma^{(2)}_{ii}}{k^2+m_i^2}-
\sum_{\begin{matrix}\scriptstyle i=u,c,t \\
                   \scriptstyle j=d,s,b   \end{matrix}}
\frac{|T_{ij}|^2}{(k^2+m_i^2)(k^2+m_j^2)},
\end{equation}
namely, terms which are a product of only diagonal elements of the matrix $(k^2+A)$,
and terms with one factor from the submatrix
$A12$, one factor from $A21$, and further 4 diagonal factors.
These two type of terms are given by the first and second contribution on the r.h.s. of Eq. (\ref{2hd24}) respectively.
The sum in the second term is over all the entries of the matrix $T$.
Inserting $c_2$ into Eq. (\ref{2hd23}) and performing the momentum integral,
we obtain

\begin{equation}\label{2hd25}
\begin{split}
\frac{dV_{\text{eff}}}{dz^{'2}}\bigg|_{z^{'2}=0}=
\:&m_{H_2}^2-\frac{N\Lambda^2}{8\pi^2}\;\sum_{i=1}^6
\bigg(1-\frac{m_i^2}{\Lambda^2}\log\bigg(\frac{\Lambda^2}{m_i^2}+1\bigg)\bigg)
\Sigma^{(2)}_{ii} \\
&+\frac{N\Lambda^2}{8\pi^2}\;
\sum_{\begin{matrix}\scriptstyle i=u,c,t \\
                   \scriptstyle j=d,s,b   \end{matrix}}
\frac{|T_{ij}|^2}{\Lambda^2}
\: I(m_i^2,m_j^2),
\end{split}
\end{equation}
with

\begin{equation}\label{2hd25b}
\begin{split}
I(m_i^2,m_j^2)&=
\int_0^{\Lambda^2}\frac{k^2dk^2}{(k^2+m_i^2)(k^2+m_j^2)},    \\
&=    \frac{m_i^2}{m_i^2-m_j^2}\log\bigg(\frac{\Lambda^2}{m_i^2}+1\bigg)+
      \frac{m_j^2}{m_j^2-m_i^2}\log\bigg(\frac{\Lambda^2}{m_j^2}+1\bigg).
\end{split}
\end{equation}
If we finally make use of the first derivative condition related to $w^{'}$
(Eq. (\ref{2hd17})) we obtain

\begin{equation}\label{2hd26}
\begin{split}
\frac{\partial V_{\text{eff}}}{\partial z^{'2}}\bigg|_{(v,w,\eta,0)}=&\;\:
\frac{N\Lambda^2}{8\pi^2}\;\sum_{i=1}^6
\bigg(1-\frac{m_i^2}{\Lambda^2}\log\bigg(\frac{\Lambda^2}{m_i^2}+1\bigg)\bigg)
    \frac{v^{'}}{w^{'}}\;
    \mathcal{R}e\Big( \Sigma^{(0)}_{ii}\;e^{i\eta^{'}}\Big)
                                        \bigg|_{(v,w,\eta,0)} \\
&\;+\frac{N\Lambda^2}{8\pi^2}
\sum_{\begin{matrix}\scriptstyle i=u,c,t \\
                   \scriptstyle j=d,s,b   \end{matrix}}
\frac{|T_{ij}|^2}{\Lambda^2}\bigg|_{v^{'}=v}
I(m_i^2,m_j^2).
\end{split}
\end{equation}
The relevant results of this Subsection are given in Eqs. (\ref{2hd14}), (\ref{2hd16}), (\ref{2hd17}) and (\ref{2hd26}).

\subsection{Analysis of the First Derivatives}

Before we calculate the second derivatives of $V_{\text{eff}}$,
we analyze the minimum conditions involving the first derivatives.
One can easily find a set of parameters which obey the minimum conditions associated with the variables $v^{'}$ and $w^{'}$,
Eqs. (\ref{2hd16}) and (\ref{2hd17}).
It suffices to choose suitable Higgs mass parameters on the
l.h.s. of these equations.
As in the previous cases which we have studied, fine-tuning is needed in each equation in order to get $m_q\ll \Lambda$.
The other two conditions, Eqs. (\ref{2hd14}) and (\ref{2hd26}),
demand a self-consistent treatment.

Let us start with the first derivative condition related to $\eta^{'}$.
Eq. (\ref{2hd14}) is equivalent to

\begin{equation}\label{2hd28}
\frac{2}{vw}\:\bigg( (a+\cot\eta\; b)\frac{\sin\eta^{'}}{\sin\eta}+
                    (c+\cot\eta \;d)\frac{\cos\eta^{'}}{\sin\eta}
              \bigg)=0,
\end{equation}
where $a$, $b$, $c$, and $d$  are real coefficients defined by

\begin{equation}\label{2hd28b}
\begin{split}
\sum_{i=1}^6 \left( 1-\frac{m_i^2}{\Lambda^2}\log(\Lambda^2/m_i^2)\right)
   \mathcal{R}e\left(\Sigma^{(0)}_{ii}\right) =
\frac{2}{vw\sin\eta}\:(a+\cot\eta\; b),   \\
\sum_{i=1}^6 \left( 1-\frac{m_i^2}{\Lambda^2}\log(\Lambda^2/m_i^2)\right)
   \mathcal{I}m\left(\Sigma^{(0)}_{ii}\right) =
\frac{2}{vw\sin\eta}\:(c+\cot\eta\; d).
\end{split}
\end{equation}
Direct expressions for these coefficients are given by

\begin{equation*}\label{}
\begin{split}
a&=\text{tr}\bigg[\mathcal{R}e\bigg(\:
  W^{u\dagger}\:\mathcal{R}e(W^u M_u U^{u\dagger})
              \:\mathcal{I}m(U^u M_u W^{u\dagger})\:W^u +
  W^{d\dagger}\:\mathcal{I}m(W^d M_d U^{d\dagger})
              \:\mathcal{R}e(U^d M_d W^{d\dagger})\:W^d \:\bigg)\bigg]\\
&\qquad\qquad -\sum_{i=u,c,t}\frac{m_i^2}{\Lambda^2}\log(\Lambda^2/m_i^2)\:
                 \mathcal{R}e\Big(\:
  W^{u\dagger}\:\mathcal{R}e(W^u M_u U^{u\dagger})
              \:\mathcal{I}m(U^u M_u W^{u\dagger})\:W^u\Big)_{ii}     \\
&\qquad\qquad -\sum_{i=d,s,b}\frac{m_i^2}{\Lambda^2}\log(\Lambda^2/m_i^2)\:
                 \mathcal{R}e\Big(\:
  W^{d\dagger}\:\mathcal{I}m(W^d M_d U^{d\dagger})
              \:\mathcal{R}e(U^d M_d W^{d\dagger})\:W^d\Big)_{ii},\\
b&=\text{tr}\bigg[\mathcal{R}e\bigg(\:
  W^{u\dagger}\:\mathcal{I}m(W^u M_u U^{u\dagger})
              \:\mathcal{I}m(U^u M_u W^{u\dagger})\:W^u +
  W^{d\dagger}\:\mathcal{I}m(W^d M_d U^{d\dagger})
              \:\mathcal{I}m(U^d M_d W^{d\dagger})\:W^d \:\bigg)\bigg]\\
&\qquad\qquad -\sum_{i=u,c,t}\frac{m_i^2}{\Lambda^2}\log(\Lambda^2/m_i^2)\:
                 \mathcal{R}e\Big(\:
  W^{u\dagger}\:\mathcal{I}m(W^u M_u U^{u\dagger})
              \:\mathcal{I}m(U^u M_u W^{u\dagger})\:W^u\Big)_{ii}     \\
&\qquad\qquad -\sum_{i=d,s,b}\frac{m_i^2}{\Lambda^2}\log(\Lambda^2/m_i^2)\:
                 \mathcal{R}e\Big(\:
  W^{d\dagger}\:\mathcal{I}m(W^d M_d U^{d\dagger})
              \:\mathcal{I}m(U^d M_d W^{d\dagger})\:W^d\Big)_{ii},\\
\end{split}
\end{equation*}

\begin{equation}\label{}
\begin{split}
c&=\text{tr}\bigg[\mathcal{I}m\bigg(\:
  W^{u\dagger}\:\mathcal{R}e(W^u M_u U^{u\dagger})
              \:\mathcal{I}m(U^u M_u W^{u\dagger})\:W^u +
  W^{d\dagger}\:\mathcal{I}m(W^d M_d U^{d\dagger})
              \:\mathcal{R}e(U^d M_d W^{d\dagger})\:W^d \:\bigg)\bigg]\\
&\qquad\qquad -\sum_{i=u,c,t}\frac{m_i^2}{\Lambda^2}\log(\Lambda^2/m_i^2)\:
                 \mathcal{I}m\Big(\:
  W^{u\dagger}\:\mathcal{R}e(W^u M_u U^{u\dagger})
              \:\mathcal{I}m(U^u M_u W^{u\dagger})\:W^u\Big)_{ii}     \\
&\qquad\qquad -\sum_{i=d,s,b}\frac{m_i^2}{\Lambda^2}\log(\Lambda^2/m_i^2)\:
                 \mathcal{I}m\Big(\:
  W^{d\dagger}\:\mathcal{I}m(W^d M_d U^{d\dagger})
              \:\mathcal{R}e(U^d M_d W^{d\dagger})\:W^d\Big)_{ii},\\
d&=\text{tr}\bigg[\mathcal{I}m\bigg(\:
  W^{u\dagger}\:\mathcal{I}m(W^u M_u U^{u\dagger})
              \:\mathcal{I}m(U^u M_u W^{u\dagger})\:W^u +
  W^{d\dagger}\:\mathcal{I}m(W^d M_d U^{d\dagger})
              \:\mathcal{I}m(U^d M_d W^{d\dagger})\:W^d \:\bigg)\bigg]\\
&\qquad\qquad -\sum_{i=u,c,t}\frac{m_i^2}{\Lambda^2}\log(\Lambda^2/m_i^2)\:
                 \mathcal{I}m\Big(\:
  W^{u\dagger}\:\mathcal{I}m(W^u M_u U^{u\dagger})
              \:\mathcal{I}m(U^u M_u W^{u\dagger})\:W^u\Big)_{ii}     \\
&\qquad\qquad -\sum_{i=d,s,b}\frac{m_i^2}{\Lambda^2}\log(\Lambda^2/m_i^2)\:
                 \mathcal{I}m\Big(\:
  W^{d\dagger}\:\mathcal{I}m(W^d M_d U^{d\dagger})
              \:\mathcal{I}m(U^d M_d W^{d\dagger})\:W^d\Big)_{ii}.
\end{split}
\end{equation}
Making use of the cyclicity of the trace, the last expressions
simplify to:

\begin{equation}\label{2hd29}
\begin{split}
a&=\text{tr}\big[
              \:\mathcal{R}e(W^u M_u U^{u\dagger})
              \:\mathcal{I}m(U^u M_u W^{u\dagger}) +
              \:\mathcal{I}m(W^d M_d U^{d\dagger})
              \:\mathcal{R}e(U^d M_d W^{d\dagger})\: \big]\\
&\qquad\qquad -\sum_{i=u,c,t}\frac{m_i^2}{\Lambda^2}\log(\Lambda^2/m_i^2)\:
                 \mathcal{R}e\Big(\:
  W^{u\dagger}\:\mathcal{R}e(W^u M_u U^{u\dagger})
              \:\mathcal{I}m(U^u M_u W^{u\dagger})\:W^u\Big)_{ii}     \\
&\qquad\qquad -\sum_{i=d,s,b}\frac{m_i^2}{\Lambda^2}\log(\Lambda^2/m_i^2)\:
                 \mathcal{R}e\Big(\:
  W^{d\dagger}\:\mathcal{I}m(W^d M_d U^{d\dagger})
              \:\mathcal{R}e(U^d M_d W^{d\dagger})\:W^d\Big)_{ii},\\
b&=\text{tr}\big[
              \:\mathcal{I}m(W^u M_u U^{u\dagger})
              \:\mathcal{I}m(U^u M_u W^{u\dagger})\: +
              \:\mathcal{I}m(W^d M_d U^{d\dagger})
              \:\mathcal{I}m(U^d M_d W^{d\dagger})\:\big]\\
&\qquad\qquad -\sum_{i=u,c,t}\frac{m_i^2}{\Lambda^2}\log(\Lambda^2/m_i^2)\:
                 \mathcal{R}e\Big(\:
  W^{u\dagger}\:\mathcal{I}m(W^u M_u U^{u\dagger})
              \:\mathcal{I}m(U^u M_u W^{u\dagger})\:W^u\Big)_{ii}     \\
&\qquad\qquad -\sum_{i=d,s,b}\frac{m_i^2}{\Lambda^2}\log(\Lambda^2/m_i^2)\:
                 \mathcal{R}e\Big(\:
  W^{d\dagger}\:\mathcal{I}m(W^d M_d U^{d\dagger})
              \:\mathcal{I}m(U^d M_d W^{d\dagger})\:W^d\Big)_{ii},\\
c&=-\sum_{i=u,c,t}\frac{m_i^2}{\Lambda^2}\log(\Lambda^2/m_i^2)\:
                 \mathcal{I}m\Big(\:
  W^{u\dagger}\:\mathcal{R}e(W^u M_u U^{u\dagger})
              \:\mathcal{I}m(U^u M_u W^{u\dagger})\:W^u\Big)_{ii}     \\
&\qquad  -\sum_{i=d,s,b}\frac{m_i^2}{\Lambda^2}\log(\Lambda^2/m_i^2)\:
                 \mathcal{I}m\Big(\:
  W^{d\dagger}\:\mathcal{I}m(W^d M_d U^{d\dagger})
              \:\mathcal{R}e(U^d M_d W^{d\dagger})\:W^d\Big)_{ii},\\
d&=0.
\end{split}
\end{equation}
Because $d$ vanishes, Eq. (\ref{2hd28}) becomes

\begin{equation}\label{2hd31}
(a+\cot\eta\; b)\frac{\sin\eta^{'}}{\sin\eta}+
        c\;     \frac{\cos\eta^{'}}{\sin\eta} =0.
\end{equation}
In order to obtain a solution for $\eta^{'}$ different from
$\sin\eta^{'}=0$ the coefficient $c$ must not vanish.
Remember that we formulate the problem in a self-consistent way.
We obtain the required quark masses and the CKM matrix only  if the
solution for $\eta^{'}$ of Eq. (\ref{2hd31}) is given by $\eta^{'}=\eta$.
The self-consistent solution of Eq. (\ref{2hd31}) is given by

\begin{equation}\label{2hd32}
\cot\eta= -\frac{a}{b+c}.
\end{equation}

It follows from Eq. (\ref{2hd31}) that if $\cot\eta\sim\mathcal{O}(1)$ then
$(a+\cot\eta\; b)$ must be of the same order than $c$ which has a suppression factor $m_i^2/\Lambda^2$.
A fine-tuning is then necessary also for this equation.
In consequence the three first derivative conditions related to the variables $v^{'}$, $w^{'}$, and $\eta^{'}$ must be fine-tuned.

Finally we rewrite the vacuum alignment condition.
Using the solution for $\cot\eta$ found in Eq.  (\ref{2hd32}), and Eqs. (\ref{2hd26}), (\ref{2hd28b}),
the first derivative condition related to $z^{'2}$ can be written as

\begin{equation}\label{2hd33}
-\frac{2\, c}{w^2 \sin^2\eta}
+\sum_{\begin{matrix}\scriptstyle i=u,c,t \\
                     \scriptstyle j=d,s,b   \end{matrix}}
\frac{|T_{ij}|^2}{\Lambda^2}
I(m_i^2,m_j^2) > 0.
\end{equation}

\subsection{Second Derivatives of the Effective Potential}

We calculate the Hessian matrix associated with the effective potential and the variables $v^{'}$, $w^{'}$, and $\eta^{'}$ in order to check the minimum condition c) of Eq. (\ref{h8a}).
We need to calculate the following derivatives:

\begin{equation}\label{3f20}
\frac{\partial^2 V_{\text{eff}}}
     {\partial\theta_a\partial\theta_b}\Big|_{(v,w,\eta,0)}
=
\frac{\partial^2}
     {\partial\theta_a\partial\theta_b}
\Big(m_{H_i}^2 H^{(i)\dagger}H^{(i)}\Big)
-
\frac{N}{8\pi^2}\int_0^{\Lambda^2} k^2\, dk^2\:
\frac{\partial^2}
     {\partial\theta_a\partial\theta_b}
\log\det(k^2+A)\Big|_{(v,w,\eta,0)} ,
\end{equation}
with $\theta_a,\theta_b=v^{'},w^{'},\eta^{'}$.
For the second term of the r.h.s. the following relation is useful

\begin{equation}\label{3f21}
\begin{split}
\frac{\partial^2 \log\det(k^2+A)}
     {\partial\theta_a\partial\theta_b}\Big|_{(v,w,\eta,0)}
&=
\;\;\sum_{i=1}^6 \:\frac{1}{k^2+m_i^2}\:
\frac{\partial^2 A_{ii}}{\partial\theta_a\partial\theta_b}\Big|_{(v,w,\eta,0)} \\
&\quad -\sum_{i,j=1}^6 \: \frac{1}{(k^2+m_i^2)(k^2+m_j^2)}\:
\frac{\partial A_{ij}}{\partial\theta_a}
\frac{\partial A_{ji}}{\partial\theta_b}\Big|_{(v,w,\eta,0)} ,
\end{split}
\end{equation}
where we again used the fact that the matrix $A$ is diagonal at the point $(v,w,\eta,0)$.
Inserting this equation into Eq. (\ref{3f20}) we obtain

\begin{equation}\label{3f22}
\begin{split}
\frac{\partial^2 V_{\text{eff}}}
     {\partial\theta_a\partial\theta_b}\Big|_{(v,w,\eta,0)}
&=
\frac{\partial^2}
     {\partial\theta_a\partial\theta_b}
\Big(m_{H_i}^2 H^{(i)\dagger}H^{(i)}\Big)
-
\frac{N\Lambda^2}{8\pi^2}\:\sum_{i=1}^6\:
\left( 1-\frac{m_i^2}{\Lambda^2}\log\Lambda^2/m_i^2\right) \:
\frac{\partial^2 A_{ii}}{\partial\theta_a\partial\theta_b}\Big|_{(v,w,\eta,0)} \\
&+
\frac{N}{8\pi^2}\:\sum_{i,j=1}^6\:I(m_i^2,m_j^2)\:
\frac{\partial A_{ij}}{\partial\theta_a}
\frac{\partial A_{ji}}{\partial\theta_b}\Big|_{(v,w,\eta,0)},
\end{split}
\end{equation}
with $I(m_i^2,m_j^2)$ given in Eq. (\ref{2hd25b}).
The first derivatives of the matrix $A$ needed in Eq. (\ref{3f22}) are given by

\begin{equation}\label{3f23}
\begin{split}
\frac{\partial A}{\partial v^{'}}\Big|_{(v,w,\eta,0)} &=
v\: \Sigma^{(1)}
+\frac{w}{2}\left(
\Sigma^{(0)} e^{i\eta}+\Sigma^{(0)\dagger} e^{-i\eta}\right) ,\\
\frac{\partial A}{\partial w^{'}}\Big|_{(v,w,\eta,0)} &=
w\: \Sigma^{(2)}
+\frac{v}{2}\left(
\Sigma^{(0)} e^{i\eta}+\Sigma^{(0)\dagger} e^{-i\eta}\right) , \\
\frac{\partial A}{\partial \eta^{'}}\Big|_{(v,w,\eta,0)} &=
\frac{ivw}{2}\left(
\Sigma^{(0)} e^{i\eta}-\Sigma^{(0)\dagger} e^{-i\eta}\right).
\end{split}
\end{equation}
Besides, the second derivatives of the diagonal elements of $A$, relevant for   Eq. (\ref{3f22}) are given by

\begin{equation}\label{3f24}
\begin{split}
\frac{\partial^2 A_{ii}}{\partial v^{'2}}\Big|_{(v,w,\eta,0)}&=\Sigma^{(1)}_{ii}, \\
\frac{\partial^2 A_{ii}}{\partial w^{'2}}\Big|_{(v,w,\eta,0)}&=\Sigma^{(2)}_{ii} ,\\
\frac{\partial^2 A_{ii}}{\partial \eta^{'2}}\Big|_{(v,w,\eta,0)}&=
            -vw\:\mathcal{R}e\left(\Sigma^{(0)}_{ii} e^{i\eta}\right) ,\\
\frac{\partial^2 A_{ii}}{\partial v^{'}\partial w^{'}}\Big|_{(v,w,\eta,0)}
&= \mathcal{R}e\left(\Sigma^{(0)}_{ii} e^{i\eta}\right) ,  \\
\frac{\partial^2 A_{ii}}{\partial v^{'}\partial \eta^{'}}\Big|_{(v,w,\eta,0)}
&= -w\:\mathcal{I}m\left(\Sigma^{(0)}_{ii} e^{i\eta}\right),  \\
\frac{\partial^2 A_{ii}}{\partial w^{'}\partial \eta^{'}}\Big|_{(v,w,\eta,0)}
&= -v\:\mathcal{I}m\left(\Sigma^{(0)}_{ii} e^{i\eta}\right).
\end{split}
\end{equation}
Using the last expressions and the first derivative conditions we found in Subsection \ref{subsection:1derivatives_3f} we finally get

\begin{equation}\label{3f25}
\begin{split}
\frac{\partial^2 V_{\text{eff}}}
     {\partial\theta_a\partial\theta_b}\Big|_{(v,w,\eta,0)}
&=
\:\:\frac{cN\Lambda^2}{4\pi^2\sin^2\eta}
\begin{pmatrix}
-1/v^2 &  1/vw  & 0 \\
1/vw   & -1/w^2 & 0 \\
0      &   0    & -1   \end{pmatrix}_{ab}  \\
&\quad+
\frac{N}{8\pi^2}\:\sum_{i,j=1}^6\:I(m_i^2,m_j^2)\:
\frac{\partial A_{ij}}{\partial\theta_a}
\frac{\partial A_{ji}}{\partial\theta_b}\Big|_{(v,w,\eta,0)},
\end{split}
\end{equation}
with $\theta_a,\theta_b=v^{'},w^{'}$, and $\eta^{'}$.

\subsection{Summary}

Here we summarize the procedure which allows to find the parameter subspace of the coupling constants $G$ of the Lagrangian given in Eq. (\ref{3f1}) for which the model predicts the correct CKM matrix and quark masses.
We restrict our analysis to the case in which the four-fermion Lagrangian can be rewritten using only two auxiliary Higgs fields.

First we choose the four unitary matrices $U^u$, $U^d$, $W^u$, and  $W^d$ with the only restriction of having a realistic CKM matrix $V_{CKM}=U^{u\dagger}U^d$.
For this it is necessary to determine the value of 32 real parameters,
9 from each unitary matrix minus the 4 physical parameters of the CKM matrix.
We also have to choose the values of the VEVs $v$ and $w$.
However, it is possible to fix the values of $v$ and $w$ without losing generality.
This is because other values of the VEVs $v$ and $w$ correspond just to rescalings of the auxiliary fields $H^{(i)}$.
Using the basis transformation matrices we chose and the quark masses, we calculate the values of the parameters $a$, $b$, and $c$ using Eqs. (\ref{2hd29}).
Next we do the first check:
The value of $c$ must be different from zero.
This is necessary in order to have a non-trivial value of $\eta$ and hence spontaneous $CP$ symmetry breaking.
If this is the case, the value of $\cot\eta$ is given by Eq. (\ref{2hd32}).
Because $\cot\eta$ is a function with period $\pi$, the last equation leads to two possible values for $\eta$, which are, however, equivalent.
After that, using Eqs. (\ref{2hd5}) and (\ref{3f8}), we find the values of the Yukawa matrices in the weak and mass bases.
Replacing these matrices in Eqs. (\ref{2hd16}) and (\ref{2hd17}) we get the appropriate values of the auxiliary field mass parameters $m_{H_1}^2$ and $m_{H_2}^2$.
At this point we make two further checks related to the conservation of the electromagnetic symmetry and the second derivatives of the effective potential.
The first of these restrictions is given in Eq. (\ref{2hd33}) and the second corresponds to having a positive definite Hessian matrix which is given in Eq. (\ref{3f25}).
Finally, if the chosen parameters pass all checks the four-fermion parameters $G$ can by obtained from Eqs. (\ref{h4}).
If not, the chosen parameters do not generate a realistic model.

\section{Quark Mass Expansion}

In order to obtain approximate expressions for the quantities
calculated in the previous Subsection,
in particular for $c$, $\cot\eta$, and $\partial V_{\text{eff}}/{\partial z^{'2}}$,
we perform an expansion in the quark mass ratios.
We profit from the fact that the top quark is much heavier than the other quarks.

In this expansion only some entries of the basis transformation matrices are relevant.
We introduce the following notation

\begin{equation}\label{2hd34}
\begin{matrix}
p_i=\mathcal{R}e(W_{i3}^u),& &r_i=\mathcal{R}e(U_{i3}^u) , \\
q_i=\mathcal{I}m(W_{i3}^u),& &s_i=\mathcal{I}m(U_{i3}^u) .
\end{matrix}
\end{equation}
The 3-dimensional vectors
$\vec{p}$, $\vec{q}$, $\vec{r}$, and $\vec{s}$ fulfill,
as a consequence of the unitarity of the transformation matrices,
the following relations

\begin{equation}\label{2hd35}
\begin{split}
p^2+q^2=1,\\
r^2+s^2=1,
\end{split}
\end{equation}
where $p=|\vec{p}\,|$, $q=|\vec{q}\,|$, $r=|\vec{r}\,|$, and $s=|\vec{s}\,|$.

First we consider the coefficients $a$ and $b$ given by Eqs. (\ref{2hd29}).
The contributions to $a$ and $b$ can be grouped,
depending on the factors $m_q$, in the following way:

\begin{equation*}
a,b= \mathcal{O}(m_t^2)+\mathcal{O}(m_t m_c)+\mathcal{O}(m_b^2)
    +\mathcal{O}(m_c^2)+\mathcal{O}(m_t m_u)+\ldots,
\end{equation*}
where the leading terms are of order $m_t^2$.
If they are not suppressed by basis transformation
matrix elements, the following approximations are valid

\begin{equation}\label{2hd36}
\begin{split}
a&=m_t^2\big[(p^2-q^2)(\vec{r}\cdot\vec{s})-(\vec{p}\cdot\vec{q})(r^2-s^2)\big]
    \; \Big(1+\mathcal{O}(m_c/m_t)\Big),\\
b&=m_t^2\big[2(\vec{p}\cdot\vec{q})(\vec{r}\cdot\vec{s})-q^2r^2-p^2s^2\big]
   \; \Big(1+\mathcal{O}(m_c/m_t)\Big).
\end{split}
\end{equation}
We give now an approximated expression for $\cot\eta$ (Eq. (\ref{2hd32})).
Because we are considering the coefficient $b$ to be of order $m_t^2$,
we can neglect the contribution of $c$
in the denominator of Eq. (\ref{2hd32}).
In this way we obtain

\begin{equation}\label{2hd37}
\cot\eta=
\frac{(p^2-q^2)(\vec{r}\cdot\vec{s})-(\vec{p}\cdot\vec{q})(r^2-s^2)}
     {q^2r^2+p^2s^2-2(\vec{p}\cdot\vec{q})(\vec{r}\cdot\vec{s})}
+\mathcal{O}\Big(\frac{m_c}{m_t}\Big).
\end{equation}
We repeat that the last expression is valid only if the numerator and
the denominator of this ratio are of order 1, and if $c\neq 0$.

Let us now consider the first derivative of the effective potential with respect to $z^{'2}$ evaluated at the minimum of the effective potential

\begin{equation}\label{2hd33b}
\frac{\partial V_{\text{eff}}}{\partial z^{'2}}\bigg|_{(v,w,\eta,0)}=
-\frac{2\, c}{w^2 \sin^2\eta}
+\sum_{\begin{matrix}\scriptstyle i=u,c,t \\
                     \scriptstyle j=d,s,b   \end{matrix}}
\frac{|T_{ij}|^2}{\Lambda^2}
I(m_i^2,m_j^2).
\end{equation}
First we examine the second term on the r.h.s. of Eq. (\ref{2hd33b}).
For $\Lambda$ much bigger than the quark masses we can make the
following approximation

\begin{equation}\label{2hd37a}
\begin{pmatrix}
\text{second term} \\ \text{in Eq.(\ref{2hd33b})}
\end{pmatrix}
\approx\;\frac{\log\Lambda^2}{\Lambda^2}\;\text{tr}(T\,T^\dagger).
\end{equation}
Using Eqs. (\ref{2hd20}), (\ref{3f8}), and (\ref{2hd5}),
and the cyclicity of the trace, we obtain for the second term of
Eq. (\ref{2hd33b}) the following expression

\begin{equation*}
\begin{pmatrix}
\text{second term} \\ \text{in Eq.(\ref{2hd33b})}
\end{pmatrix}
\approx
\frac{\log\Lambda^2}{\Lambda^2}\;
\frac{v^2}{2}\;
\frac{4}{v^2 w^2 \sin^2\eta}\; K,
\end{equation*}
where the factor $K$ depends only on the quark masses and on the basis
transformation matrices

\begin{equation}\label{2hd38}
\begin{split}
K&=\frac{v^2 w^2 \sin^2\eta}{4}\;
   \text{tr}\Big[(g^{(2)T}h^{(1)}-g^{(1)T}h^{(2)})
                (h^{(1)T}g^{(2)}-h^{(2)T}g^{(1)})\Big],\\
&=\text{tr}\Big(\:\Big[
-\mathcal{I}m(W^u M_u U^{u\dagger})\:\mathcal{R}e(U^d M_d W^{d\dagger})
+\mathcal{R}e(W^u M_u U^{u\dagger})\:\mathcal{I}m(U^d M_d W^{d\dagger})
   \:\Big]                      \\
& \qquad \quad\;\:      \Big[\:
 \mathcal{R}e(W^d M_d U^{d\dagger})\:\mathcal{I}m(U^u M_u W^{u\dagger})
-\mathcal{I}m(W^d M_d U^{d\dagger})\:\mathcal{R}e(U^u M_u W^{u\dagger})
   \:\Big] \;\Big).
\end{split}
\end{equation}
Assuming that the basis transformation matrices do not suppress the factor $K$,
we can consider its absolute value to be of order $m_t^2\,m_b^2$.
In this case we obtain

\begin{equation}\label{2hd_second}
\begin{pmatrix}
\text{second term} \\ \text{in Eq.(\ref{2hd33b})}
\end{pmatrix}
\approx
\pm
\frac{m_t^2\,m_b^2}{w^2 \Lambda^2}\;\log(\Lambda^2)\;
\frac{2}{\sin^2\eta},
\end{equation}
where the sign of the contributions is equal to the sign of the factor $K$.
We estimate now the first term on the r.h.s. of Eq. (\ref{2hd33b}).
For that, we need an approximate expression for $c$.
The leading contribution to $c$ is proportional to $m_t^4/\Lambda^2$:

\begin{equation}\label{2hd39}
\begin{split}
c&=-\frac{m_t^4}{\Lambda^2}\log(\Lambda^2/m_t^2)\:
\Bigg[              \mathcal{I}m\Big(\:
W^{u\dagger}\:\frac{\mathcal{R}e(W^u M_u U^{u\dagger})}{m_t}
            \:\frac{\mathcal{I}m(U^u M_u W^{u\dagger})}{m_t}\:W^u\Big)_{33}
+\mathcal{O}\Big(\frac{m_c}{m_t}\Big) \Bigg]   \\
&=\;\;\frac{m_t^4}{\Lambda^2}\log(\Lambda^2/m_t^2)\:
\Big(
p^2q^2-(\vec{p}\cdot\vec{q})^2
+\mathcal{O}\Big(\frac{m_c}{m_t}\Big) \Big),
\end{split}
\end{equation}
where we used Eqs. (\ref{2hd35}).
The last result is only valid if $(p^2q^2-(\vec{p}\cdot\vec{q})^2)=\mathcal{O}(1)$.
Using it, we obtain for the first term on the r.h.s. of Eq. (\ref{2hd33b}) the following approximate expression

\begin{equation}\label{2hd_first}
\begin{pmatrix}
\text{first term} \\ \text{in Eq.(\ref{2hd33b})}
\end{pmatrix}
\;\approx
-\frac{m_t^4}{w^2 \Lambda^2}\;\log(\Lambda^2)\;
\frac{2}{\sin^2\eta}\:(p^2q^2-(\vec{p}\cdot\vec{q})^2).
\end{equation}
Due to the Schwarz inequality Eq. (\ref{2hd_first}) is always negative,
while the sign of the second contribution to
$\partial V_{\text{eff}}/{\partial z^{'2}}$, given in Eq. (\ref{2hd_second}),
is equal to the sign of the factor $K$.
The ratio between their absolute values is
$(p^2q^2-(\vec{p}\cdot\vec{q})^2)(m_t^2/m_b^2)$.
In order to have a ratio smaller than 1 (smaller first term), and thus,
if $K$ is positive, alignment,
the leading contribution to $c$ must be suppressed,
i.e. $p^2q^2-(\vec{p}\cdot\vec{q})^2 \ll 1$.
Besides, if we suppress the contribution to $c$ proportional to $m_t^4/\Lambda^2$,
we also suppress the whole term with $i=t$ in the sum for $c$ in Eq. (\ref{2hd29}),
and the term proportional to $m_c^2\, m_t^2/\Lambda^2$ coming from $i=c$ in this sum.
Thus, the leading contribution to $c$ are of the order of the terms proportional to $m_b^4/\Lambda^2$, $m_c^3\, m_t/\Lambda^2$.

\subsection{Example}

In order to be more specific we give an example in which the transformation matrices $W^u$ and $U^d$ are both real.
(The two transformation matrices for the left-handed fields cannot be simultaneously real because the CKM matrix is complex.
In a similar way, the transformation matrices for the right-handed fields cannot be both real because $c$ must be different from zero, see Eq. (\ref{2hd29}).)
As we did in Eq. (\ref{2hd34}) for the up-sector, we introduce a convenient notation for the relevant matrix elements of the down-type transformation matrices:

\begin{equation}\label{3f40}
\begin{split}
k_i &=\mathcal{R}e(W_{i3}^d),  \\
l_i &=\mathcal{I}m(W_{i3}^d),  \\
n_i &=\mathcal{R}e(U_{i3}^d)=U_{i3}^d  .
\end{split}
\end{equation}
Because $\vec{q}=0$ we obtain from Eq. (\ref{2hd37}):

\begin{equation}\label{3f41}
\cot\eta \approx
\frac{\vec{r}\cdot\vec{s}}{s^2}.
\end{equation}
The last approximation is valid only if
$\vec{r}\cdot\vec{s},\: s^2\sim\mathcal{O}(1)$.
On the other, hand imposing the constraint $|(V_{ckm})_{33}|\approx 1$, we get a restriction on the left transformation matrix elements:
The three vectors $\vec{n}$, $\vec{r}$, and $\vec{s}$ must be parallel to each other.
Using Eq. (\ref{3f41}) and $|\vec{n}|=1$ we find

\begin{equation}\label{3f42}
\begin{split}
\vec{r} &=\cot\eta\;\vec{s}, \\
\vec{n} &=\pm\frac{\vec{s}}{s},
\end{split}
\end{equation}
and also $s=|\sin\eta|$, $r=|\cos\eta|$.
Therefore the approximation we did in Eq. (\ref{3f41}) is valid if $|\cot\eta|\sim\mathcal{O}(1)$.
Besides, the factors $c$ and $K$ are given by

\begin{eqnarray}
c &\approx & \frac{m_b^4}{\Lambda^2}\: \log\Lambda^2/m_b^2\:
             [l^2k^2-(\vec{l}\cdot\vec{k})^2],               \label{3f43}  \\
K &\approx & m_b^2 m_t^2 \:|(V_{CKM})_{33}|^2
            [k^2\sin^2\theta +l^2\cos^2\theta
             +2(\vec{l}\cdot\vec{k})\sin\theta\cos\theta],   \label{3f44}
\end{eqnarray}
where $l=|\vec{l}|$, $k=|\vec{k}|$, and the angle $\theta$ is given by $(U^{u\dagger}U^d)_{33}=|(V_{CKM})_{33}| e^{i\theta}$.
Choosing $l=k=1/\sqrt{2}$ and $\vec{l}\cdot\vec{k}=0$ we get

\begin{eqnarray}
c &\approx & \frac{m_b^4}{4\Lambda^2} \log\Lambda^2/m_b^2,  \label{3f45}  \\
K &\approx & \frac{m_b^2 m_t^2}{2}.                         \label{3f46}
\end{eqnarray}
With this choice of parameters we obtain $c\neq 0$ and
$\partial V_{\text{eff}}/{\partial z^{'2}}>0$ (see Eq. (\ref{2hd33b})).
It is necessary to check also the second derivative condition.
Calculating the Hessian, Eq. (\ref{3f25}) up to order $m_b^4$, one finds that two eigenvalues of this matrix are real and one is equal to zero.
In order to obtain the third eigenvalue more precisely and to see whether it is positive, a more detailed evaluation of the Hessian is needed.
We verified this numerically for some specific cases.

We give approximate, non-fine-tuned expressions for the coupling constants $G$ of Eq. (\ref{3f1}). For our example, using Eqs. (\ref{2hd5}) and (\ref{int6}), we get

\begin{equation}\label{3f47}
\begin{split}
G_{ijkl}   &\approx \frac{8\pi}{N\Lambda^2}\;\frac{1}{\sin^2\eta}\; s_i p_j s_k p_l,\\
G_{ijkl}^{'} &\approx \frac{8\pi}{N\Lambda^2}\;
            2 s_i (k_j-\cot\eta\: l_j) s_k (k_l-\cot\eta\: l_l),  \\
G_{ijkl}^{''} &\approx \frac{8\pi}{N\Lambda^2}\;\frac{m_c}{m_b}
            \;2\sin\eta\; s_i (k_j-\cot\eta \: l_j)
    [(r_k^{'}p_l^{'}+s_k^{'}q_l^{'})-\cot\eta(s_k^{'}p_l^{'}-r_k^{'}q_l^{'})],
\end{split}
\end{equation}
where

\begin{equation}\label{}
\begin{matrix}
p_i^{'}=\mathcal{R}e(W_{i2}^u),& &r_i^{'}=\mathcal{R}e(U_{i2}^u) , \\
q_i^{'}=\mathcal{I}m(W_{i2}^u),& &s_i^{'}=\mathcal{I}m(U_{i2}^u) .
\end{matrix}
\end{equation}
The corrections to Eqs. (\ref{3f47}) are by a factor $\sim m_c^2/m_b^2$ smaller for each of the three types of coupling constants $G_{ijkl}$, $G_{ijkl}^{'}$, and $G_{ijkl}^{''}$.

\section{Masses of the Composite Higgs Bosons}

We calculate in this Section the neutral and charged two-point proper vertices associated with the Lagrangian given in Eq. (\ref{3f1}) which involves the three quark families.
As before we restrict ourselves to the case when the Lagrangian can be rewritten using only two auxiliary fields which condense and  break the EW symmetry in the correct way.
The situation is analogous to the one of Section \ref{section:masses_1fam} where only one family of quarks was considered.
The values of the flowing momentum, for which an eigenvalue of the proper vertex matrix becomes zero,
correspond to the composite Higgs masses we are looking for.

For the calculation we use the neutral and charged auxiliary field bases which we defined in Subsection \ref{subsection:new_bases}.
In this Subsection we also found the auxiliary field mass terms in the new bases.
We need also the Yukawa terms in these bases.
From Eq. (\ref{int11}) we obtain the following Yukawa couplings for the neutral auxiliary fields
$\varphi^1$, $\varphi^2$, $\varphi^3$, and $G$:

\begin{equation}\label{3f30}
\begin{split}
\mathcal{L}_{\text{Yukawa-neutral}} &=
-\frac{\varphi^1}{\sqrt{v^2+w^2}}\:
    \bar{u}^{'}(K_u\,\projectorr + K_u^\dagger\,\projectorl)u^{'}
-\frac{\varphi^2}{\sqrt{v^2+w^2}}\:
    \bar{u}^{'}(i K_u\,\projectorr -i K_u^\dagger\,\projectorl)u^{'}  \\
&\quad\quad
-\frac{\varphi^3}{\sqrt{v^2+w^2}}\:
    \bar{u}^{'}\sqrt{2}M_u u^{'}
-\frac{G}{\sqrt{v^2+w^2}}\:
    \bar{u}^{'}\sqrt{2}M_u\, i\gamma_5\, u^{'}+\;\dots,
\end{split}
\end{equation}
where the dots represent analogous terms for the down sector.
In the charged sector we find for the fields $\varphi^\pm$ and $G^\pm$:

\begin{equation}\label{3f31}
\begin{split}
\mathcal{L}_{\text{Yukawa-charged}} &=
-\frac{\varphi^+\; e^{i\eta}}{\sqrt{v^2+w^2}}\: \bar{u}^{'}
    (K_u^\dagger\, V_{CKM}\,\projectorl -V_{CKM} K_d^\dagger\,\projectorr)d^{'} \\
&\quad
-\frac{G^+}{\sqrt{v^2+w^2}}\: \bar{u}^{'}
    (\sqrt{2}M_u V_{CKM}\,\projectorl -V_{CKM}\sqrt{2} M_d\,\projectorr)d^{'}+h.c.,
\end{split}
\end{equation}
where $\projectorl$ and $\projectorr$ are the left and right projectors.
$V_{CKM}$, $M_q$, and $K_q$ are $3\times 3$ matrices in flavor space.
$V_{CKM}$ is the CKM matrix, the matrices $M_q$ with $q=u,d$ are the diagonal quark mass matrices of the up and down sector, and the matrices $K_q$ are given by

\begin{equation}\label{3f32}
K_q\equiv w\,\lambda_q^{(1)}-v\,e^{i\eta}\lambda_q^{(2)},
\end{equation}
with $q=u,d$.

\begin{figure}[tbp]
\begin{center}
\psfig{file=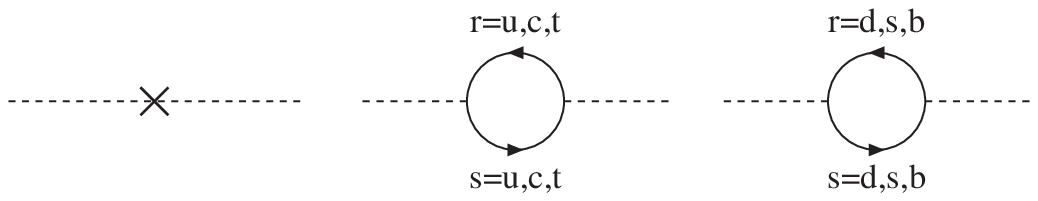,width = 9.9cm}
\end{center}
\caption{Diagrammatic representation of the contributions to the neutral two-point proper vertices.}
\label{fig:3fam_masses_neutral}
\end{figure}

With these interaction terms and the mass terms of the auxiliary fields which are given in Subsection \ref{subsection:new_bases} we calculate the two-point proper vertices.
In the calculation we also use the fine-tuned first derivative conditions found in Subsection \ref{subsection:1derivatives_3f}.
For the neutral sector the relevant diagrams are shown in Fig. \ref{fig:3fam_masses_neutral}.
The related proper vertices are given by

\begin{equation}\label{2pf_neutral1}
\begin{split}
i\Gamma_{(\varphi^1,\varphi^1)/(\varphi^2,\varphi^2)}&(p^2)
 = \frac{i N}{8\pi^2(v^2+w^2)} \Bigg\{
    \; \frac{4\, c\, \Lambda^2}{\sin^2\eta}
        \left(\frac{w^2}{v^2}+\frac{v^2}{w^2} +2\right)   \\
& +\sum_{i=1}^6\: m_i^2\log\Lambda^2/m_i^2
    \big[ w^2(\Sigma_{ii}^{(1)}-\tilde{\Sigma}_{ii}^{(1)})
          +v^2(\Sigma_{ii}^{(2)}-\tilde{\Sigma}_{ii}^{(2)})     \\
&  \qquad\qquad\qquad\qquad\qquad\qquad
 -2vw \,\mathcal{R}e((\Sigma_{ii}^{(0)}-\tilde{\Sigma}_{ii}^{(0)})e^{i\eta})
      \big]                                                    \\
&  + \sum_{r,s=1}^3 \sum_{q=u,d}\: I(m_r^2,m_s^2;p^2)
 \Big[(p^2-m_r^2-m_s^2) (K_q)_{rs}(K_q^\dagger)_{sr}         \\
&  \qquad\qquad\qquad\qquad
   \mp m_r m_s ((K_q)_{rs}(K_q)_{sr}+(K_q^\dagger)_{rs}(K_q^\dagger)_{sr})
 \Big] \Bigg\},
\end{split}
\end{equation}
where the minus sign in the last term corresponds to $i\Gamma_{\varphi^1,\varphi^1}$ and the plus sign to $i\Gamma_{\varphi^2,\varphi^2}$.
The matrices $\tilde{\Sigma}^{(i)}$ are defined by

\begin{equation}\label{}
\tilde{\Sigma}^{(0)}=
\begin{pmatrix}
\lambda_u^{(2)}\lambda_u^{(1)\dagger} & 0 \\
0 & \lambda_d^{(1)\dagger}\lambda_d^{(2)}
\end{pmatrix},
\end{equation}

\begin{equation}\label{}
\tilde{\Sigma}^{(i)}=
\begin{pmatrix}
\lambda_u^{(i)}\lambda_u^{(i)\dagger} & 0 \\
0 & \lambda_d^{(i)\dagger}\lambda_d^{(i)}
\end{pmatrix}\:,
\;\qquad\text{for }i=1,2.
\end{equation}
Further neutral proper vertices are given by

\begin{eqnarray}
i\Gamma_{\varphi^1,\varphi^2}(p^2)&=&
 \frac{2i N}{8\pi^2(v^2+w^2)} \sum_{r,s=1}^3 \sum_{q=u,d}\:
 m_r m_s \, \mathcal{I}m[(K_q)_{rs}(K_q)_{sr}]\, I(m_r^2,m_s^2;p^2),
                                                    \label{2pf_neutral2}    \\
i\Gamma_{\varphi^3,\varphi^3}(p^2)&=&
 \frac{2i N}{8\pi^2(v^2+w^2)} \sum_{i=1}^6 \:
 (p^2-4m_i^2)m_i^2\, I(m_i^2,m_i^2;p^2)  ,       \label{2pf_neutral3}          \\
i\Gamma_{G,G}(p^2)&=&
 \frac{2i N}{8\pi^2(v^2+w^2)}\,p^2\, \sum_{i=1}^6 \:
 m_i^2\, I(m_i^2,m_i^2;p^2)    ,                   \label{2pf_neutral4}      \\
i\Gamma_{\varphi^1,\varphi^3}(p^2)&=&
 \frac{i N}{8\pi^2(v^2+w^2)} \sum_{i=1}^6 \:
 (p^2-4m_i^2)\,\sqrt{2}\,m_i\,\mathcal{R}e[(K_{u/d})_{ii}]\, I(m_i^2,m_i^2;p^2) ,
                                                   \label{2pf_neutral5}  \\
i\Gamma_{\varphi^2,G}(p^2)&=&
 \frac{i N}{8\pi^2(v^2+w^2)}\,p^2\, \sum_{i=1}^6 \:
 \sqrt{2}\,m_i\,\mathcal{R}e[(K_{u/d})_{ii}]\, I(m_i^2,m_i^2;p^2)  ,
                                                      \label{2pf_neutral6}   \\
i\Gamma_{\varphi^1,G}(p^2)&=&
 \frac{i N}{8\pi^2(v^2+w^2)}\,p^2\, \sum_{i=1}^6 \:
 \sqrt{2}\,m_i\,\mathcal{I}m[(K_{u/d})_{ii}]\, I(m_i^2,m_i^2;p^2),
                                                      \label{2pf_neutral7}   \\
i\Gamma_{\varphi^2,\varphi^3}(p^2)&=&
 \frac{i N}{8\pi^2(v^2+w^2)} \sum_{i=1}^6 \:
 (4m_i^2-p^2)\,\sqrt{2}\,m_i\,\mathcal{I}m[(K_{u/d})_{ii}]\, I(m_i^2,m_i^2;p^2) ,
                                                           \label{2pf_neutral8}
\end{eqnarray}
where in Eqs. (\ref{2pf_neutral5})-(\ref{2pf_neutral8}) $K_{u/d}$ stands for $K_u$ for $i=1,2,3$ and for $K_d$ for $i=4,5,6$.

\begin{figure}[tbp]
\begin{center}
\psfig{file=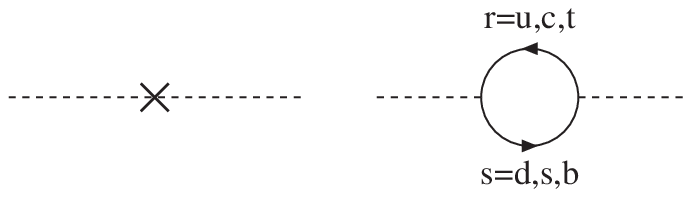,width = 6.6cm}
\end{center}
\caption{Diagrammatic representation of the contributions to the charged two-point proper vertices.}
\label{fig:3fam_masses_charged}
\end{figure}

In the charged sector the three relevant two-point proper vertices,
with associated diagrams shown in Fig. \ref{fig:3fam_masses_charged}, are given by

\begin{equation}\label{2pf_charged1}
\begin{split}
i\Gamma_{\varphi^+,\varphi^-}(p^2)
&= \,\frac{2i N}{8\pi^2(v^2+w^2)} \frac{c\,\Lambda^2}{\sin^2\eta}
      \left(\frac{w^2}{v^2}+\frac{v^2}{w^2} +2\right)             \\
&  -\frac{i N}{16\pi^2(v^2+w^2)}
 \sum_{\begin{matrix}\scriptstyle r=u,c,t \\
                     \scriptstyle s=d,s,b   \end{matrix}} \, \Big\{
 (K_u^\dagger V_{CKM})_{rs} (V_{CKM}^\dagger K_u)_{sr}
        \,\big[ J(m_s^2,m_r^2;p^2) + 2 m_s^2         \big]     \\
&\qquad\qquad\qquad\qquad\qquad
+(V_{CKM} K_d^\dagger)_{rs} (K_d V_{CKM}^\dagger)_{sr}
        \,\big[ J(m_r^2,m_s^2;p^2) + 2 m_r^2         \big]     \\
&\qquad-\big[ (K_u^\dagger V_{CKM})_{rs} (K_d V_{CKM}^\dagger)_{sr}
        +(V_{CKM} K_d^\dagger)_{rs} (V_{CKM}^\dagger K_u)_{sr} \big]
  2 m_r m_s\, I(m_r^2,m_s^2;p^2)   \Big\},
\end{split}
\end{equation}

\begin{equation}\label{2pf_charged2}
i\Gamma_{G^+,G^-}(p^2)
= \frac{-i N}{8\pi^2(v^2+w^2)}
 \sum_{\begin{matrix}\scriptstyle r=u,c,t \\
                     \scriptstyle s=d,s,b   \end{matrix}}
 \;(V_{CKM})_{rs} (V_{CKM}^\dagger)_{sr} \,  \Big\{
      m_r^2 \, J(m_s^2,m_r^2;p^2) +m_s^2\, J(m_r^2,m_s^2;p^2)        \Big\},
\end{equation}

\begin{equation}\label{2pf_charged3}
\begin{split}
i\Gamma_{\varphi^+,G^-}(p^2)
= \frac{-\sqrt{2}\, i N e^{i\eta}}{16\pi^2(v^2+w^2)}
 \sum_{\begin{matrix}\scriptstyle r=u,c,t \\
                     \scriptstyle s=d,s,b   \end{matrix}}
  \;(V_{CKM})_{rs} \, \Big\{&m_r (V_{CKM}^\dagger K_u)_{sr}\, J(m_s^2,m_r^2;p^2)  \\
                           &+m_s (K_d V_{CKM}^\dagger)_{sr}\, J(m_r^2,m_s^2;p^2)
                      \Big\},
\end{split}
\end{equation}
where $J(m_r^2,m_s^2;p^2)$ defined by
\begin{equation}\label{2pf_charged4}
J(m_r^2,m_s^2;p^2)=m_r^2\log\Lambda^2/m_r^2 - m_s^2\log\Lambda^2/m_s^2
                 -(p^2+m_r^2-m_s^2)\, I(m_r^2,m_s^2;p^2),
\end{equation}
vanishes at $p^2=0$.

In all the two-point proper vertices the quadratic divergences cancel.
As expected, the neutral and charged Goldstone bosons correspond to eigenvectors of the proper vertex matrix at $p^2=0$ with vanishing eigenvalues.
Besides these three Goldstone bosons there are three neutral and one charged Higgs particles.
The masses of these particles are roughly $\sim 2m_q$.
We find that one neutral mass is $\sim 2m_t$ and the rest of the Higgs masses are much smaller.

\chapter{Compositeness Condition and Renormalization Group Analysis}
\label{chapter:composite_condition}

If the scale associated with the new four-fermion interactions is much higher than the scale of EW symmetry breaking, i.e. $\Lambda\gg v$, it is possible to make a (perturbative) renormalization group (RG) analysis of the theory.
This analysis leads to much better predictions for the top-quark and Higgs-boson masses than the crude NJL approach
because it includes effects of the gauge boson interactions which become important at the EW scale.
In this Chapter we first review the compositeness condition and the related RG analysis for the one-Higgs doublet case.
After that we consider the case of having two composite Higgs doublets which is relevant for our model.
Finally we mention the possibility of including Majorana neutrinos.\\

In Chapter \ref{chapter:eff_potential} we rewrote the four-fermion interaction
$\mathcal{L}_{\text{4f}}$ in terms of auxiliary fields with the quantum numbers of the Higgs boson.
We saw that for a range of values of the coupling constants $G$ it is possible to rewrite the theory with the help of only one (auxiliary) Higgs doublet.
The condensation of this field provides all quarks with masses.
In the following analysis, however, we neglect fermion masses other than $m_t$.
In the unitarity gauge the Lagrangian of the theory
(without including gauge bosons) is given by

\begin{equation}\label{rg1}
\mathcal{L}^{(\Lambda)}=\mathcal{L}^{(\Lambda)}_\text{kin}
		-\frac12 m_{H_0}^2 \phi^2-\frac{g_{t_0}}{\sqrt{2}}\:\bar{t}t\phi,
\end{equation}
where $\mathcal{L}_\text{kin}$ denotes the fermionic kinetic term and the coupling constant $G_t$ of the four-fermion interaction is given by $G_t=g_{t_0}^2/m_{H_0}^2$
(see  Eq. (\ref{min2})).
The Lagrangian  $\mathcal{L}^{(\mu)}$ is defined at the scale $\mu$, with all field components related to higher scales being integrated out.

The Lagrangian given in Eq. (\ref{rg1}) is very similar to the one of the SM
(neglecting gauge bosons and fermion masses other than $m_t$).
The only difference is that in our case the running parameters are constrained at the scale $\Lambda$:
The field $\phi$ has neither a kinetic term nor a $\phi^4$ self-interaction term at this scale.
This is known as the ``compositeness condition".

However, when integrating out high momentum degrees of freedom of the fields,
the $\phi$ kinetic and self-interaction terms are generated.
At energies $\mu< \Lambda$, $\phi$ becomes a dynamical field.
As we see in Chapter \ref{chapter:eff_potential},
this field describes a scalar particle with mass of about $2m_t$.
The Lagrangian at energies lower than $\Lambda$ is then given by

\begin{equation}\label{rg2}
\mathcal{L}^{(\mu)}=\mathcal{L}^{(\mu)}_\text{kin}
		+\frac12 Z_H \partial_\mu\phi\:\partial^\mu\phi
		-\frac12 m_H^2 \phi^2-\frac{g_t}{\sqrt{2}}\:\bar{t}t\phi
		-\frac{\lambda}{8} \phi^4,
\end{equation}
where $Z_H$, $m_H^2$, $g_t$, and $\lambda$ are running couplings.
The  $t_L$- and $t_R$-fields possess also wave-function normalization constants
$Z_{tL}$ and  $Z_{tR}$, respectively. 
They are equal to one in our approximation
(gauge couplings being turned off and $N\rightarrow\infty$).

The compositeness condition is given by

\begin{equation}\label{rg3}
Z_H=\lambda =0,
\end{equation} 
at $\mu=\Lambda$.
Besides we have at this scale
\begin{equation}\label{rg4}
m_H^2=m_{H_0}^2,\qquad g_t=g_{t_0}.
\end{equation} 

By calculating the Higgs two- and four-1PI point functions
(see Eq. (\ref{eff30})) we obtain the running couplings as a function of the scale $\mu$:

\begin{equation}\label{rg4b}
\begin{split}
Z_H(\mu)&\approx\frac{g_{t_0}^2 N}{16\pi^2}\;\log\Lambda^2/\mu^2,
\\
m_H^2(\mu)&\approx m_{H_0}^2-\frac{g_{t_0}^2 N}{8\pi^2}(\Lambda^2-\mu^2),
\\
g_t(\mu)&\approx g_{t_0},
\\
\lambda(\mu)& \approx \frac{g_{t_0}^4 N}{8\pi^2}\;\log\Lambda^2/\mu^2.
\end{split}
\end{equation} 
The running parameter $m_H^2(\mu)$ receives a quadratic contribution from the fermion loop.
This contribution makes $m_H^2(\mu)$ negative if 
$G_t>G_{\text{crit}} \equiv 8\pi^2/(N\Lambda^2)$, leading to EWSB.
We recover in this way the critical coupling $G_{\text{crit}}$ found in Chapter \ref{chapter:min}.

Before we make the crucial step of
abstracting the compositeness condition for the Higgs boson given in Eq. (\ref{rg3}) to boundary conditions in a perturbative RG analysis,
it is convenient to change the normalizations of the fields.
The RG equations are normally written using  different field normalizations as in Eq. (\ref{rg2}).
In the conventional normalization the fermionic and bosonic kinetic terms are normalized as in the free theory.
We redefine the fields in order to apply the RG equations written in the conventional normalization:

\begin{equation}\label{rg5}
\begin{split}
\phi&\longrightarrow \phi/\sqrt{Z_H},\\
t_L&\longrightarrow t_L/\sqrt{Z_{tL}},\\
t_R&\longrightarrow t_R/\sqrt{Z_{tR}}.
\end{split}
\end{equation} 
The Lagrangian is now given by

\begin{equation}\label{rg6}
\mathcal{L}^{(\mu)}=\mathcal{L}^{(\mu)}_\text{kin}
		+\frac12 \partial_\mu\phi\:\partial^\mu\phi
		-\frac12 \bar{m}_H^2 \phi^2-\frac{\bar{g}_t}{\sqrt{2}}\:\bar{t}t\phi
		-\frac{\bar{\lambda}}{8} \phi^4,
\end{equation}
where the overbar stands for the parameters in the conventional normalization:

\begin{equation}\label{rg7}
\begin{split}
\bar{m}_H^2&= \frac{m_H^2}{Z_H},
\\
\bar{g}_t&=\frac{g_t}{\sqrt{Z_H Z_{tL} Z_{tR}}},
\\
\bar{\lambda}&=\frac{\lambda}{Z_H^2}.
\end{split}
\end{equation} 
The compositeness condition Eq. (\ref{rg3}) is now given by

\begin{eqnarray}
\bar{g}_t \;\;    &\longrightarrow & \infty,      \label{rg8}\\
\bar{\lambda}/\bar{g}_t^4 &\longrightarrow & 0,      \label{rg9}
\end{eqnarray} 
as $\mu\rightarrow\Lambda$.\\

Now we demand Eqs. (\ref{rg8}) and (\ref{rg9}) to be the boundary conditions for the perturbative RG equations of the full theory.
The one-loop $\beta$ functions of the SM
(neglecting light quark masses and mixing angles) 
are given by

\begin{eqnarray}
16\pi^2\, \frac{\partial\bar{g}_t}{\partial t}&=&
 \left(\frac{9}{2}\,\bar{g}_t^2-8\,\bar{g}_3^2
	 -\frac{9}{4}\,\bar{g}_2^2
	 -\frac{17}{12}\,\bar{g}_1^2\right) \bar{g}_t,\label{rg10}\\
16\pi^2\, \frac{\partial\bar{g}_i}{\partial t}&=&
 -c_i\,\bar{g}_i^3,  \label{rg11}\\
16\pi^2\, \frac{\partial\bar{\lambda}}{\partial t}&=&
 12(\bar{\lambda}^2+(\bar{g}_t^2-A) \bar{\lambda}+B-\bar{g}_t^4),\label{rg12}
\end{eqnarray} 
where $t=\log\mu$ and $\bar{g}_i$, $i=1,2,3$, are the gauge running coupling constants.
The coefficients $c_i$, $A$, and $B$ are given by

\begin{equation}\label{rg13}
\begin{split}
&\qquad c_1=-41/6,\quad c_2=19/6,\quad c_3=7,\\
&A=\frac14\, \bar{g}_1^2+\frac34 \,\bar{g}_2^2,\quad
 B=\frac{1}{16}\, \bar{g}_1^4+\frac18\,\bar{g}_1^2\,\bar{g}_2^2
	+\frac{3}{16}\, \bar{g}_2^4.
\end{split}
\end{equation} 
In the original works \cite{Bardeen:1989ds,Luty:1990bg},
the RG equation related to the variable $m_H^2$ was not used in order to find its low energy value $m_H^2(0)$.
Instead of that,
motivated by the results obtained with the NJL approach to the model,
it was assumed that at $\mu= 0$  $m_H^2$ takes the necessary negative value in order to break the EW symmetry and to lead to the observed gauge boson masses, i.e. $v=246\;GeV$.
We follow the same reasoning here.

First we consider the RG equation for $\bar{g}_t$, Eq. (\ref{rg10}),
with the boundary condition $\bar{g}_t\rightarrow\infty$ as $\mu\rightarrow\Lambda$.
Solving this equation yields the running coupling $\bar{g}_t(\mu)$ and hence a prediction for the top-quark mass.
This is given by $m_t=\bar{g}_t(m_t)v/\sqrt{2}$, with $v=246\;GeV$.
If $\bar{g}_t$ is much bigger than the gauge couplings we can approximate  Eq. (\ref{rg10}) to

\begin{equation}\label{rg14}
16\pi^2\, \frac{\partial\bar{g}_t}{\partial t}=\frac{9}{2}\,\bar{g}_t^3.
\end{equation} 
This equation shows that starting at a scale $\mu\approx\Lambda$ with a big value of $\bar{g}_t$, this parameter decreases rapidly when evolving towards the infrared.
For numerical computations a large but finite value of $\bar{g}_t$ is chosen at the compositeness scale $\Lambda$. Because of Eq. (\ref{rg14}), the Landau pole is assumed to be located very near to this point.

Once $\bar{g}_t$ gets of the order of the gauge couplings the strong coupling $\bar{g}_3$ becomes also important.
Taking it into account we obtain

\begin{equation}\label{rg15}
16\pi^2\, \frac{\partial\bar{g}_t}{\partial t}=
 \Big(\frac{9}{2}\,\bar{g}_t^2-8\,\bar{g}_3^2\Big) \bar{g}_t.
\end{equation} 
From this equation we see that the SM infrared attractive fixed point \cite{Pendleton:1980as,Hill:1980sq}
is approximately given by the condition

\begin{equation}\label{rg16}
\bar{g}_t^2(\mu)\approx \frac{16}{9}\,\bar{g}_3^2(\mu).
\end{equation} 
The solution of the RG equations are attracted to this point.
Its location is modified once one considers all gauge couplings and their evolution equations.
The fixed point makes the low-energy prediction for $\bar{g}_t$ stable against variations of the scale $\Lambda$.
Solving the full one-loop RG equations for  $\bar{g}_t$ numerically,
one obtains predictions for the top-quark mass \cite{Bardeen:1989ds}.
These are shown in Table \ref{table:bhl_masses} for different values of the scale of new physics $\Lambda$.

Now we turn to the evolution of $\bar{\lambda}$.
This leads to  a prediction for the Higgs-boson mass which is given by
$m_h=v \sqrt{\bar{\lambda}(m_h)}$. 
For big values of  $\bar{\lambda}$ and $\bar{g}_t$ it is possible to neglect gauge couplings in Eqs. (\ref{rg10}) and (\ref{rg12}).
From these two equations we obtain the following evolution equation for 
$\bar{\lambda}/\bar{g}_t^2$:

\begin{equation}\label{rg17}
16\pi^2\, \frac{\partial\;}{\partial t} 
\left( \frac{\bar{\lambda}}{\bar{g}_t^2}\right) =
12 \bar{g}_t^2
\left( \frac{\bar{\lambda}}{\bar{g}_t^2}-x_-\right) 
\left( \frac{\bar{\lambda}}{\bar{g}_t^2}-x_+\right) ,
\end{equation} 
with $x_\pm = (-1\pm\sqrt{65})/8$.
This indicates the presence of an infrared attractive fixed point located at
$\bar{\lambda}/\bar{g}_t^2=x_+\approx 0.88$ 
(besides an ultraviolet attractive fixed point located at
$\bar{\lambda}/\bar{g}_t^2=x_-\approx -1.33$).
As for the running coupling  $\bar{g}_t$,
this ultraviolet fixed point makes the low-energy prediction for $\bar{\lambda}/\bar{g}_t^2$, and therefore for $\bar{\lambda}$,
stable against variations of $\Lambda$ and against variations of the initial 
 value (at $\mu\approx\Lambda$) of the ratio $\bar{\lambda}/\bar{g}_t^2$,
as long as the trajectories correspond to the infrared fixed point.
Numerical results for the Higgs-boson mass are shown in Table \ref{table:bhl_masses}.

\begin{table}[tbp]
\begin{equation*}
\begin{array}{lllllllllllll}
\hline \hline 
\Lambda\, (GeV) & 
10^{19}&10^{17}&10^{15}&10^{13}&10^{11}&
10^{10}&10^{9}&10^{8}&10^{7}&10^{6}&10^{5}&10^{4}\\
\hline
m_t \, (GeV) &  
218 & 223 & 229 & 237 & 248 & 255 & 264 & 277 & 293 & 318 & 360 & 455 \\
m_H \, (GeV) &  
239 & 246 & 256 & 268 & 285 & 296 & 310 & 329 & 354 & 391 & 455 & 605 \\
\hline\hline
\end{array} 
\end{equation*}
\caption{Predictions \cite{Bardeen:1989ds} for the top-quark and Higgs-boson masses using the SM one-loop RG equations and the compositeness condition for the Higgs boson at the scale $\Lambda$.}
\label{table:bhl_masses}
\end{table}

We make now some comments about the validity of the approach.
The RG equations which we use are obtained perturbatively and thus valid only for small values of the coupling constants.
On the other hand our boundary conditions, Eqs. (\ref{rg8}) and  (\ref{rg9}),
demand values of the coupling constants far away from this regime.
This in principle invalidates the whole analysis.
However, using the perturbative RG equations we see that only for a small ``running time" the coupling constants are not in the perturbative regime.
In addition, the predictions of the model are robust due to the presence of infrared attractive fixed points.
Lattice simulations also in general confirm the results of the one-loop RG analysis for the case when the trajectories correspond to these fixed points \cite{Shigemitsu:1989xb}.
Due to these considerations it is believed \cite{Hill:2002ap} that the non-perturbative effects are small and that the whole approach is reliable.

Unfortunately the predictions of the model are phenomenologically not acceptable.
In order to get the right value for the top mass the energy related to the appearance of new physics $\Lambda$ should be (by extrapolating Table \ref{table:bhl_masses})
much bigger than the Planck scale. \\

Next we consider models including a second Higgs doublet.
The RG equations related to these models allow to obtain a top mass 
$m_t\sim 180\; GeV$ and lead to predictions for the Higgs-boson masses which are phenomenologically acceptable.
The following analysis corresponds to a perturbative RG treatment of the model
we studied in Chapter \ref{chapter:3fam}.
The original four-fermion interaction given in Eq. (\ref{3f1}) can,
for a range of values of the coupling constants $G$,
be rewritten with help of two auxiliary Higgs doublets as shown in Eq. (\ref{3f2}).
In analogy to Eq. (\ref{rg2}),
in this case the low-energy Lagrangian is given by

\begin{equation}\label{rg20}
\begin{split}
\mathcal{L}^{(\mu)}&=\mathcal{L}^{(\mu)}_\text{kin}
		+ Z_{H_1} \partial_\mu H^{(1)\dagger}\:\partial^\mu H^{(1)}
		+ Z_{H_2} \partial_\mu H^{(2)\dagger}\:\partial^\mu H^{(2)}\\
  &\qquad	- m_{H_1}^2 H^{(1)\dagger}H^{(1)}
		- m_{H_2}^2 H^{(2)\dagger}H^{(2)}
		+\mathcal{L}_{\text{Yukawa}}    - V,
\end{split}
\end{equation}
where, neglecting Yukawa interactions related to the first and second quark generation, $\mathcal{L}_{\text{Yukawa}}$ is given by

\begin{equation}\label{rg21}
\mathcal{L}_{\text{Yukawa}} = -\sum_{i=1}^2\;
(\;g_t^{(i)}\:\bar{\psi}_{L}t_{R}\:H^{(i)}+
 h_b^{(i)}\:\epsilon^{ac}\:\bar{\psi}_{L}^a d_{R}\:H^{c(i)*}
\;+\; h.c.\;).
\end{equation}
The Higgs potential $V$ is given by (see Section \ref{section:lee_model})

\begin{equation}\label{rg22}
\begin{split}
V&=  
 \lambda_1 (H^{(1)\dagger}H^{(1)})^2
+\lambda_2 (H^{(2)\dagger}H^{(2)})^2    
+\lambda_3 (H^{(1)\dagger}H^{(1)})(H^{(2)\dagger}H^{(2)})   \\
&\;
+\lambda_4 (H^{(1)\dagger}H^{(2)})(H^{(2)\dagger}H^{(1)}) 
+[\lambda_5 (H^{(1)\dagger}H^{(2)})^2     
+\lambda_6 (H^{(1)\dagger}H^{(2)})(H^{(1)\dagger}H^{(1)})   \\
&\;\qquad\qquad\qquad
+\lambda_7 (H^{(1)\dagger}H^{(2)})(H^{(2)\dagger}H^{(2)})   + h.c.].
\end{split}
\end{equation}
The compositeness condition for the two Higgs doublet model is given by
the generalization of Eq. (\ref{rg3}):

\begin{equation}\label{rg23}
Z_{H_1}=Z_{H_2}=\lambda_i =0,
\qquad\qquad \text{with }i=1,\dots,7,
\end{equation} 
at $\mu=\Lambda$.
Besides, at this scale the Higgs mass parameters and Yukawa couplings are finite.

As far as we know the compositeness conditions have not been studied in a RG group analysis of the general two Higgs doublet model given in Eq. (\ref{rg20}).
Only a special case, the two Higgs doublet type II model has been studied
\cite{Luty:1990bg,Froggatt:1990wa,Froggatt:1992wt,Mahanta:1991jc,Cvetic:1993xg}. 
The fact that in the type II model acceptable top-quark and Higgs masses can be obtained, strongly support that in the general case which correspond to the model studied in Chapter \ref{chapter:3fam}  this is also the case.
The Lagrangian of the two Higgs doublet type II model is given by

\begin{equation}\label{rg24}
\begin{split}
\mathcal{L}^{(\mu)}&=\mathcal{L}^{(\mu)}_\text{kin}
		+ Z_{H_t} \partial_\mu H_t^{\dagger}\:\partial^\mu H_t
		+ Z_{H_b} \partial_\mu H_b^{\dagger}\:\partial^\mu H_b\\
  &\qquad	- \mu_t^2 H_t^{\dagger}H_t
		- \mu_b^2 H_b^{\dagger}H_b
		- \mu_{tb}^2 (H_t^{\dagger}H_b+h.c.)
		+\mathcal{L}_{\text{Yukawa}}    - V,
\end{split}
\end{equation}
where $\mu_t^2$, $\mu_b^2$, and $\mu_{tb}^2$ are the Higgs mass parameters.
The fields $H_t$, $H_b$ are linear combinations of the fields 
$H^{(1)}$, $H^{(2)}$. The Yukawa interaction is given by

\begin{equation}\label{rg24b}
\mathcal{L}_{\text{Yukawa}} = -(
 g_t\:\bar{\psi}_{L}t_{R}\:H_t
+h_b\:\epsilon^{ac}\:\bar{\psi}_{L}^a d_{R}\:H_b^{c*}
\;+\; h.c.\;).
\end{equation}
The field $H_t$ couples only to the quark field $t_{R}$, while
the field $H_b$ couples only to the quark field $b_{R}$,
leading to FCNC suppression.

At the scale $\Lambda$ the Lagrangian of the theory almost respects the PQ symmetry
(see Section \ref{section:pq_sym}).
Only the term proportional to $\mu_{tb}^2$ explicitly violates it.
In a mass independent renormalization scheme the RG equations for the dimensionless parameters do not depend on  $\mu_{tb}^2$ and therefore these parameters evolve as if the PQ symmetry were exact.
In this way some parameters of the Higgs potential are zero ($\lambda_5=\lambda_6=\lambda_7=0$) for all energies\footnote{Note
that the previous argument relies strongly on the applicability of the one-loop RG equations at all energies.} 
$\mu \leq\Lambda$,
simplifying the analysis.

An important point arises by the definition of the compositeness condition as a boundary condition for the RG analysis.
In the conventional normalization the conditions for the Yukawa couplings are
$\bar{g}_t$, $\bar{h}_b\rightarrow\infty$ as $\mu\rightarrow\Lambda$.
As we mentioned before, in the numerical calculations large but finite values are chosen for the diverging couplings at the compositeness scale $\Lambda$.
It happens, however, that low-energy predictions,
in particular the top mass, are very sensitive to the ratio
$\bar{g}_t(\Lambda)/\bar{h}_b(\Lambda)$. For example for\footnote{For
arguments supporting this choice see \cite{Cvetic:1997eb}.}
$\bar{g}_t(\Lambda)/\bar{h}_b(\Lambda)=1$,
values for $m_t$ which are about 10\% lower than the ones of the one Higgs doublet analysis (Table \ref{table:bhl_masses}) are obtained \cite{Luty:1990bg}.
These top mass predictions are still too high in order to be phenomenologically acceptable.
Nevertheless, allowing other values for the ratio
$\bar{g}_t(\Lambda)/\bar{h}_b(\Lambda)$
a realistic top mass can be obtained
\cite{Froggatt:1990wa,Froggatt:1992wt,Cvetic:1993xg}, see also \cite{Cvetic:1997eb}.
For example choosing a Higgs VEV ratio $v_u/v_d=0.5$, $1.0$, or $5$,
a top mass $m_t\approx 180\; GeV$ is found
(with different values of the scale $\Lambda$ for each Higgs VEV ratio) \cite{Cvetic:1993xg}.
In this calculation the boundary conditions at the scale $\Lambda$ are given by
$Z_u(\Lambda)=0$ and $Z_d(\Lambda)\sim 1$.
The Yukawa couplings in the conventional normalization are related to the Higgs wave-function normalization constants by the relations 
$\bar{g}_t^2\sim 1/Z_t$ and $\bar{h}_b^2\sim 1/Z_b$.
Thus both Yukawa couplings become non-perturbative but not exactly at the same point\footnote{Due to the fact that $Z_d(\Lambda)$ is not zero, doubts about the composite nature of the Higgs doublet $H_d$ can arise \cite{Cvetic:1997eb}. However it is difficult to give definitive statements about that, because at this scale the perturbative treatment is not fully reliable.}.
Predictions\footnote{Corrections due to heavy Higgs boson threshold effects were not taken into account.} \cite{Luty:1990bg} for the masses of the 5 physical Higgs particles 
for $\Lambda=10^{15}\; GeV$ and the parameter $\alpha=0.012$
(see \cite{Luty:1990bg}) are given by

\begin{equation}
\begin{split}
m_{H_1}=& \; m_{H_P}= 230\; GeV,\\
m_{H_2}&= 270\; GeV,\\
m_{H^\pm}&= 350\; GeV.
\end{split}
\end{equation}

Another possibility investigated in the literature consists in including Majorana neutrinos \cite{Antusch:2002xh,Martin:1991xw}.
The essential idea of these models is the following.
One assumes a four-fermion interaction that involves not only quarks but also leptons, and large Majorana mass terms for the right-handed neutrinos.
For critical four-fermion couplings, condensation of neutrino bilinears (and also of top-quark bilinears) occurs.
In this way Dirac neutrino mass terms of the order of the EW scale are dynamically generated.
Then by means of the see-saw mechanism small physical neutrino Majorana masses are obtained.
The compositeness condition has been  studied in models of this type considering one and two Higgs doublets \cite{Antusch:2002xh,Martin:1991xw}.
The RG equations are modified in comparison to the ones we considered before, 
due to the presence of the large Dirac neutrino mass terms
(and hence large neutrino Yukawa couplings),
and due to the inclusion of the dimension 5 Majorana neutrino operator.
The value of the top mass is obtained in models having one or two Higgs doublets.
In the one Higgs case the value of the Higgs mass is predicted to lay between $170 \; GeV$ and $195 \; GeV$,
 for $\Lambda=10^{16}\; GeV$ and for values of the see-saw scale
 between $10^{14} \; GeV$ and $10^{15.5} \; GeV$.

\chapter{Summary and Conclusions}
\label{chapter:conclusion}

We studied in this thesis top-condensation models of electroweak symmetry breaking.
In these models a four-fermion interaction term involving  SM fermions is postulated at some high energy scale which we called $\Lambda$.
The strong, attractive four-fermion interaction induces the formation of color-singlet fermion-antifermion bound states,
in particular in spin-zero states which correspond to composite Higgs bosons, and their condensation.
These condensates transform non-trivially under $SU(2)_L\times U(1)_Y$ transformations and therefore break the EW symmetry.
As a consequence gauge bosons and fermions acquire masses.
The major goals of this type of models are
to explain the origin of the EW scale,
to dynamically generate the gauge boson and fermion masses,
and, as we investigated in this thesis,
to generate quark mixing angles and the $CP$ violating phase of the CKM matrix.
Specifically, we started with a $CP$-invariant Lagrangian,
and we showed that in our model the $CP$-violating phase of the CKM matrix was generated through complex VEVs of the composite Higgs fields.

In order to study top-condensation models the NJL approach is widely adopted in the literature.
In this thesis we applied this method first to models involving only quarks of the third generation, and then to the physically relevant model with three quark generations.\\

Two possible scenarios concerning the origin of the four-fermion interaction term can be considered.
In the traditional one,
one starts with a non-abelian gauge theory at very high energies $E\gg \Lambda$ which
becomes strongly-coupled at scales $\sim\Lambda$.
For energies below $\Lambda$ the new interaction is effectively described by operators constructed with the fields corresponding to the light
($m_\text{particle}<\Lambda$) degrees of freedom of the theory.
The new interaction must violate the flavor symmetry,
i.e. must be non-universal, in order to generate the observed fermion mass pattern.
Topcolor models \cite{Hill:1991at} are examples of  a theory of this type.

At low energies the most important non-renormalizable operators are the ones having the lowest mass dimension.
Therefore dimension-six four-fermion operators were considered.
Besides, only four-fermion operators made of (pseudo)scalar fermion bilinears were taken into account
(they lead to (pseudo)scalar composite fields),
ignoring the ones made of (axial)vector fermion bilinears.
Note that the distinction between (pseudo)scalar and (axial)vector bilinears is ambiguous due to Fierz identities.

In the second scenario the four-fermion interaction term acquires a more fundamental status.
It is assumed that
the SM with the Higgs sector being replaced by a general dimension-six four-fermion interaction is a (non-perturbatively) renormalizable theory \cite{Gies:2003dp}.
This is the case if one or more non-Gaussian ultraviolet stable fixed points are established  beyond the point-like approximation  \cite{Gies:2003dp}.\\

In this thesis the four-fermion interaction term was used as a starting point.
Therefore, besides the four-fermion and the SM gauge couplings,
the scale $\Lambda$ at which the whole Lagrangian was defined, is also a parameter of the theory.
If $\Lambda$ is much bigger than the electroweak scale,
complementary to the NJL approach
a perturbative RG analysis can be made.
This method, which incorporates the SM gauge interactions, provides reliable values for the top-quark and Higgs-boson masses.
In this approach the information that composite Higgs doublets appear at the scale $\Lambda$ is encoded in the compositeness condition.
This condition is used as a boundary condition at the scale $\Lambda$ in the RG analysis.

In the two-Higgs doublet type II model the RG analysis leads to phenomenologically acceptable Higgs masses, with values of about $230\; GeV$ and higher (see end of Chapter \ref{chapter:composite_condition}).
In addition the correct mass of top quark is obtained
if the ratio of the Yukawa couplings at the scale $\Lambda$ is taken different from 1.
These results strongly support that for the general two-Higgs doublet model
the compositeness condition in a RG analysis also leads to phenomenologically acceptable mass values.
If the $CP$-violating phase of the CKM matrix is spontaneously generated by the condensation of two-Higgs doublet fields,
then this necessary happens within the general two-Higgs doublet model.

To have a very high scale $\Lambda$ is, however, not very attractive because the  theory suffers from fine-tuning in exactly the same way as the SM.
Another important point is the one related to the distinction between fundamental and  composite Higgs particles.
If the compositeness scale $\Lambda$ is very high,
the composite nature of these particles cannot be directly verified by experiments in the near future.

A more interesting possibility is to have a scale $\Lambda$ not very much higher than the EW scale, $\Lambda\sim 5-10\; TeV$.
In this case no fine-tuning problem appears.
Besides, the generation of the scale  $\Lambda$ could be explained from dimensional transmutation.
This would solve the hierarchy problem.
A perturbative RG analysis cannot be justified in this case.
Topcolor assisted technicolor \cite{Hill:1994hp},
and top-quark see-saw \cite{Dobrescu:1997nm,Chivukula:1998wd} are examples of  theories of this type.
In these theories the NJL approach is widely used. \\

We found that if the dimension-six four-fermion Lagrangian includes only quarks of the third family, then  spontaneous $CP$ symmetry breaking does not occur.
This model possesses three four-fermion coupling constants.
We considered the case where the EW symmetry is spontaneously broken,
giving both the top and bottom quarks their respective masses.
We distinguished between two possibilities.
For $G_{tb}=0$ the Lagrangian has a Peccei-Quinn symmetry and therefore the theory predicts an axion.
An axion related to the EW scale is, however, ruled out.
For $G_{tb}\neq 0$ the spectrum of the theory includes only one scalar Higgs particle with mass about $2m_t$.
It corresponds to a one-Higgs doublet model and not to a two-Higgs doublet model as one could naively assume.
In this model there are modes which are associated with masses of order  $\Lambda$,
which is, however, the upper limit of the validity range of the model.
Therefore these modes cannot be interpreted as particles.\\

Finally we summarize the results obtained in a model with a dimension-six four-quark interaction term involving the three quark generations.
All four-fermion couplings were considered to be real in order to have a $CP$-invariant Lagrangian.
We restricted our analysis to the case where the theory can be rewritten in terms of two auxiliary Higgs fields.
In this case the number of independent four-quark coupling constants is 36.
We found by means of a self-consistent approach a range for the four-quark couplings  where spontaneous $CP$ symmetry breaking occurs
together with EWSB.
Specifically, complex VEVs of the two composite Higgs doublets lead to a realistic $CP$-violating CKM matrix.
Besides, masses for all quarks and weak gauge bosons are generated.
We also showed that the vacuum respects the electromagnetic $U(1)$ symmetry.

The model predicts five physical spin-zero bound states:
three neutral composite Higgs bosons and one charged boson and its charge conjugated state.
The masses of the  composite Higgs bosons were also calculated leading in part to very low (non-acceptable) values.
Using the Nambu-Jona-Lasinio method,
we obtained for one neutral state a mass value of about $2m_t$.
However, the masses of the other states obtained with this approximation turn out to be unacceptably low.
This is not surprising because this approximation which is the method of choice for investigating the ground state of the model is not reliable for calculating the Higgs masses, because the effect of SM gauge interactions is not and the effect of Higgs self-interactions is only partly
(i.e. only the part induced by fermion loops in the $N\rightarrow\infty$ approximation) taken into account.
If $\Lambda$ is very high the RG analysis,
which incorporates these interactions, provides a more reliable prediction for these masses.
As we saw an analysis of this type leads to acceptable Higgs boson masses.
In models with $\Lambda$ being not too much higher than the EW scale,
theories like topcolor assisted technicolor provide a mechanism for increasing the value of the composite Higgs boson masses \cite{Hill:1994hp}.

Another aspect of our model is related to
flavor-changing neutral currents (FCNCs).
We saw that Higgs mediated FCNCs are necessarily present if the $CP$-violating CKM phase
is spontaneously generated.
On the other hand FCNCs are strongly constrained especially in the down-quark sector.
It is possible, however, to construct a model having non-diagonal Yukawa matrices only in the up-sector.
An open question is whether a model with FCNC connected only to the top quark (or better, only between the top and charm quarks) can be consistent.
More importantly, the question how FCNCs are suppressed in top-condensation models remains open.

\begin{appendix}
\chapter{Surface Terms in the Spherical Cutoff Regularization}\label{appendix:surface_terms}

In this Appendix we show,
using as example the minimal scheme treated in Chapter \ref{chapter:min}, 
how surface terms affect the position of the poles
in some correlation functions.
In particular we see that the mass of the composite Higgs boson is ambiguous, 
while the Goldstone bosons remain massless, as they should. \\
A Feynman integral defined with a spherical cutoff as regulator, 
is in general not invariant under a shift of the integration variables.
Convergent or logarithmically divergent integrals are invariant under this operation, 
while linearly or more than linearly divergent integrals are not.
In this case additional surface terms appear \cite{JAUCH-ROHRLICH55,Willey:1993cp}.
For linearly and quadratically divergent integrals the following surface terms appear

\begin{eqnarray}
\int \frac{d^4l}{(2\pi)^4}\;\frac{l^\mu}{((l-p)^2 -a^2)^2}&=&     \label{ap:st1} 
\int \frac{d^4l}{(2\pi)^4}\;\frac{l^\mu + p^\mu}{(l^2 -a^2)^2}-\frac{ip^\mu}{32\pi^2}, \\
\int \frac{d^4l}{(2\pi)^4}\;\frac{1}{(l-p)^2 -a^2}&=&         \label{ap:st2} 
\int \frac{d^4l}{(2\pi)^4}\;\frac{1}{l^2 -a^2}-\frac{ip^2}{32\pi^2}.
\end{eqnarray} 
We now calculate the poles of the correlation functions 
(calculated in Section \ref{section:min_poles}),
considering this time the surface terms.
In the scalar channel the loop integral $I_s$ defined in Eq. (\ref{min16}) 
is in general given by

\begin{figure}[b]
\begin{center}
\psfig{file=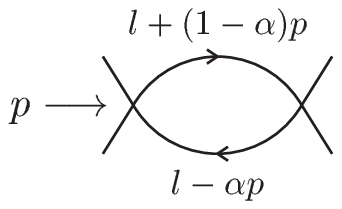,width = 4.5cm}
\end{center}
\caption{$\alpha$ dependence of the momentum flow in the internal lines.}
\label{fig:alpha}
\end{figure}

\begin{equation}\label{ap:st3}
I_s=\int\frac{d^4l}{(2\pi)^4}\tr\left( \frac{i}{l\contract -\alpha p\contract -m_t} 
\;\;\frac{i}{l\contract +(1-\alpha)p\contract -m_t}\right),
\end{equation}
where the value of $\alpha$ in the loop, see Fig. \ref{fig:alpha}, 
is not determined by the theory.
For a regulator which allows a shift in the integration variable $l$ there is no ambiguity
in the definition of $I_s$, 
but in the case of a spherical cutoff regulator the integral $I_s$ depends on $\alpha$.
Using Eq. (\ref{ap:st2}) we obtain a new expression for the scalar four-point function
(assuming that all the momentum loops in Fig. \ref{fig:four-point} have the same $\alpha$-prescription) 

\begin{equation}\label{ap:st4}
i\mathcal{M}_s= \frac{-i}{2N} \left( 
\frac{1}{\Delta(\alpha) p^2+(p^2-4m_t^2) B(p^2)}  \right) ,
\end{equation} 
with

\begin{equation}\label{ap:st5}
\Delta(\alpha)=\frac{1-2\alpha+2\alpha^2}{32\pi^2},
\end{equation} 
and
\begin{equation}\label{ap:st6}
B(p^2)= -i\int \frac{d^4l}{(2\pi)^4}\;\frac{1}{(l^2 -m_t^2)((l+p)^2 -m_t^2)}
\equiv c(p^2)\log(\Lambda^2/m_t^2),
\end{equation} 
with $c(p^2)>0$.
Because $\Delta(\alpha)$ is always bigger than $1/(64\pi^2)$ the pole in (\ref{ap:st4}) is never at $p^2=4m_t^2$.
The position of the pole is now given by

\begin{equation}\label{ap:st7}
p^2=4m_t^2 
	\left( 1+\frac{\Delta(\alpha)}{c(p^2)\log(\Lambda^2/m_t^2)}\right)^{-1}.
\end{equation}
The mass of the composite boson is therefore always smaller than $4m_t^2$ if one takes
surface terms into account.\\
Now we turn to the other channels, which are associated with the Goldstone bosons.
The pseudoscalar amplitude is given now by

\begin{equation}\label{ap:st8}
i\mathcal{M}_p= \frac{-i}{2Np^2} \left( 
\frac{1}{\Delta(\alpha) +B(p^2)} \right) .
\end{equation} 
One sees that the position of the pole does not change.
Finally in the charged channel we find that the position of the pole also
remains at $p^2=0$.
After summing the geometric series we obtain (see Fig.\ref{fig:four-point})

\begin{equation}\label{ap:st9}
i\mathcal{M}_c= \frac{i G_t}{4} \left(1+\frac{i G_t N I_c}{4}\right)^{-1}.
\end{equation} 
The integral $I_c$ is now given by

\begin{eqnarray}
I_c &=& \int\frac{d^4l}{(2\pi)^4}\tr\left(
(1+\gamma_5) \frac{i}{l\contract -\alpha p\contract -m_t} 
(1-\gamma_5) \frac{i}{l\contract +(1-\alpha)p\contract }\right), \nonumber \\
&=&  -8 \int\frac{d^4l}{(2\pi)^4}\frac{1}{l^2-m_t^2}
+4 i p^2\Delta(\alpha)                             \label{ap:st10}
+\int\frac{d^4l\, dx}{(2\pi)^4}\frac{8x p^2}{(l^2-xm_t^2+x(1-x)p^2)^2},
\end{eqnarray}
where we introduced a Feynman parameter $x$.
Replacing $I_c$ in Eq. (\ref{ap:st9}) we obtain

\begin{equation}\label{ap:st11}
i\mathcal{M}_c= \frac{-i}{4 N p^2} \left(
\Delta(\alpha)
-2i \int\frac{d^4l\, dx}{(2\pi)^4}\frac{x}{(l^2-xm_t^2+x(1-x)p^2)^2}\right)^{-1}.
\end{equation} 
In conclusion, taking the surface terms (\ref{ap:st1}) and (\ref{ap:st2}) into account
does not alter the position of the poles at $p^2=0$
in the pseudoscalar and charged amplitudes.  
However, in the scalar amplitude the position of the pole changes.

\end{appendix}

\thispagestyle{empty}

\LARGE
\noindent
Danksagung\\

\normalsize

\vspace{2.5cm}

Ich m\"ochte mich in erster Linie und
vor allem  bei Herrn Prof. Dr. W. Bernreuther
f\"ur seine Betreuung bedanken.
Seit unseren ersten e-mails vor etwa f\"unf Jahren
habe ich stets Unterst\"utzung von ihm bekommen.
W\"ahrend meiner Promotion hatte ich die M\"oglichkeit,
wissenschaftlich viel von ihm zu lernen.
Ich muss auch sagen, da{\ss} seinerseits eine grosse Dosis Geduld n\"otig war.\\

Ich bedanke mich auch bei Herrn  Prof. Dr. L. Sehgal f\"ur die \"Ubernahme des Korreferats.\\

Zahlreiche Diplomanden, Doktoranden  und Postdocs begleiteten mich w\"ahrend meines Aufenthalts in Aachen.
Ich bedanke mich bei ihnen f\"ur die netten, zusammen verbrachten Stunden.\\

Herzlichen Dank auch den Sekret\"arinnen des Lehrstuhls E,
Frau S. Chang und Frau A. Bachtenkirch,
f\"ur ihre Hilfe und Sympathie.\\

Zuletzt  m\"ochte ich mich beim Deutschen Akademischen Austauschdienst (DAAD)
f\"ur die finanzielle F\"orderung bedanken.


\end{document}